\documentclass[twoside]{cernyrep}

\usepackage[utf8]{inputenc}
\DeclareUnicodeCharacter{2212}{-}
\counterwithin{table}{section}
\counterwithin{figure}{section}
\counterwithin{equation}{section}

\usepackage{comment} 
\usepackage{amsmath}
\usepackage{color}
\usepackage{xcolor}

\usepackage{graphicx}
\usepackage{caption}
\usepackage{subfig}
\usepackage{subfloat}
\usepackage{multirow}
\usepackage{booktabs}
\usepackage{adjustbox}
\usepackage{rotating}
\usepackage{placeins}
\usepackage{makecell}

\usepackage{siunitx}
\DeclareSIUnit\clight{\text{c}}
\DeclareSIUnit\barn{b}
\DeclareSIUnit\bar{bar}

\usepackage{wasysym}

\usepackage[backend=biber,sorting=none]{biblatex}

\usepackage{pdfpages}

\usepackage{epstopdf}
\usepackage{acronym}
\usepackage{balance}
\usepackage{authblk}
\usepackage{fancyhdr}
\usepackage{afterpage}
\usepackage{csvsimple}
\usepackage{pgfsys}

\usepackage{todonotes}

\usepackage{tcolorbox}

\usepackage{emptypage}
\usepackage{lipsum}

\usepackage{texnames}
\usepackage{ctable}
\usepackage[T1]{fontenc}

\usepackage{blindtext}

\usepackage[bookmarks, colorlinks=true, linktoc=page, pdftex, linkcolor=blue, citecolor=blue, urlcolor=blue]{hyperref}
\usepackage{float}
\usepackage{xcolor}
\usepackage[toc,page]{appendix}
\usepackage[bookmarks, colorlinks=true, pdftex, linkcolor=blue, citecolor=blue, urlcolor=blue]{hyperref}

\usepackage{times,lipsum}
\usepackage[margin=1in]{geometry}
\usepackage[onehalfspacing]{setspace}

\usepackage{arydshln} %for dashed lins in tables         % Includes all packages from this folder

\usepackage[hang,flushmargin]{footmisc}
\fancypagestyle{plain}{}
\newlength{\oddmarginwidth}
\newlength{\evenmarginwidth}
\fancyhfoffset[LO,RE]{\oddmarginwidth}
\fancyhfoffset[LE,RO]{\evenmarginwidth}
\bibliography{main} % Add citations to go at the end of the document
\newcommand{\xdownarrow}[1]{%
  {\left\downarrow\vbox to #1{}\right.\kern-\nulldelimiterspace}
}

\begin{document}
\pagenumbering{roman}
\setcounter{page}{1}
\thispagestyle{empty}
\setlength{\unitlength}{1mm}



# IMCC authors

Carlotta Accettura[1], Simon Adrian[2], Rohit Agarwal[3], Claudia Ahdida[1], Chiara Aime'[4,5], Avni Aksoy[6,1], Gian Luigi Alberghi[7], Simon Albright[1], Siobhan Alden[8], Luca Alfonso[9], Muhammad Ali[10,11], Anna Rita Altamura[12,13], Nicola Amapane[13,12], Kathleen Amm[14], David Amorim[15,1], Paolo Andreetto[16], Fabio Anulli[17], Ludovica Aperio Bella[18], Rob Appleby[19], Artur Apresyan[20], Pouya Asadi[21], Mohammed Attia Mahmoud[22], Bernhard Auchmann[23,1], John Back[24], Anthony Badea[25], Kyu Jung Bae[26], E.J. Bahng[27], Lorenzo Balconi[28,29], Fabrice Balli[30], Laura Bandiera[31], Carmelo Barbagallo[1], Daniele Barducci[32,5], Roger Barlow[33], Camilla Bartoli[34], Nazar Bartosik[12], Emanuela Barzi[20], Fabian Batsch[1], Matteo Bauce[17], Michael Begel[35], J. Scott Berg[35], Andrea Bersani[9], Alessandro Bertarelli[1], Francesco Bertinelli[1], Alessandro Bertolin[16], Pushpalatha Bhat[20], Clarissa Bianchi[34], Michele Bianco[1], William Bishop[24,36], Kevin Black[37], Fulvio Boattini[1], Alex Bogacz[38], Maurizio Bonesini[39], Bernardo Bordini[1], Patricia Borges de Sousa[1], Salvatore Bottaro[40], Luca Bottura[1], Steven Boyd[24], Johannes Braathen[18], Marco Breschi[34,7], Francesco Broggi[29], Matteo Brunoldi[41,4], Xavier Buffat[1], Laura Buonincontri[11,16], Marco Buonsante[42,10], Philip Nicholas Burrows[43], Graeme Campbell Burt[44,45], Dario Buttazzo[5], Barbara Caiffi[9], Ilkay Turk Cakir[6], Orhan Cakir[6], Rama Calaga[1], Sergio Calatroni[1], Marco Calviani[1], Simone Calzaferri[41], Daniele Calzolari[1,16], Kyle Capobianco-Hogan[146], Vieri Candelise[46,46], Silvio Candido[1], Ali Can Canbay[6], Claudio Cantone[47], Rodolfo Capdevilla[20], Christian Carli[1], Carlo Carrelli[48], Fausto Casaburo[49,17], Massimo Casarsa[46], Luca Castelli[49,17], Maria Gabriella Catanesi[10], Lorenzo Cavallucci[34,7], Gianluca Cavoto[49,17], Francesco Giovanni Celiberto[50], Luigi Celona[51], Alessia Cemmi[48], Sergio Ceravolo[47], Alessandro Cerri[52,53,5], Francesco Cerutti[1], Gianmario Cesarini[47], Cari Cesarotti[54], Antoine Chancé[30], Nikolaos Charitonidis[1], mauro chiesa[4], Paolo Chiggiato[1], Vittoria Ludovica Ciccarella[47,49], Pietro Cioli Puviani[55], Anna Colaleo[42,10], Francesco Colao[48], Francesco Collamati[17], Marco Costa[56], Nathaniel Craig[57], David Curtin[58], Laura D'Angelo[59], Giacomo Da Molin[60], Heiko Damerau[1], Sridhara Dasu[37], Jorge de Blas[61], Stefania De Curtis[62], Herbert De Gersem[59], Andre de Gouvea[63], Tommaso Del Moro[49,48], Jean-Pierre Delahaye[1], Dmitri Denisov[35], Haluk Denizli[64], Radovan Dermisek[65], Paula Desiré Valdor[1], Charlotte Desponds[1], Luca Di Luzio[16], Elisa Di Meco[47], Karri Folan Di Petrillo[25], Ilaria Di Sarcina[48], Eleonora Diociaiuti[47], Tommaso Dorigo[16,66], Karlis Dreimanis[67], Tristan du Pree[68,69], Hatice Duran Yildiz[6], Juhi Dutta[70], Thomas Edgecock[33], Mamad Eshraqi[71,72], Siara Fabbri[1], Marco Fabbrichesi[46], Stefania Farinon[9], Davide Fazioli[1], Javier Fernandez Roncal[1], Guillaume Ferrand[30], Samuel Ferraro[73], Jose Antonio Ferreira Somoza[1], Marco Ferrero[12], Max Fieg[74], Frank Filthaut[75,68], Patrick Fox[20], Roberto Franceschini[76,77], Rui Franqueira Ximenes[1], Frank Gaede[18], Simone Galletto[12,13], Michele Gallinaro[60], Maurice Garcia-Sciveres[3], Luis Garcia-Tabares[78], Rocky Bala Garg[79], Ruben Gargiulo[49], Cedric Garion[1], Maria Vittoria Garzelli[80], Marco Gast[81], Lisa Generoso[42,10], Cecilia E. Gerber[82], Luca Giambastiani[11,16], Alessio Gianelle[16], Eliana Gianfelice-Wendt[20], Stephen Gibson[8], Simone Gilardoni[1], Dario Augusto Giove[29], Valentina Giovinco[1], Carlo Giraldin[16,11], Alfredo Glioti[17], Arkadiusz Gorzawski[71,1], Mario Greco[77], Christophe Grojean[18], Alexej Grudiev[1], Edda Gschwendtner[1], Emanuele Gueli[17,17], Nicolas Guilhaudin[1], Tao Han[83], Chengcheng Han[84], John Michael Hauptman[27], Matthew Herndon[37], Adrian D Hillier[36], Micah Hillman[85], Gabriela Hoff[86], Tova Ray Holmes[85], Samuel Homiller[87], Walter Hopkins[88], Lennart Huth[18], Sudip Jana[89], Laura Jeanty[21], Sergo Jindariani[20], Sofia Johannesson[71], Benjamin Johnson[85], Owain Rhodri Jones[1], Paul-Bogdan Jurj[90], Yonatan Kahn[20], Rohan Kamath[90], Anna Kario[69], Ivan Karpov[1], David Kelliher[36], Wolfgang Kilian[91], Ryuichiro Kitano[92], Felix Kling[18], Antti Kolehmainen[1], K.C. Kong[93],

Jaap Kosse[23], Jakub Kremer[18], Georgios Krintiras[93], Karol Krizka[94], Nilanjana Kumar[95], Erik Kvikne[1], Robert Kyle[96], Stephan Lachnit[18], Emanuele Laface[71], Elleanor Lamb[1], Kenneth Lane[97], Andrea Latina[1], Anton Lechner[1], Lawrence Lee[85], Junghyun Lee[26], Seh Wook Lee[26], Thibaut Lefevre[1], Emanuele Leonardi[17], Giuseppe Lerner[1], Gabriele Levati[98], Filippo Levi[9], Peiran Li[99], Qiang Li[100], Tong Li[101], Wei Li[102], Roberto Li Voti[49,47], Giulia Liberalato[46], Mats Lindroos[†,71], Ronald Lipton[20], Da Liu[83], Zhen Liu[99], Miaoyuan Liu[103], Alessandra Lombardi[1], Shivani Lomte[37], Kenneth Long[90,36], Luigi Longo[10], José Lorenzo[104], Roberto Losito[1], Ian Low[63,88], Xianguo Lu[24], Donatella Lucchesi[11,16], Tianhuan Luo[3], Anna Lupato[11,16], Yang Ma[105], Shinji Machida[36], Edward MacTavish[1], Thomas Madlener[18], Lorenzo Magaletti[106,10,106], Marcello Maggi[10], Tommaso Maiello[9], Helene Mainaud Durand[1], Abhishikth Mallampalli[37], Fabio Maltoni[105,34,7], Jerzy Mikolaj Manczak[1], Marco Mandurrino[12], Claude Marchand[30], Francesco Mariani[29,49], Stefano Marin[1], Samuele Mariotto[28,29], Simon Marsh[1], Stewart Martin-Haugh[36], David Marzocca[46], Maria Rosaria Masullo[107], Giorgio Sebastiano Mauro[51], Anna Mazzacane[20], Andrea Mazzolari[31,108], Patrick Meade[109], Barbara Mele[17], Federico Meloni[18], Xiangwei Meng[110], Matthias Mentink[1], Rebecca Miceli[34], Natalia Milas[71], Abdollah Mohammadi[37], Dominik Moll[59], Francesco Montagno Bozzone[111,112], Alessandro Montella[113], Manuel Morales-Alvarado[46], Mauro Morandin[16], Marco Morrone[1], Tim Mulder[1], Riccardo Musenich[9], Toni Mäkelä[74], Elias Métral[1], Krzysztof Mękała[114,18], Emilio Nanni[79,115], Marco Nardecchia[49,17], Federico Nardi[11], Felice Nenna[11,10], David Neuffer[20], David Newbold[36], Daniel Novelli[9,49], Maja Olvegård[116], Yasar Onel[117], Domizia Orestano[76,77], Inaki Ortega Ruiz[1], John Osborne[1], Simon Otten[69], Yohan Mauricio Oviedo Torres[86], Daniele Paesani[47,1], Simone Pagan Griso[3], Davide Pagani[7], Kincso Pal[1], Mark Palmer[35], Leonardo Palombini[16], Alessandra Pampaloni[9], Paolo Panci[5,32], Priscilla Pani[18], Yannis Papaphilippou[1], Rocco Paparella[29], Paride Paradisi[11,16], Antonio Passeri[77], Jaroslaw Pasternak[90,36], Nadia Pastrone[12], Kevin Pedro[20], Antonello Pellecchia[10], Fulvio Piccinini[4], Henryk Piekarz[20], Tatiana Pieloni[15], Juliette Plouin[30], Alfredo Portone[104], Karolos Potamianos[24], Joséphine Potdevin[15,1], Soren Prestemon[3], Teresa Puig[118], Ji Qiang[3], Lionel Quettier[30], Tanjona Radonirina Rabemananjara[119,68], Emilio Radicioni[10], Raffaella Radogna[10,42], Ilaria Carmela Rago[17], Angira Rastogi[3], Andris Ratkus[67], Elodie Resseguie[3], Juergen Reuter[18], Pier Luigi Ribani[34], Cristina Riccardi[41,4], Stefania Ricciardi[36], Caroline Riggall[85], Tania Robens[120], Youri Robert[1], Chris Rogers[36], Juan Rojo[68,119], Marco Romagnoni[108,31], Kevin Ronald[96,45], Benjamin Rosser[25], Carlo Rossi[1], Lucio Rossi[28,29], Leo Rozanov[25], Maximilian Ruhdorfer[79], Richard Ruiz[121], Farinaldo S. Queiroz[86,122], Saurabh Saini[52,1], Filippo Sala[34,7], Claudia Salierno[34], Tiina Salmi[123], Paola Salvini[4,41], Ennio Salvioni[52], Nicholas Sammut[124], Carlo Santini[29], Alessandro Saputi[31], Ivano Sarra[47], Giuseppe Scarantino[29,49], Hans Schneider-Muntau[125], Daniel Schulte[1], Jessica Scifo[48], Sally Seidel[126], Claudia Seitz[18], Tanaji Sen[20], Carmine Senatore[127], Abdulkadir Senol[64], Daniele Sertore[29], Lorenzo Sestini[62], Vladimir Shiltsev[128], Ricardo César Silva Rêgo[86,122], Federica Maria Simone[106,10], Kyriacos Skoufaris[1], Elise Sledge[129], Valentina Sola[12,13], Gino Sorbello[130,51], Massimo Sorbi[28,29], Stefano Sorti[28,29], Lisa Soubirou[30], Simon Spannagel[18], David Spataro[18], Anna Stamerra[42,10], Marcel Stanitzki[18], Steinar Stapnes[1], Giordon Stark[131], Marco Statera[29], Bernd Stechauner[132,1], Shufang Su[133], Wei Su[84], Ben Suitters[36], Xiaohu Sun[100], Alexei Sytov[31], Yoxara Sánchez Villamizar[86,135], Jingyu Tang[136,110], Jian Tang[84], Rebecca Taylor[1], Herman Ten Kate[69,1], Pietro Testoni[104], Leonard Sebastian Thiele[2,1], Rogelio Tomas Garcia[1], Max Topp-Mugglestone[1], Toms Torims[67,1], Riccardo Torre[9], Luca Tortora[77,76], Ludovico Tortora[77], Luca Tricarico[34,48], Sokratis Trifinopoulos[54], Donato Troiano[42,10], Alexander Naip Tuna[137], Sosoho-Abasi Udongwo[2,1], Ilaria Vai[41,4], Riccardo Umberto Valente[29], Giorgio Vallone[3], Ursula van Rienen[2], Rob Van Weelderen[1], Marion Vanwelde[1], Gueorgui Velev[20], Rosamaria Venditti[42,10], Adam Vendrasco[85], Adriano Verna[48], Gianluca Vernassa[1,138], Arjan Verweij[1], Piet Verwilligen[10], Ludovico Vittorio[135], Paolo Vitulo[41,4],



\newpage

\vskip 10cm
\begin{center}
{\huge\bfseries
IMCC authors}

\end{center}

\vspace{5mm}
{\small
\noindent
Carlotta~Accettura$^{1}$, 
Simon~Adrian$^{2}$, 
Rohit~Agarwal$^{3}$, 
Claudia~Ahdida$^{1}$, 
Chiara~Aime'$^{4, 5}$, 
Avni~Aksoy$^{6, 1}$, 
Gian Luigi~Alberghi$^{7}$,
Simon~Albright$^{1}$,
Siobhan~Alden$^{8}$, 
Luca~Alfonso$^{9}$
% ...
} 

\vspace{3mm}

\begin{flushleft}

{\em\footnotesize
$^{1}$ CH - CERN, \\ 
$^{2}$ DE - UROS, University of Rostock, \\ 
$^{3}$ US - LBL, Lawrence Berkely National Laboratory, \\ 
$^{4}$ IT - INFN - Pavia,  Istituto Nazionale di Fisica Nucleare Sezione di Pavia, \\ 
$^{5}$ IT - INFN - Pisa, Instituto Nazionale Di Fisica Nucleare - Sezione di Pisa, \\ 
$^{6}$ TR - AU, Ankara University, \\ 
$^{7}$ IT - INFN - Bologna, Instituto Nazionale Di Fisica Nucleare - Sezione di Bologna, \\ 
$^{8}$ UK - RHUL, Royal Holloway and Bedford New College, \\ 
$^{9}$ IT - INFN - Genova, Istituto Nazionale di Fisica Nucleare Sezione di Genova, \\ 
% ...
}
\end{flushleft}
\tableofcontents
\listoftables
\clearpage
\pagenumbering{arabic}
\setcounter{page}{1}

\section{Introduction}
\label{intro:sec}
This document contains updated parameters for the MuCol study.
This is the third and final iteration of the parameters, and is developed from the preliminary parameters report of 2024~\cite{parameters2024} and tentative parameters report of 2023~\cite{parameters2023}.\\
This consolidated collection of parameters considers a high-level baseline overview for a \SI{10}{\tera\electronvolt} centre-of-mass collider, and the resulting parameters of each sub-system.
Additional details from the baseline parameter set are written in detail within the appendix. The appendix also contains variations on designs and site-based parameters that have been developed bottom-up by the teams that work on the different parts of the complex and different technologies.
These parameters are already the fruit of the R\&D of each team, or the goals that the team considers realistic based on their expertise and studies carried out so far.

\subsection{Muon Collider Design}
\label{intro:sec:design}
The design effort focuses on a high energy stage at 10 TeV with a luminosity of \SI[per-mode=reciprocal]{18E34}{\per\centi\meter\squared\per\second}.
This would match approximately the physics reach of a \SI{100}{\tera\electronvolt} energy FCC-hh design.\\
Whilst it is possible to reach this through intermediary stages, such as with a \SI{3}{\tera\electronvolt} collider, or a lower-luminosity \SI{10}{\tera\electronvolt} collider, this report only considers the final product.

\subsection{Structure of the Document}
\label{intro:sec:structure}
\begin{figure}[!h]
    \centering
    \includegraphics[width=1\textwidth]{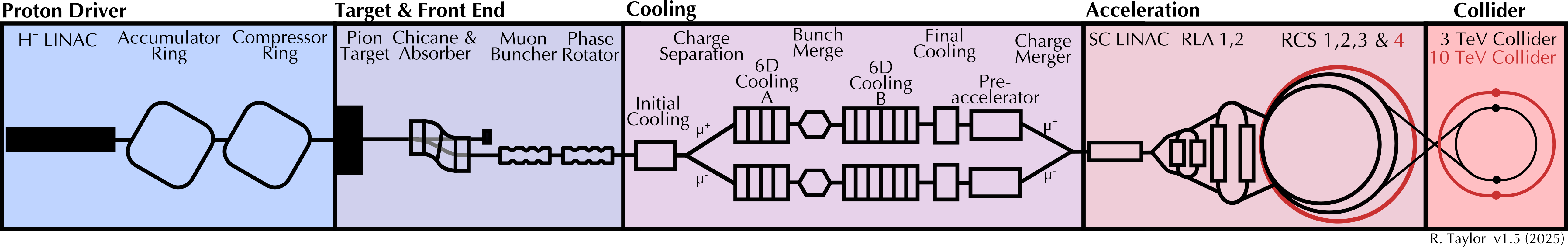}
    \caption{Simplified overview of the proton driver and muon collider accelerator complex.}
    \label{intro:fig:schematic}
\end{figure}

The high-level baseline parameters are listed in Section \ref{top:sec} followed by parameters for each subsystem split by section.
Figure \ref{intro:fig:schematic} demonstrates the present systems and subsystems of the complex, starting with the proton driver (blue) in Section \ref{sec:proton}, passing through to the front end (purple) in Section \ref{sec:target}, the ionization cooling (pink) in Section \ref{cool:sec}, acceleration (light red) in Section \ref{sec:highE} and finally the collider ring (red) in Section \ref{col:sec}.
The Detector and Machine-Detector Interface (MDI) designs are described in Section \ref{sec:detector} and \ref{sec:MDI} respectively.
Details of underlying technologies are given in subsequent sections, including magnets (Section \ref{mag:sec}) and RF cavities (Section \ref{sec:RF}).
Collective effects throughout the complex are described in Section \ref{sec:imp}, and the radiation shielding and protection considerations throughout the complex are described in Section \ref{sec:radshield} and \ref{sec:radpro} respectively.
Finally details of the demonstrator cooling cell can be found in Section \ref{demo:sec}.
The appendix of this report contains additional details of the lattice designs of each section, including system variations, such as site-specific designs.

%\subsection{Differences from Preliminary and Tentative Parameter Report 2023}
%\label{intro:sec:differences}
%The preliminary parameters considered a wide range of options, sites and variations, with details on each.
%The consolidated parameters will focus on the baseline muon collider option.
%In addition, significant progress has been made, including the civil engineering studies, the integration of a full demonstrator facility. %\clearpage
\subsection{Top-Level Parameters}
\label{top:sec}
The top-level parameters for the Muon Collider are shown in Table \ref{top:tab:highlevel}.
These are the ideal design specifications that each subsystem aims to achieve.
The parameters have been adapted as of the ESPPU update \cite{ESPPU}, whereby the baseline collider arc peak field is \SI{14}{\tesla}, based on feedback within the magnet community. This provides a 12\% linear scaling to the circumference and estimated luminosity.\\
Table \ref{top:tab:transmission_target} provides an overview of the whole complex, with the key numbers including the length and outgoing beam energy for each system.
It also indicates the required transmission performance per system to achieve the target transmission values.
An additional estimate of transmission and emittance is provided in Table \ref{top:tab:transmission_estimate}, based on the current efficiency of each simulated system.

The total estimated muon charge after the front-end and cooling is \num{2.5E12}, which is less than the required \num{3.6E12}. We aim to increase the transmission in the different subsystems to achieve this.
In addition, we are studying a variation of the muon production target design with an increase proton beam power of up to \SI{4}{\mega\watt}, the design of which is featured in Appendix \ref{target:sec:4target}. These two approaches should enable us to reach the bunch charge target in the collider ring.

A commentary on the assumptions in the luminosity calculations is in Appendix \ref{app:top:lumi}. Additional information on the decay throughout the facility is in Appendix \ref{app:top:decay}.

We notice that the transverse emittance target seems to be in reach, if further studies confirm the preservation along the accelerator chain. For the longitudinal emittances that indications are that we might be able to achieve a better value than the target. However, at this moment we do not change the target for the collider ring and accelerator chain to better understand the boundaries. The better value thus serves as a margin and we will later review how to integrate it into the design.

\begin{table}[!h]
\centerline{
  \begin{tabular}{l|ccc}
    \hline
    Parameter & Symbol & Unit & Baseline \\
\hline
    Centre-of-mass energy & $E_{\mathrm{cm}}$ & TeV &  10 \\
    Target integrated luminosity & $\int{\cal L}_{\mathrm{target}}$ & $\rm ab^{-1}$ &  10\\
    Estimated luminosity & ${\cal L}_{\mathrm{estimated}}$ & $10^{34}\rm cm^{-2}s^{-1}$ & 18\\
    Collider circumference& $C_{\mathrm{coll}}$ & $\rm km$ &  11.4\\
    Collider arc peak field& $B_{\mathrm{arc}}$ & $\rm T$ &  14\\
    Luminosity lifetime & $N_{\mathrm{turn}}$ &turns& 1363\\
    \hline
    Muons/bunch & $N$ & $10^{12}$ &  1.8 \\
    Repetition rate & $f_{\mathrm{r}}$ & $\rm Hz$ & 5\\
    Beam power  & $P_{\mathrm{coll}}$ & $\rm MW$  & 14.4 \\
    RMS longitudinal emittance& $\varepsilon_\parallel$ & $\rm eVs$  & 0.025 \\
    Norm.\,RMS transverse emittance& $\varepsilon_\perp$ & \textmu m & 25\\
    \hline
    IP bunch length& $\sigma_z $ & $\rm mm$ & 1.5*\\
    IP betafunction& $\beta $ & $\rm mm$ &  1.5*\\
    IP beam size& $\sigma $ & \textmu m & 0.9\\
    \hline
    Protons on target/bunch & $N_{\mathrm{p}}$ & $10^{14}$ &  5 \\
    Proton energy on target  & $E_{\mathrm{p}}$ & $\rm GeV$ &  5 \\
    \hline
  \end{tabular}
}
\caption[Consolidated target parameters for a muon collider at \SI{10}{\tera\electronvolt}.]{Consolidated target parameters for a muon collider at \SI{10}{\tera\electronvolt}. The estimated luminosity refers to the value that can be reached if all target specifications can be reached, including beam-beam effects. *Relaxed $\beta$ options are displayed in Table \ref{col:tab:param}}
\label{top:tab:highlevel}
\end{table}

\begin{table}[!h]
\centering
\resizebox{0.9\columnwidth}{!}{%
\begin{tabular}{lccccccc} \hline \hline
 Subsystem & Energy Out& Length& Tar. & Tar. & Target& Target &Target\\
 & & & $\varepsilon_T$ & $\varepsilon_L$ & Transm.& Cumulative &$\mu^-$/bunch\\
 & \si{\giga\electronvolt} & \si{\meter}& um& \si{\electronvolt\milli\second}& \si{\percent} & Transm. \%&$10^{12}$\\
\hline
Proton Driver & 5 ($p^+$) & 1500 &  &  & --  &  & 500 ($p^+$) \\
Front End & 0.17 & 150 & 17000 & 16.0& 9 & 100.0 & 45 \\
\hline
Charge Sep. & 0.2 & 12 & 17000 & 16.0& \multirow{6}{*}{\raisebox{-7ex}[0pt][0pt]{8}}&\multirow{6}{*}{\raisebox{-7ex}[0pt][0pt]{8}} & \multirow{6}{*}{\raisebox{-7ex}[0pt][0pt]{3.6}}\\
Rectilinear A & 0.2 & 363 & 1240 & 0.6&  &  &  \\
Bunch Merge & 0.13 & 134 & 5130 & 3.5&  &  &  \\
Rectilinear B & 0.124 & 424 & 300 & 0.5&  &  &  \\
Final Cooling  & 0.005 & 100 & 22.5 & 22.9&  &  &  \\ 
Pre-Acc. & 0.25 & 140 & 22.5 & 22.9& & & \\ \hline 
LINAC & 1.25 & 500 & \multirow{7}{*}{$\xdownarrow{1.75cm}$} & \multirow{7}{*}{$\xdownarrow{1.75cm}$} & \multirow{2}{*}{90} & \multirow{2}{*}{7.3} & \multirow{2}{*}{3.3}\\
RLA1 & 5 & $\circ$500 &  & & & & \\ \hline
RLA2 & 62.5 & $\circ$2400 &  &  & 92.6& 6.7& 3.0\\
RCS1  & 314 & $\circ$5990 &  &  & 90& 6.1& 2.7\\
RCS2  & 750 & $\circ$5990 &  &  & 90& 5.5& 2.5\\
RCS3  & 1500 & $\circ$10700 & &  & 90& 4.9& 2.2\\
3 TeV Collider & 1500 & $\circ$4500 & &  & -- & -- & 2.2 \\
RCS4  & 5000 & $\circ$35000 & 25 & 25.2& 90* & 4.4& 1.8\\
10 TeV Collider & 5000 & $\circ$11400 &  25&  25.2& -- & -- & 1.8 \\
 \hline \hline
\end{tabular}
}
\caption[Target Lengths, energies and transmission of each subsystem]{Target beam parameters at the end of each section of the acceleration chain for the baseline muon collider. Lengths are approximate and $\circ$ refers to the circumference. The 9\% transmission in the front-end refers to the yields from Option 1 as per Table \ref{target:tab:yield}. For $\mu^+$ the yield at the Front End is 12\% but the additional charge will be reduced via collimation to provide the same bunch charge and beam-loading in both beams.
*A 10\% transmission budget is added to encompass potential additional losses throughout acceleration.
%Currently, the achieved muon transmission is lower than the target value in the cooling and somewhat higher than the target value in the muon accelerator part. Further improvement is expected. A 4 MW target would provide almost twice as many muons at the beginning. $^*$ For the initial muon acceleration no design exists at this moment, the target value is given.
}    
\label{top:tab:transmission_target}
\end{table}

\begin{table}[!h]
\centering
\resizebox{0.69\columnwidth}{!}{%
\begin{tabular}{lccccc} \hline \hline
 Subsystem & Est. & Est. & Target& Cumulative&Estimated\\
 & $\varepsilon_T$ & $\varepsilon_L$ & Transm.& Estimated &$\mu^-$/bunch\\
 & um& \si{\electronvolt\milli\second}& \si{\percent} & Transm.&$10^{12}$\\
\hline
Proton Driver &  &  & --  &  & 500 ($p^+$) \\
Front End & 17000 & 16.0& 9& 100.0& 45\\
\hline
Charge Sep. & 17000 & 16.0& 95& 95.0& 42.8\\
Rectilinear A & 1240 & 0.6&  49.6&  47.1&  21.2\\
Bunch Merge & 5130 & 3.5&  78&  36.8&  16.5\\
Rectilinear B & 300 & 0.5&  28.6&  10.5&  4.7\\
Final Cooling  & 22.5 & 7.7&  61.4&  6.5&  2.9\\ 
Pre-Acc. & 22.5 & 7.7& 86& 5.6& 2.5\\ \hline
LINAC & \multirow{7}{*}{$\xdownarrow{1.75cm}$} & \multirow{7}{*}{$\xdownarrow{1.75cm}$} & \multirow{2}{*}{90}& \multirow{2}{*}{5.0}& \multirow{2}{*}{2.2}\\
RLA1 &  & & & & \\ \hline
RLA2 &  & & 92.6& 6.7& 2.1\\
RCS1  &  & & 90& 6.1& 1.9\\
RCS2  &  & & 90& 5.5& 1.7\\
RCS3  &  & & 90& 4.9& 1.5\\
3 TeV Collider & & & -- & -- & 1.5\\
RCS4  & 25 & 7.7 -- 8.5& 90& 4.4& 1.4\\
10 TeV Collider &  25&  7.7 -- 8.5& -- & -- & 1.4\\
 \hline \hline
\end{tabular}
}
\caption[Estimated emittance and transmission of each subsystem]{Estimated beam parameters at the end of each section of the acceleration chain, based on best-available simulations. Currently, the achieved muon transmission is lower than the target value and further improvement is expected. A \SI{4}{\mega\watt} target (Appendix \ref{target:sec:4target}) would provide almost twice as many muons at the beginning.
Longitudinal emittance range assumes $\approx 10 \%$ emittance growth throughout the acceleration complex.
%$^*$ For the initial muon acceleration no design exists at this moment, the target value is given.
}    
\label{top:tab:transmission_estimate}
\end{table}

 \clearpage
\section{Proton Driver}
\label{sec:proton}
Main overview of the proton driver complex, for the \SI{5}{\giga\electronvolt}, \SI{2}{\mega\watt} option can be found in Table \ref{proton:tab:H-main}. 
Information on the parameter sets for the accumulator ring and compressor ring designs are in Table \ref{proton:tab:acc_comp_main}.
The parameter set for the \SI{10}{\giga\electronvolt}, \SI{4}{\mega\watt} option can be seen in the appendix Table \ref{proton:tab:H-alt}. Accumulator ring and compressor ring designs for this alternative parameter set is in Table \ref{proton:tab:acc_comp_alt}.

\begin{table}[!h]
\centering

\begin{tabular}{ l l l}
Parameters & Unit & main \\
\hline
Final Energy & GeV & 5 \\
Repetition Rate & Hz & 5 \\
Max. source pulse length & ms & 2.5 \\
Max. source pulse current & mA & 80 \\
Source emittance & mm.mrad & 0.25 \\
Power & MW & 2 \\
Linac length & m & 670 \\
RF frequency & MHz & 352, 704 \\
\end{tabular}\
\caption[H- LINAC parameters]{H- LINAC parameters for the baseline option considering 1)  LINAC single use for muon production; 2) Chopping will later reduce the average current.}
\label{proton:tab:H-main}
\end{table}

\begin{table}[!h]
\centering
\begin{tabular}{l l c c}
Parameters & Unit &  Accumulator & Compressor  \\
\hline
Energy & GeV & 5 & 5 \\
Circumference & m & 180 & 314 \\
Final rms bunch length & ns & 120 & 2 \\
Geo. rms. emit & $\pi$.mm.mrad & 9.5 & 9.5 \\
number of bunches &  & 2 & 2 \\
Number of turns &  & 4167 & 41 \\
\hline
RF voltage  & MV & - & 4 \\
RF harmonic &  & - & 2 \\
initial mom. spread & \% & 0.025 & 0.025 \\
final mom. spread & \% & 0.025 & 1.5 \\
Protons on target & $10^{14}$ & - & 5 \\
\end{tabular}
\caption{Baseline Accumulator and Compressor parameters}
\label{proton:tab:acc_comp_main}
\end{table}

 \FloatBarrier
\section{Target \& Front-End}
\label{sec:target}
The deep inelastic interactions of the proton beam with the target produces kaons and pions, which eventually decay into muons.
To capture the produced particles and keep the emittance under control, the production target and the subsequent line has to be kept in a strong solenoidal magnetic field, which confines the charged particles along helical trajectories.
The baseline case considers a graphite target as the most suitable option.
This material allows operation at high temperatures and has a high thermal-shock resistance.
Therefore the majority of studies performed to optimize the pion-yield and estimate the radiation load on the front-end magnets have taken this target as baseline.
An overview of the proton driver parameters being used in the studies of the front-end target systems is shown in Table~\ref{target:sec:target-main}.
Different ranges of these parameters have been considered in order to optimise both the physics and engineering design.
Additional target geometries and higher power alternatives can be found in Appendix \ref{app:target}.

\begin{table}[!h]
\centering
\begin{tabular}{l c | c c}
Parameters & Unit & Baseline & Range \\ \hline
Beam power & MW & 2 & 1.5-3.0 \\
Beam energy & GeV & 5 & 2-10 \\
Pulse frequency & Hz & 5 & 5-50 \\
Pulse intensity & p+ $10^{14}$ & 5 & 3.7-7.5 \\
Bunches per pulse &  & 1 & 1-2 \\
Pulse length & ns & 2 & 1-2 \\
Beam size & mm & 5 & 1-7.5 \\
Impinging angle & \textdegree  & 0 & 0-10 \\
\end{tabular}
\caption{Assumed beam from proton driver via carbon target used in studies}
\label{target:sec:target-main}
\end{table}

To assess the most suitable conditions to operate the proton driver and to design the target, several FLUKA simulations were conducted, calculating the muon and the pion yield in each setting.
For this purpose, it was assumed that all the muon and pions going in the chicane can be captured if their momentum is below \SI[per-mode=symbol]{500}{\mega\electronvolt\per\clight}.
 The obtained yields are summarized as a function of beam energy in Table~\ref{target:tab:yield}, assuming a transverse beam sigma of 5~mm and a graphite target rod with a radius of 15~mm. 
 
\begin{table}[!h]
\centering
\begin{tabular}{l | c  c  c  c c c c c}
Yield [$10^{-2} $\si{\giga\electronvolt} $/p^+$] & 3 & 4 & 5 & 6 & 7 & 8 & 9 & 10 \\
\hline
$\mu^+$  & 2.8 & 2.6 & 2.4 & 2.3 & 2.2 & 2.1 & 1.9 & 1.9 \\
$\mu^-$  & 1.8 & 1.8 & 1.8 & 1.8 & 1.7 & 1.7 & 1.7 & 1.7 \\
$\pi^+$  & 1.3 & 1.2 & 1.1 & 1.1 & 1 & 0.98 & 0.92 & 0.9 \\
$\pi^-$  & 0.84 & 0.81 & 0.84 & 0.82 & 0.83 & 0.8 & 0.8 & 0.81 \\

\end{tabular}
\caption{Yield per unit energy proton beam [$10^{-2} $\si{\giga\electronvolt} $/p^+$]}
\label{target:tab:yield}
\end{table}

   \FloatBarrier
\section{Cooling}
\label{cool:sec}
The cooling channel is defined from the end of the RF capture system to the beginning of acceleration.
Five sub-systems are part of the cooling apparatus. Details of the sub-systems can be found in Appendix \ref{app:cool}.
\begin{enumerate}
\item Charge separation, which splits the positive and negative muon species into separate beamlines;
\item Rectilinear cooling (A and B lattices) which cools the beam in 6D phase space. Detailed parameters on each of these stages are in Appendix \ref{app:6d};
\item Bunch merge after the A lattice which merges the microbunches produced by the front end into a single bunch;
\item Final cooling, which cools in 4D phase space and produces the final low transverse emittance beam, at the cost of a larger longitudinal emittance. Detailed parameters on this sub-system including design alternatives are in Appendix \ref{app:final};
\item Re-acceleration, which accelerates the low energy beam up to 339 MeV/c momentum which is 250 MeV kinetic energy. Potential performance for re-acceleration is estimated in Table \ref{cool:tab:preacc}.
\end{enumerate}
For this iteration, parameters are listed in Table \ref{cool:tab:summary} for the principal subsystems: rectilinear cooling and final cooling. This is for two alternative stagings: One with a shorter rectilinear cooling, finishing at stage 8 but with a longer final cooling. The other with a longer rectilinear cooling until stage 10, but with a shorter final cooling.
%Additional details are available for the charge separation \cite{Yoshikawa:2013eba} and bunch merge \cite{Bao:2016fmz} subsystems.
A potential initial cooling stage is described in Section \ref{app:initial}. This system would be integrated prior to the charge separation.

\begin{table}
\centering
\begin{tabular}{l | c c c c c c }
& $\varepsilon_T$  & \textit{$\varepsilon_T$ target} & $\varepsilon_L$   & \textit{$\varepsilon_L$ target } & Mean $p_z $ & Transm. \\
& \si{\micro\meter} &  \si{\micro\meter} & \si{\milli\meter} & \si{\milli\meter}  & MeV/c & \% \\
\hline
End of charge separation & 17000 &  & 46 &  & 288 & 95 \\
6D Cooling end of \textbf{Stage 8}{\color{white}0}  & 260 & \textit{300} & 1.86 & \textit{1.5} & 200 & 14.9 \\
End of Final Cooling & 22.5 & \textit{22.5} & 42 -- 72 & \textit{64} & 28 & 6.4 \\
End of Reacceleration & 22.5 & \textit{22.5} & 64 & \textit{64} & 339 & 5.8 \\
\end{tabular}

\begin{tabular}{l | c c c c c c }
& $\varepsilon_T$  & \textit{$\varepsilon_T$ target} & $\varepsilon_L$   & \textit{$\varepsilon_L$ target } & Mean $p_z $ & Transm. \\
& \si{\micro\meter} &  \si{\micro\meter} & \si{\milli\meter} & \si{\milli\meter}  & MeV/c & \% \\
\hline
End of charge separation & 17000 &  & 46 &  & 288 & 95 \\
6D Cooling end of\textbf{ Stage 10} & 140 & \textit{300} & 1.56 & \textit{1.5} & 200 & 10.5 \\
End of Final Cooling & 22.5 & \textit{22.5} & {\color{white} ex} 22 {\color{white} ex} & \textit{64} & 26.5 & 6.4 \\
End of Reacceleration & 22.5 & \textit{22.5} & 22 & \textit{64} & 339 & 5.8 \\

\end{tabular}
\caption[Beam parameters of the cooling system for short- and long-rectilinear options.]{Beam parameters entering and leaving the cooling system for short-rectilinear (top) and long-rectilinear (bottom) options. The target emittances are listed. They are 10 \% more demanding than the nominal emittances in the RCS and collider, allowing for some emittance growth at some point in the acceleration chain. Note: 64 mm = 0.0225 eVs}
\label{cool:tab:summary}
\end{table}

   \FloatBarrier
\section{Acceleration}
\label{sec:highE}
\subsection{Low Energy Acceleration}
The low energy acceleration chain brings the muon beams from \SI{250}{\mega\electronvolt} after the pre-accelerator to \SI{62.5}{\giga\electronvolt} for injection into the high energy acceleration chain described in Section \ref{sec:highE}.\\
Details of the Low Energy Acceleration systems are in Appendix \ref{app:lowE}.
They are composed of a single-pass superconducting LINAC outlined in Table \ref{low:tab:linac}, followed by two recirculating linear accelerators (RLA), described in Table \ref{low:tab:RLA}.\\
RLA2 has a preliminary optics design. No optics design exists for LINAC and RLA1.
Both RLAs have an assumed racetrack geometry.
The transmission through RLA2 is 92.6\%. The target transmission for LINAC and RLA1 is 90\%, which corresponds to an effective average gradient of $4.1\;\rm MV/m$.

\subsection{High Energy Acceleration}
Below is the main overview of the high energy acceleration system.
Table \ref{high:tab:RCS_key} shows the general RCS parameters. 
We assume a survival rate of 90\,\% per ring and linear ramping only considering losses due to muon decay, even though these values are subject to further adjustments to optimize the RF and magnet powering parameters with respect to total costing, ramp shape, bunch matching, and the overall transmission of the entire chain.
The lattice parameters based on the key parameters are shown in Table \ref{high:tab:RCS_lattice}.
The high energy acceleration parameters for site-based variations are summarized in Appendix \ref{app:highE}.
Parameters for high energy acceleration performed with a fixed field accelerator are shown in Appendix \ref{app:FFA}.
\begin{table}[!h]
    \centering
    \begin{tabular}{lc|cccc} 
         Parameter&  Unit&  RCS1&  RCS2&  RCS3&  RCS4\\ \hline
         Hybrid RCS&  -& no & yes & yes & yes \\
         Repetition rate&  Hz&  5&  5&  5&  5\\
         Circumference& m& 5990& 5990& 10700& 35000\\
         Injection energy& GeV& 63& 314& 750& 1500\\
         Extraction energy& GeV& 314& 750& 1500& 5000\\
         Energy ratio&  &  5.0&  2.4&  2.0&  3.33\\
         Assumed survival rate& & 0.9& 0.9& 0.9&0.9\\
         Cumulative survival rate& &  0.9&  0.81&  0.73&  0.66\\     
         Acceleration time&  ms&  0.34&  1.10&  2.37&  6.37\\
         Revolution period&  \textmu s&  20&  20&  36&  117\\
         Number of turns& -& 17& 55& 66& 55\\
         Required energy gain/turn& GeV& 14.8& 7.9& 11.4& 63.6\\
         Average accel.~gradient& MV/m& 2.44& 1.33& 1.06& 1.83\\ \hline
         Number of bunches& & 1& 1& 1& 1\\
         Inj.~bunch population& \num{E12}& 2.7& 2.4& 2.2& 2\\
         Ext.~bunch population& \num{E12}& 2.4& 2.2& 2& 1.8\\
         Beam current per bunch& mA& 21.67& 19.5& 9.88& 2.75\\
         Peak RF power&  MW&  640&  310&  225&  350\\
         Vert.~norm.~emittance& \textmu m& 25& 25& 25& 25\\
         Horiz.~norm.~emittance&  \textmu m&  25&  25&  25&  25\\
         Long.~norm.~emittance&  eVs&  0.025&  0.025&  0.025&  0.025\\
         Bunch length at injection & ps & 31& 22& 18& 14\\
         Bunch length at ejection & ps & 22& 18& 14& 10\\ \hline
         Straight section length&  m & 1012.4&  536.4&  793.6&  4385.6\\
         Length with pulsed dipole magnets& m & 3654& 2539 & 4366 & 20376\\
         Length with steady dipole magnets& m & -& 1115 & 2358 & 4257\\
         Max.~pulsed dipole field& T& 1.8& 1.8& 1.8&1.8\\
         Max.~steady dipole field& T& -& 10& 10&16\\
         Ramp rate& T/s& 4232& 3272& 1519&565\\
         Main RF frequency& GHz& 1.3& 1.3& 1.3&1.3\\
         Harmonic number&  &  25900&  25900&  46300&  151400\\
    \end{tabular}
    \caption{RCS acceleration chain key parameters}
    \label{high:tab:RCS_key}
\end{table}

\begin{table}[!h]
\centering
\begin{tabular}{l c | c c c c}
Parameter & Unit & RCS1 & RCS2 & RCS3 & RCS4 \\
\hline
Fill ratio dipole & \% & 61 & 61 & 62.8 & 70.4 \\
Cells per arc &  & 17 & 17 & 29 & 36 \\
Number of arcs &  & 8 & 8 & 8 & 8 \\
Cell length & m & 36.6 & 40.1 & 42.7 & 106.3 \\
Total Arc length & m & 4977.6 & 5453.6 & 9906.4 & 30614.4 \\
Arc Ratio & -  & 0.83 & 0.91 & 0.93 & 0.87 \\
Relative path length difference & \num{E-06} & 0 & 5.1 & 0.9 & 1 \\
Horizontal aperture & mm  & 91.3 & 83.4 & 49.7 & 88.6 \\
Vertical aperture  & mm  & 24.8 & 21.6 & 21.1 & 21.1 \\
Transition gamma &  & 31.02 & 33.36 & 55.85 & 68.78 \\
%Momentum compaction factor & \num{E-04} & 10.395 & 8.986 & 3.206 & 2.114 \\
% too many digits - overprecision - rounded up - V.Shiltsev
Momentum compaction factor & \num{E-04} & 10.4 & 9.0 & 3.2 & 2.1 \\
Horizontal tune (ring) &  & 41.02 & 39.56 & 64.54 & 82.40 \\
Vertical tune (ring) &  & 42.36 & 36.08 & 62.44 & 81.43 \\
Mean horizontal beta & m & 36.54 & 35.42 & 38.31 & 97.94 \\
Mean vertical beta & m & 38.31 & 36.02 & 37.84 & 92.75 \\
Horizontal natural chromaticity (ring) &  & -60.91 & -53.96 & -86.58 & -112.94 \\
Vertical natural chromaticity (ring) &  & -61.11 & -46.88 & -82.11 & -102.39 \\
\end{tabular}
\caption{RCS acceleration chain lattice parameters}
\label{high:tab:RCS_lattice}
\end{table}

% Taylor, please decide whether the paragraph below should remain in the main document or move to the appendices. Many thanks!

Recent longitudinal tracking studies in both RLA2 and RCS1 revealed an important longitudinal mismatch at injection into the high-energy acceleration chain at \SI{62.5}{GeV}. This mismatch, mainly due to the important difference in the aspect ratios of the RF buckets, could be mitigated by lowering the RF frequency of RCS1, presently \SI{1.3}{GHz}, closer towards the frequency of the accelerating sections at \SI{352}{MHz} in RLA2.

To progress with the baseline optics study, an eightfold symmetry with eight long straight sections has been assumed. As this would be insufficient from the longitudinal beam dynamics point of view, due to the large energy gain per long straight section, alternative options are being investigated to either lower the momentum compaction factor of the lattice or to introduce more straight sections for acceleration. Both approaches would smooth the impact of the discrete energy kicks by the RF system.

The present assumption to reach a \SI{90}{\%} survival rate per RCS with an initial bunch intensity of $2.7\cdot10^{12}$ muons injected at \SI{62.5}{GeV} simplified the preliminary choice of parameters given in this report. However, a global optimization of the beam transfer energies between the different RCS in the chain will be performed, with the objective to maximize the overall transmission for muons up to \SI{1.5}{TeV} or \SI{5}{TeV}. This optimization is also expected to adjust RF voltages, average accelerating gradients and horizontal apertures of all four synchrotrons. Of course, variations cannot be completely avoided in the upstream part of the chain due to the constraint of installing the first two RCS in the same accelerator tunnel.
\section{Collider}
\label{col:sec}

%% ==== START TentativeChapters/Author6-Collider ==== %%
The present work concentrates on the design of a \SI{10}{\tera\electronvolt} center-of-mass collider.
The aim is to maximize the luminosity to the two possible experiments, introduced in Section \ref{sec:detector}.
The basic luminosity assumptions (in Section \ref{app:top:lumi}) are extrapolations from lower energy starting with a relative rms momentum spread of $\sigma_\delta = 1 \cdot 10^{-3}$.
Together with the longitudinal emittance, this fixes the rms bunch length $\sigma_z = 1.5$ \si{\milli\meter} and the $\beta^* = 1.5$ \si{\milli\meter} to the same value, such that the hour glass luminosity reduction factor $f_{hg} = 0.758$ starts to become significant.
Maximization of the luminosity requires to choose the shortest possible circumference $C$ compatible with feasibility of the magnets (average bending field assumed to be $\bar B \approx$ \SI{10.48}{\tesla} leading to $C \approx$ \SI{10}{\kilo\meter}).
Note that extrapolation of parameters to higher energies lead to very large chromatic effects further increasing with energy.

The main parameters are described in Table \ref{col:tab:param}, which contains a set of target parameters which meet the performance of Table \ref{top:tab:highlevel}.
The set of relaxed parameters considers a lattice with reduced beta oscillations and chromatic aberrations, to study imperfections and the effects of movers.\\
The radial build of arc dipoles is described in Table~\ref{col:tab:arcs}. The radial build assumes a radiation shielding thickness of 3~cm, which can be accepted from a cryogenics point of view if the operating temperature is 20~K. The estimated heat load and radiation damage in arc dipoles is summarized in Table~\ref{rad:tab:colliderarcdipoles}. 
%% ==== END TentativeChapters/Author6-Collider ==== %%

\begin{table}[!h]
    \centering
    \begin{tabular}{l|c|c|c}
         &  & \multicolumn{2}{c}{version} \\
         Parameter&  Unit& relaxed & target \\ \hline
         Center of mass energy&  TeV &\multicolumn{2}{c}{10}  \\
         Geometric Luminosity\footnotemark & \SI[per-mode=reciprocal]{E34}{\per\centi\meter\squared\per\second} & 5.77 & 19.2 \\ 
         Beam energy& TeV & \multicolumn{2}{c}{5}\\
         Relativistic Lorentz factor& & \multicolumn{2}{c}{47322}\\
         Circumference\footnotemark &  km & \multicolumn{2}{c}{$\approx$ 10}\\
         Dist. of last magnet to IP& m& \multicolumn{2}{c}{6}\\
         Repetition rate& Hz& \multicolumn{2}{c}{5}\\
         Bunch intensity (one bunch per beam)& \num{E12} & \multicolumn{2}{c}{1.80}\\
         Injected beam power per beam&  MW& \multicolumn{2}{c}{7.2} \\
         Normalized transverse rms emittance&  \textmu m& \multicolumn{2}{c}{25} \\
         Longitudinal norm. rms emittance&  eVs& \multicolumn{2}{c}{0.025}\\
         Relative rms momentum spread& $10^{-3}$ & 0.3 & 1 \\
         RMS bunch length in space&  mm& 5 & 1.5 \\
         RMS bunch length in time domain&  ns& .017 & 0.005 \\
         Twiss betatron function at the IP&  mm& 5 & 1.5 \\
         Energy loss per turn\footnotemark & MeV & \multicolumn{2}{c}{$\approx$ 27.2}  \\
         Integrated RF gradient\footnotemark
          & MV & \multicolumn{2}{c}{30}
    \end{tabular}
    \caption{Assumptions for the main parameters used in the design of a \SI{10}{\tera\electronvolt} muon collider.}
    \label{col:tab:param}
\end{table}%

\begin{table}[!ht]
    \centering
    \begin{tabular}{l|c|cc}
         Parameter&  Unit&  Thickness& Outer radius\\ \hline
         Beam aperture&  mm&  23.49& 23.49\\
         Coating (copper)&  mm&  0.01& 23.5\\
         Radiation absorber (tungsten alloy)&  mm&  30& 53.5\\
         Shielding support and thermal insulation&  mm&  11& 64.5\\
         Cold bore&  mm&  3& 67.5\\
         Insulation (Kapton)&  mm&  0.5& 68\\
         Clearance to coils&  mm&  1& 69\\
    \end{tabular}
    \caption[Collider arcs, coil inner aperture.]{Collider arcs, coil inner aperture. For options using low temperature superconductor, i.e. at 3 TeV, the shielding thickness should be 40 mm and the other parameters changed accordingly.}
    \label{col:tab:arcs}
\end{table}

\setcounter{footnote}{\value{footnote}-3}
\footnotetext[\value{footnote}]{%
Luminosities for Gaussian beams with hour glass reduction factor and without beam-beam effect. Multiturn beam simulations with the correct lattice and tunes are needed in addition to first single pass simulations resulting in a modest luminosity increase.}%
\stepcounter{footnote}%
\footnotetext[\value{footnote}]{%
The approximate circumference depends on the maximum achievable dipole field. The current lattice design assumes a collider arc peak field of \SI{16}{\tesla}, but will likely be updated later to reduce the field to \SI{14}{\tesla}. This change should not significantly affect the optics and will mainly result in an increased collider circumference. }%
\stepcounter{footnote}%
\footnotetext[\value{footnote}]{%
Assuming constant bending field of \SI{16}{\tesla}. The exact value will depend on the detailed lattice design and likely be lower.}%
\stepcounter{footnote}%
\footnotetext[\value{footnote}]{%
Assuming that only the synchrotron radiation losses have to be compensated. Some margin and no particular frequency requirements as long as the RF voltage does not vary too much over the bunch length of few 10s of ns.} \FloatBarrier
\section{Detectors}
\label{sec:detector}

\begin{table}[h!]
\begin{center}
\begin{tabular}{lccc}
\hline\hline
 \textbf{Requirement} & \multicolumn{2}{c}{\textbf{Baseline}} & \textbf{Aspirational}\\
  & \textbf{$\sqrt{s}=3$~TeV} & \textbf{$\sqrt{s}=10$~TeV} & \\
\hline
Angular acceptance & $|\eta|<2.5$ & $|\eta|<2.5$ & $|\eta|<4$ \\
Minimum tracking distance [cm] & $\sim 3$ & $\sim 3$ & $< 3$\\
Forward muons ($\eta> 5$) & -- & tag & $\sigma_{p}/p \sim 10$\%\\
Track $\sigma_{p_T}/p^{2}_{T}$ [GeV$^{-1}$] & $4 \times 10^{-5}$ & $4 \times 10^{-5}$ & $1 \times 10^{-5}$ \\
Photon energy resolution & $0.2/\sqrt{E\text{ [GeV]}}$ & $0.2/\sqrt{E\text{ [GeV]}}$ & $0.1/\sqrt{E\text{ [GeV]}}$\\
Neutral hadron energy resolution & $0.5/\sqrt{E\text{ [GeV]}}$ & $0.4/\sqrt{E\text{ [GeV]}}$ & $0.2/\sqrt{E\text{ [GeV]}}$ \\
Timing resolution (tracker) [ps] & $\sim 30-60$ & $\sim 30-60$ & $\sim 10-30$ \\
Timing resolution (calorimeters) [ps] & 100 & 100 & 10 \\
Timing resolution (muon system) [ps] & $\sim 50$ for $|\eta|>2.5$ & $\sim 50$ for $|\eta|>2.5$ & $< 50$ for $|\eta|>2.5$ \\
Flavour tagging & $b$ vs $c$ & $b$ vs $c$ & $b$ vs $c$, $s$-tagging \\
Boosted hadronic resonance ID & $h$ vs W/Z & $h$ vs W/Z & W vs Z\\
\hline\hline
\end{tabular}
\end{center}
\caption[Detector baseline and aspirational targets]{Preliminary summary of the ``baseline'' and ``aspirational'' targets for selected key metrics, reported separately for machines taking data at $\sqrt{s}=3$ and $10$~TeV. The reported performance targets refer to the measurement of the reconstructed objects in physics events after, for example, background subtraction and not to the bare detector performance.}
\label{tab:detector_req}
\end{table}

The design of the detector for $\sqrt{s}=10$ TeV follows the concept already developed for $\sqrt{s}=3$ TeV with modifications to account for the higher energy. Two distinct detector concepts are presented, MAIA (Muon Accelerator Instrumented Apparatus) and MUSIC (MUon System for Interesting Collisions), to fully exploit the two interaction points of the collider.
Both designs share a similar structure, a cylinder $11.4$ m long with a diameter of $12.8$ m. The main detector components are:
\begin{itemize}
    \item Tracking system
    \item Electromagnetic calorimeter (ECAL)
    \item Hadron calorimeter (HCAL)
    \item A superconducting solenoid 
    \item A muon sub-detector 
\end{itemize}
The origin of the space coordinates is the beam interaction point at the center of the detector. The $z$-axis follows the direction of the clockwise-circulating $\mu^+$ beam, the $y$-axis is parallel to gravity acceleration, and the $x$-axis is defined as perpendicular to both the $y$ and $z$ axes.

Table~\ref{tab:detector_req} summarises the baseline and aspirational performance and acceptance targets for the muon collider detectors.
Table~\ref{tab:detector_general} summarises the detector parameters sub-system by sub-system for the two concepts. While the tracking system has a similar structure, the MAIA detector has the solenoid just after the tracker, before the ECAL while MUSIC places the solenoid magnet between ECAL and HCAL.

\begin{table}[h!]
\centering
\begin{tabular}{l|ccc} 
Detector Concept & \textbf{MuColl} & \textbf{MUSIC} & \textbf{MAIA} \\
 & $\sqrt{s}=3~\textrm{TeV}$ & $\sqrt{s}=10~\textrm{TeV}$ & $\sqrt{s}=10~\textrm{TeV}$ \\
\hline
\textbf{Inner Trackers} & & & \\
R$_\textrm{min}$ -- R$_\textrm{max}$ [\si{\milli\meter}] & 30 -- 1486 & 29 -- 1486 & 30 -- 1486 \\
z$_\textrm{min}$ -- z$_\textrm{max}$ [\si{\milli\meter}] & 0 -- 2190 & 0 -- 2190 & 0 -- 2190 \\
Angular Acceptance [\si{\degree}] & 10 -- 170 & 10 -- 170 & 10 -- 170 \\
$X / X_0$ & 0.3 & 0.1 & 0.1\\
$L / L_0$ & 0.1 & 0.04 & 0.04\\
\hline
\textbf{EM Calorimeters} & & & \\
R$_\textrm{min}$ -- R$_\textrm{max}$ [\si{\milli\meter}] & 1500 -- 1702 & 1690 -- 1960 & 1857 -- 2125 \\
z$_\textrm{min}$ -- z$_\textrm{max}$ [\si{\milli\meter}] & 2307 -- 2210 & 2307 -- 2577 & 2307 -- 2575 \\
Angular Acceptance [\si{\degree}] & 10 -- 170 & 10 -- 170 & 10 -- 170 \\
$X / X_0$ & 26 -- 32 & 33 -- 38 & 40 -- 42\\
$L / L_0$ & 1.2 -- 1.5 & 1.4 -- 1.7 & 1.8 -- 1.9\\
\hline
\textbf{Hadron Calorimeters} & & & \\
R$_\textrm{min}$ -- R$_\textrm{max}$ [\si{\milli\meter}] & 1740 -- 3330 & 2902 -- 4756 & 2125 -- 4113 \\
z$_\textrm{min}$ -- z$_\textrm{max}$ [\si{\milli\meter}] & 2539 -- 4129 & 2579 -- 4434 & 2575 -- 4562 \\
Angular Acceptance [\si{\degree}] & 10 -- 170 & 10 -- 170 & 10 -- 170 \\
$X / X_0$ & 82 -- 87 & 89 -- 116 & 100 -- 114 \\
$L / L_0$ & 8.8 -- 9.3  & 9.5 -- 12.5 & 10.9 -- 12.3\\
\hline
\textbf{Muon Systems} & & & \\
R$_\textrm{min}$ -- R$_\textrm{max}$ [\si{\milli\meter}] & 4461 -- 6450 & 4806 -- 6800 & 4150 -- 7150 \\
z$_\textrm{min}$ -- z$_\textrm{max}$ [\si{\milli\meter}] & 4179 -- 5638 & 4444 -- 5903 & 4565 -- 6025 \\
Angular Acceptance [\si{\degree}] & 10 -- 170 & 10 -- 170 & 10 -- 170 \\
\hline
\textbf{Solenoid} & & & \\
R$_\textrm{min}$ -- R$_\textrm{max}$ [\si{\milli\meter}] & 3483 -- 4290 & 2000 -- 2807 & 1500 -- 1857 \\
z$_\textrm{min}$ -- z$_\textrm{max}$ [\si{\milli\meter}] & 0 -- 4129 & 0 -- 2500 & 0 -- 2307 \\
$X / X_0$ & -- & 18 & 6\\
$L / L_0$ & -- & 2.7 & 1.4\\
$B_z$ [T] & 3.6 & 5 & 5 \\
\hline
\textbf{Nozzles} & & & \\
R$_\textrm{min}$ -- R$_\textrm{max}$ [\si{\milli\meter}] & 10 -- 600 & 10 -- 550 & 10 -- 550 \\
z$_\textrm{min}$ -- z$_\textrm{max}$ [\si{\milli\meter}] & 60 -- 6000 & 60 -- 6000 & 60 -- 6000 \\
\end{tabular}
\caption[Detector parameters]{Detector parameters for the MuColl (v1), MUSIC (v2) and MAIA (v0) concepts. Values that are left empty ("--") are not relevant for the specific detector. $X / X_0$ and $L / L_0$ are for a particle travelling from the nominal beam interaction point (IP). The origin of the space coordinates is the IP. The $z$-axis has direction parallel to the beam pipe, the $y$-axis is parallel to gravity acceleration and the $x$-axis is defined as perpendicular to both the $y$ and $z$ axes.}
\label{tab:detector_general}
\end{table}

\subsection{Tracking System}
\label{sec:trackingsistem}
The tracking detector is composed of the vertex and tracker sub-detectors, both of them structured in barrels and end-caps. The barrels consist of sensor modules arranged in cylindrical configurations with varying lengths and radii, whose axes align with the beamline, covering the central region of the detector.
%The barrels are cylindrical surfaces with variable lengths and radii, whose axes coincide with the beamline and cover the central part of the detector.
The endcaps are annuli centered on the $z$ axis, with variable distance from the interaction point and radii which cover the forward part of the detector.
The major characteristics of this sub-system are described in Table~\ref{det:tab:tracker}.\\
The vertex detector is close to the interaction point in order to allow a good resolution on track impact parameter. The building blocks of the barrel detection layers are rectangular staves of sensors, arranged to form a cylinder, while the endcaps are constituted by trapezoidal modules of sensors, arranged as "petals" to form a disk.
The MAIA detector has 5 layers, with the first two structured as a double layer, while MUSIC has 5 distinct layers. The length of the MUSIC barrel is 26 cm, which is double that of MAIA.

The barrel layers of the vertex detector have silicon pixels of size $25 \times 25$ \textmu m$^2$, and thickness $50$ \textmu m. % whose radius and geometrical 
%The eight endcaps layers, four for each side of the interaction point are composed of silicon pixels of size $25 \times 25$ \textmu m$^2$ and thickness 50\,\textmu m and 16 modules. 

The inner and outer trackers are based on the same technology for MAIA and MUSIC, single layer of silicon macro-pixels sensors of 100 \textmu m thickness.
%Strips on the barrels are oriented with the long side parallel to the beam axis while the end-caps are composed of radial modules composed by rectangular pads. 

\begin{table}[h!]
    \centering
    \begin{tabular}{l|c|c|c}
                        &  \textbf{Vertex Detector} & \textbf{Inner Tracker} & \textbf{Outer Tracker} \\
    \hline\hline
    Sensor type           &  pixels & macro-pixels & macro-pixels \\
    Barrel Layers  &  5  & 3  & 3 \\
    Endcap Layers (per side)  &  4   &  7 & 4 \\
    Cell Size           &  \qty{25}{\um} $\times$ \qty{25}{\um} & \qty{50}{\um} $\times$ \qty{1}{\mm} & \qty{50}{\um} $\times$ \qty{10}{\mm} \\
    Sensor Thickness    &  \qty{50}{\um} & \qty{100}{\um} & \qty{100}{\um} \\
    Time Resolution     &  \qty{30}{\ps} & \qty{60}{\ps} & \qty{60}{\ps} \\
    Spatial Resolution  &  \qty{5}{\um} $\times$ \qty{5}{\um} & \qty{7}{\um} $\times$ \qty{90}{\um} & \qty{7}{\um} $\times$ \qty{90}{\um} \\
    \end{tabular}
    \caption[Assumed spatial and time resolution for MAIA and MUSIC tracking detector sub-systems]{Assumed spatial and time resolution for MAIA and MUSIC Tracking Detector sub-systems. There is no resolution difference between the barrel and end-cap regions. The first layer of the Vertex barrel and all Vertex endcap layers of MAIA are implemented as double layers.}
    \label{det:tab:tracker}
\end{table}

\subsection{Calorimeter System}
The calorimeter system is composed of the electromagnetic (ECAL) and hadronic (HCAL) sub-detectors. A summary of the main characteristics are in Tables~\ref{det:tab:calo_MAIA} and \ref{det:tab:calo_MUSIC}.

The MAIA ECAL configuration is inspired by CLIC. It consists of a dodecagonal barrel and two endcap systems. It is composed of 50 alternating layers of Tungsten as absorber material $2.2$ mm thick and Si sensor as active material with $5.1 \times 5.1$ mm$^{2}$ silicon detector cells. It is located outside of the superconducting solenoid.

The MUSIC ECAL, has the same geometry as MAIA's, but is positioned immediately after the tracking system and within the superconducting solenoid. It is a semi-homogeneous longitudinally-segmented calorimeter based on lead-fluoride (PbF$_2$) crystals read out by Silicon Photomultipliers. It represents a modern design approach that aims to combine the intrinsic high-energy resolution of homogeneous calorimeters with the longitudinal segmentation typically found in sampling calorimeters.

MAIA and MUSIC currently share the same technology for HCAL. It consists of a dodecagonal barrel and two endcap systems, structured in alternating layers of iron absorber 20 mm thick and plastic scintillating tiles with cell size $30 \times 30 $ mm$^2$, 75 layers in MAIA and 70 in MUSIC.
It allows the reconstruction of hadronic jets and helps in particle identification, to separate hadrons from leptons and photons.

\begin{table}[h!]
    \centering
    \begin{tabular}{l|c|c}
                        &  \textbf{Electromagnetic Calorimeter} & \textbf{Hadron Calorimeter} \\
    \hline\hline
    Cell type           &  Silicon - Tungsten & Iron - Scintillator \\
    Cell Size           &  \qty{5.1}{\mm} $\times$ \qty{5.1}{\mm} & \qty{30.0}{\mm} $\times$ \qty{30.0}{\mm}  \\
    Sensor Thickness    &  \qty{0.5}{\mm} & \qty{3.0}{\mm}  \\
    Absorber Thickness  &  \qty{2.2}{\mm} & \qty{20.0}{\mm}  \\
    Number of layers  &  50 & 75  \\
    \end{tabular}
    \caption[MAIA calorimeter systems]{Cell and absorber sizes in the MAIA calorimeter systems, describing both the barrel and end-cap regions.}
    \label{det:tab:calo_MAIA}
\end{table}

\begin{table}[h!]
    \centering
    \begin{tabular}{l|c|c}
                        &  \textbf{Electromagnetic Calorimeter} & \textbf{Hadron Calorimeter} \\
    \hline\hline
    Cell type           & PbF$_{2}$ crystal  & Iron - Scintillator \\
    Cell Size           &  \qty{10}{\mm} $\times$ \qty{10}{\mm} $\times$ \qty{40}{\mm} & \qty{30.0}{\mm} $\times$ \qty{30.0}{\mm}  \\
    Sensor Thickness    &  - & \qty{3.0}{\mm}  \\
    Absorber Thickness  &  - & \qty{20.0}{\mm}  \\
    Number of layers    &  6 & 70  \\
    \end{tabular}
    \caption[MUSIC calorimeter systems]{MUSIC calorimeter systems, describing both the barrel and end-cap regions.}
    \label{det:tab:calo_MUSIC}
\end{table}

\subsection{Muon System}
The current configuration of the two detector concepts does not include a magnetic field outside the calorimetric system, so the role of the muon detector must be reconsidered. In particular, for high-energy muons, new methods based on machine learning, which combine tracking detector and calorimeter information, could be employed. In this case, the muon detector would primarily serve to identify that the particle is a muon.

\section{Machine-Detector Interface}
\label{sec:MDI}
This section contains the main overview of the Machine-Detector Interface (MDI). An overview of the detectors can be found in Section \ref{sec:detector}.
An indication of the ionization dose and neutron-equivalent fluence of both detector geometries can be found in Table \ref{mdi:tab:raddamagedetector}.
The results for the vertex detector, the inner tracker, as well as the electromagnetic calorimeter correspond to one year of operation, assuming 1.2$\times$10$^7$ seconds of operation (139~days). The studies considered only muon decay, while neglecting the contribution of collision products and beam halo losses. The results were computed for IR lattice version 0.8.

Table \ref{tab:MDI_bib_particles} indicates the species of secondary particles that enter the detectors.
Additional information and variations can be found in Appendix \ref{app:MDI}.

\begin{table}[!h]
\centering
\begin{tabular}{l | c | c | c | c}
 & \multicolumn{2}{c|}{Dose} & \multicolumn{2}{c}{1 MeV neutron-equivalent} \\
Unit & \multicolumn{2}{c|}{kGy} & \multicolumn{2}{c}{fluence in Si $10^{14}$ n/\si{\centi\meter\squared}} \\
 & MAIA & MUSIC & MAIA & MUSIC \\
\hline
Vertex (barrel) & \multicolumn{2}{c|}{900} & \multicolumn{2}{c}{2} \\
Vertex (endcaps) & \multicolumn{2}{c|}{1800} & \multicolumn{2}{c}{7} \\
Inner trackers (barrel) & \multicolumn{2}{c|}{61} & 4 & 3.5 \\
Inner trackers (endcap) & \multicolumn{2}{c|}{26} & 10 & 8.8 \\
ECAL & 0.51 & 1.2 & 0.13 & 0.8 \\

\end{tabular}
\caption[Ionizing dose and neutron-equivalent fluence in MUSIC and MAIA detectors]{Maximum values of the ionizing dose and the 1~MeV neutron-equivalent fluence (Si) in the MAIA and MUSIC detectors. All values are per year of operation (10~TeV) and include only the contribution of muon decay. The updated values assume a collider ring circumference of 11.4 km.}

\label{mdi:tab:raddamagedetector}
\end{table}

\begin{table}[!h]
\centering

\begin{tabular}{l | c | c}
\textbf{Particle type}  & \textbf{Particles entering detector}  & \textbf{Threshold}  \\
\hline
Photons  & \SI{1.0e8}{}  & 100 keV  \\
Neutrons  & \SI{1.1e8}{}  & 0.01 meV  \\
Electron/positrons  & \SI{1.2e6}{}  & 100 keV  \\
Muons  & \SI{1.1e4}{}  & 100 keV  \\
Charged hadrons  & \SI{4.0e4}{}  & 100 keV  \\
\end{tabular}
\caption[Number of secondary particles entering the detector volume]{Number of secondary particles (muon decay) entering the detector volume (10~TeV). Only particles above the threshold values were included. The multiplicities include only the contribution of one beam and correspond to one bunch crossing.}
\label{tab:MDI_bib_particles}
\end{table}\FloatBarrier
\section{Magnets}
\label{mag:sec}
The below table summarizes the latest studies of the most challenging magnets of the muon collider.
The main performance targets and target ranges (i.e., not yet to specification) of the most challenging magnets of the muon collider are shown Table \ref{mag:tab:developments}.
Though these targets are bound to adapt as the study proceeds, they already provide a good basis to feedback on beam optics and accelerator performance, and to identify outstanding issues to be addressed by future work and dedicated R\&D. 
The whole accelerator complex functions in steady state, apart from the fast ramped magnets in the rapid cycling synchrotrons.

Specific details on the 6D cooling solenoids can be found in Appendix \ref{app:mag}.

\begin{table}[!h]
    \centering
    \resizebox{\textwidth}{!}{
    \begin{tabular}{l|cccccccc}
         Complex&  Magnet &No.&  Aper.&  Length&  Field&Grad.&  Ramp rate&  Temp.\\
 Unit& & & [mm]& [m]& [T]&[T/m]& [T/s]&[K]\\ \hline
         Target, capture&  Solenoid Coils &23&  1380&   $\approx$ 0.4 -- 0.8&  2 -- 20 &&  SS&  20\\
         6D cooling&  Solenoid Coils & $\approx$ 6000&  90-1500&  0.08 -- 0.5&  2 -- 17 &&  SS&  4.2-20\\
 Final cooling&  Solenoid Coils & 20 & 50& 0.5& >40 && SS&4.2\\
 RCS&  NC dipole & $\approx$ 1500& 30x100& 5& $\pm$ 1.8 && 4200&300\\
 &  SC dipole & $\approx$ 2500& 30x100& 1.5& 10 && SS&4.2-20\\
         Collider arc&  Dipoles& $\approx$ 1050&  140&  5&  14* &&  SS&  \\
         &  CF& $\approx$ 628&  140&  5 -- 10&  4 -- 8&$\pm$100--$\pm$150*&  SS&  4.2-20\\
         IR &  quadrupoles&  $\approx$ 20&  100 - 280&  5 -- 10&  &$\pm$110 -- $\pm$330**&  SS&  4.2-20\\
    \end{tabular}
    }
    \caption[Summary of main magnet development targets]{Summary of main magnet development targets. For the collider magnet values marked with a * slightly higher values are assumed in the lattice design but no important changes are expected adjusting to the specified performances. The values marked with ** correspond to the lattice design but might be too high for the magnets; the lattice design will be updated accordingly. Specific configurations still need to be evaluated and this is a work in progress. CF stands for combined-function magnets.}
    \label{mag:tab:developments}
\end{table}\FloatBarrier
\section{RF Cavities}
\label{sec:RF}

The RF parameters which should be considered in the design are listed in Table \ref{rf:tab:RFparameters}.
In the other sub-systems of the muon cooling complex: capture, bunch merge, final cooling, etc many different RF frequencies are necessary. It is recommended to keep these RF frequencies as high as reasonable possible from the beam dynamics point of view, since the size of the achievable gradient scales approximately as $\sqrt{(f_{RF})}$.\\
Further details on the designs of the 6D cooling cavities and the RCS cavities can be found in Appendix \ref{app:rf}.

\begin{table}[!h]
\centering
\begin{tabular}{lll|lll}

\multicolumn{6}{l}{\textbf{Proton driver}} \\
LINAC & RF frequencies & MHz & 352 & 704 &  \\ \hline
\multicolumn{6}{l}{\textbf{Muon cooling complex}} \\
\multirow{3}{*}{6D Cooling Channels} & RF frequencies & MHz & 352 & 704 & 1056 \\
 & Max accelerating field (conservative) & MV/m & 22 & 30 & 30 \\
 & Max accelerating field (optimistic) & MV/m & 35 & 50 & 50 \\\hline
\multicolumn{6}{l}{\textbf{Acceleration complex}} \\
\multirow{3}{*}{LINAC} & RF frequencies & MHz & 352 & 704 & 1056 \\
 & Max accelerating field (conservative) & MV/m & 20 & 25 & 30 \\
 & Max accelerating field (optimistic) & MV/m & 30 & 38 & 45 \\
\multirow{3}{*}{RCS} & RF frequency & MHz &  & 704 & 1056, 1300 \\
 & Max accelerating field (conservative) & MV/m &  & 25 & 30 \\
 & Max accelerating field (optimistic) & MV/m &  & 38 & 45 \\
\end{tabular}
\caption{RF frequencies and gradients to be used in the beam dynamics studies.}
\label{rf:tab:RFparameters}
\end{table}\FloatBarrier
\section{Impedance}
\label{sec:imp}
This section is devoted to beam intensity limitations that could be encountered in the different machines due to collective effects.
\subsection{Impedance model for the Rapid Cycling Synchrotrons}
\label{imp:sec:rcs}
Impedance models for the four RCS of the acceleration chain were developed.
The Rapid Cycling Synchrotrons (RCS) will be comprised of many RF cavities to provide the large acceleration voltage needed to reach the muon survival target, as developed in Section \ref{sec:highE}.
It is assumed that the RCS 1, 2, 3 and 4 have respectively 700, 380, 540 and 3000 cavities.
Because of their number, the cavities are expected to be a large contributor to the RCS impedance model.
The models assume that superconducting TESLA cavities~\cite{rf:TESLA_cavity} are used for the RF system, and include the High-Order Modes (HOMs) generated by these cavities~\cite{impedance:bib:tesla_homs}.
Additional details on the RCS impedance models, including the HOMs parameters for a single cavity, are reported in the Appendix, Table~\ref{impedance:tab:HOMS_t}.

The RCS parameters relevant for the impedance and coherent stability simulations are reported in Table~\ref{impedance:tab:collective}.

A second important contributor to the impedance model of the RCS is the normal conducting magnets vacuum chamber.
Because of the high ramping rate, a large eddy current would appear if a fully metallic chamber was used~\cite{bib:kvikne_eddy_currents_imcc_2024}.
A ceramic chamber with a thin metallic coating on the inner surface would therefore be used~\cite{impedance:bib:rcs_vacuum_chamber}.
Its dimension and characteristics are reported in Table~\ref{impedance:tab:rcs2_nc}.

\begin{table}[!h]
\centering
\begin{tabular}{lc|c}
Parameter & Unit & Value \\
\hline
Inner dimension width, height & mm, mm & 30, 20 \\
Inner RF shield (copper stripes) thickness & um & 500 \\
Copper resistivity at 300 K & \si{\nano\ohm\meter} & 17.9 \\
Ceramic thickness & mm & 5 \\
Ceramic type &  & HA-997 \\
Outer dimension width, height & mm, mm & 40, 30 \\
\end{tabular}
\caption{RCS normal conducting magnets vacuum chamber used in simulations}
\label{impedance:tab:rcs2_nc}
\end{table}

\begin{table}[!h]
\centering

\begin{tabular}{l c |c}
Parameter & Unit & Value \\
\hline
Average Twiss beta horizontal/vertical & m & 50, 50 \\
Chromaticity Q' horizontal/vertical &  & +20, +20 \\
Transverse damper & turns & 20 \\
Maximum transverse offset admissible & um & 100 \\
Detuning from octupoles horizontal/vertical & m$^{-1}$ & 0, 0 \\

\end{tabular}
\caption{RCS Collective Effects Parameters used in simulations.}
\label{impedance:tab:collective}
\end{table}

\subsection{Impedance model for the 10 TeV collider ring}
\label{imp:sec:coll}

In the \SI{10}{\tera\eV} collider ring, the main impedance source would be the resistive-wall contribution from the magnets' vacuum chamber.
To protect the superconducting magnet coils from muon decay induced heating and radiation damage, a tungsten shield is proposed to be the inserted in the magnet cold bore as detailed in Section~\ref{sec:radshield} and described in Ref.~\cite{bib:shielding_requirements}.

Previous parametric studies performed with Xsuite and PyHEADTAIL showed that a minimum chamber radius of \SI{13}{\milli\metre}, together with a copper coating on the inner diameter are required to ensure coherent transverse beam stability.
The current dipole magnet radial build detailed in Table~\ref{col:tab:arcs} foresees a \SI{23.5}{\milli\metre} inner radius, with a \SI{10}{\micro\metre} copper coating.
The vacuum chamber properties used for the impedance model computation are summarized in Table~\ref{impedance:tab:collider}.

\begin{table}[!h]
    \centering
    \begin{tabular}{cc|c}
         Parameter&  Unit& Value\\ \hline
         Chamber geometry&  & circular\\
         Chamber length&  \si{\meter} & 10000\\
         Copper coating thickness&  \si{\micro\meter}& 10\\
         Copper resistivity at \SI{300}{\kelvin} &  \si{\nano\ohm\meter}& 17.9\\
         Tungsten resistivity at \SI{300}{\kelvin}&  \si{\nano\ohm\meter}& 54.4\\
         % Min. chamber radius required (50-turn damper)&  \si{\milli\meter}& \\
         Chamber radius (from magnet radial build)&  \si{\milli\meter}& 23\\
 Min. chamber radius required (50-turn damper)& mm&13\\
    \end{tabular}   
    \caption{\SI{10}{\tera\electronvolt} collider parameters for impedance model simulations.}
    \label{impedance:tab:collider}
\end{table}

Transverse coherent beam stability simulations were performed with Xsuite~\cite{Iadarola:2023fuk} and PyHEADTAIL, including the effect of muon beam decay~\cite{impedance:bib:amorim_coll10tev_imcc_2024}.
The beam parameters used for these simulations are summarized in Table~\ref{impedance:tab:collider_param}.

\subsection{Impedance models in cooling}
Studies are ongoing to understand impedance models and other collective effects in the cooling stages.
These are challenging, due to the extra step of studying these behaviours within the absorbers.
In addition, the beam conditions change significantly through each stage.
The beam-loading, wakefields, space-charge and inter-beam scattering in the 6D cooling and final cooling lattices listed in Section \ref{app:cool} will be modelled with RF-Track~\cite{Latina2021RFTrackManual}.\FloatBarrier
\section{Radiation Shielding}
\label{sec:radshield}
The below tables summarize the latest studies of radiation shielding in the muon collider for two particular systems: The target solenoids considering proton impact on a Graphite target; and magnets in the collider arcs and interaction regions due to muon decay.

\subsection{Radiation load on the target superconducting solenoids}
\label{sec:radshield:target}
Generic radiation load studies for the superconducting solenoid were performed by means of  FLUKA Monte Carlo simulations. A 5\,GeV proton beam with a beam sigma of 5\,mm and a beam power of 2\,MW was assumed to impinge on a graphite target rod (see Table~\ref{target:tab:radialbuild} for the target dimensions). The target was centered along the beam axis and therefore no dependence on the azimuthal angle can be expected. The simulation results for the coils are presented in Table \ref{rad:tab:DPA_coils}, showing the maximum displacement per atom (DPA) per year and the maximum yearly absorbed dose. The studies were carried out for different target shielding thicknesses and shielding compositions. The shielding inner radius in the area of the target vessel is fixed at \SI{17.8}{cm}. The gap between the shielding outer radius and the magnet coils is always kept at \SI{7.5}{cm}. The shielding outer radius can be read from the table by subtracting \SI{7.5}{cm} from the magnet coils' inner radius. The target shielding was either assumed to be made of pure tungsten or tungsten with an outer, neutron-absorbing layer made of water combined with boron-carbide.
 
\begin{table}[h!]
\centering\resizebox{1.0\textwidth}{!}{%
\begin{tabular}{cccc}
\hline\hline
%\multicolumn{4}{c}{\textbf{Pure Tungsten}}  \\
%\hline
& \textbf{Tungsten + Water  + Boron-Carbide} & & \\
Coil inner radius & Shielding thickness around the target & DPA/year [$10^{-3}$] & Dose [MGy/year] \\
\hline
%\SI{60}{cm} &  W \SI{34.7}{cm}  & &  \\
%\SI{65}{cm} & W \SI{39.7}{cm} & &  \\
%\SI{70}{cm} & W \SI{44.7}{cm} & &  \\
%\SI{75}{cm} & W \SI{49.7}{cm} & & \\
%\SI{80}{cm} & W \SI{54.7}{cm} & &  \\
%\SI{85}{cm} & W \SI{59.7}{cm} & &  \\
%\hline 
%\multicolumn{4}{c}{} \\
\hline
\SI{60}{cm} & \textbf{(B)W \SI{31.2}{cm} + H\textsubscript{2}O  \SI{2}{cm} + B\textsubscript{4}C \SI{0.5}{cm} + W \SI{1}{cm}} & 1.70 ± 0.02&  10.0 ± 0.3\\
\SI{65}{cm} & W \SI{36.2}{cm} + H\textsubscript{2}O  \SI{2}{cm} + B\textsubscript{4}C \SI{0.5}{cm} + W \SI{1}{cm} &  0.90 ± 0.02&  5.6 ± 0.2\\
\SI{70}{cm} & W \SI{41.2}{cm} + H\textsubscript{2}O  \SI{2}{cm} + B\textsubscript{4}C \SI{0.5}{cm} + W \SI{1}{cm} & 0.49 ± 0.01&  3.1 ± 0.1\\
\SI{75}{cm} & W \SI{46.2}{cm} + H\textsubscript{2}O  \SI{2}{cm} + B\textsubscript{4}C \SI{0.5}{cm} + W \SI{1}{cm} & 0.29 ± 0.01& 1.9 ± 0.1\\
\SI{80}{cm} & W \SI{51.2}{cm} + H\textsubscript{2}O  \SI{2}{cm} + B\textsubscript{4}C \SI{0.5}{cm} + W \SI{1}{cm} &  0.16 ± 0.01&  1.0 ± 0.1\\
\SI{85}{cm} & W \SI{56.2}{cm} + H\textsubscript{2}O  \SI{2}{cm} + B\textsubscript{4}C \SI{0.5}{cm} + W \SI{1}{cm} & 0.09 ± 0.01&  0.6 ±  0.1\\
\hline\hline
\end{tabular}
}
\caption[Radiation load on target solenoids]{Radiation load on the target superconducting magnet coils in terms of the maximum displacement per atom (DPA) and the maximum absorbed dose per year of operation for various shielding configurations.}
\label{rad:tab:DPA_coils}
\end{table}

\subsection{Muon decay in the collider ring}
\label{sec:radshield:collider}
The radiation-induced power load and radiation effects in collider equipment are dominated by the products of muon decay.
While decay neutrinos yield a negligible contribution to the radiation load on the machine, the decay electrons and positrons induce secondary particle showers, which dissipate their energy in the surrounding materials.
A continuous shielding is therefore needed, which dissipates the induced heat and protects the superconducting magnets against long-term radiation damage.
Shielding studies for muon colliders have been previously carried out within MAP~\cite{Mokhov2011PAC,Kashikhin2012IPAC,Mokhov2014IPAC}.
In particular, the shielding must: 
\begin{itemize}
    \item prevent magnet quenches,
    \item reduce the thermal load to the cryogenic system (by reducing the heat load to the cold mass of magnets),
    \item prevent magnet failures due to the ionizing dose in organic materials (e.g. insulation, spacers) and atomic displacements in the superconductor. 
\end{itemize}

The assumed beam parameters and operational scenarios for the radiation studies are summarized in Table~\ref{rad:tab:ring}.

\begin{table}[h!]
\begin{center}
\begin{tabular}{ll|cc}
 &Units& \textbf{\SI{3}{\tera\electronvolt}} & \textbf{\SI{10}{\tera\electronvolt}} \\
\hline
Particle energy  &\si{\tera\electronvolt}& 1.5& 5\\
Bunches/beam  && 1& 1\\
Muons per bunch  &\num{E12}& 2.2& 1.8\\
Circumference  &\si{\kilo\meter}& 4.5& 11.4\\\hline
Muon decay rate per unit length  &\SI[per-mode=reciprocal]{E9}{\per\meter\per\second}& 4.9&  1.8\\
Power ($e^{\pm}$)/meter  &\si{\kilo\watt\per\meter}& 0.411& 0.443\\\hline
Operational years  &years& \multicolumn{2}{c}{5 -- 10}\\
Operational time per year (average)  &days& \multicolumn{2}{c}{139}\\
Operational time per year (average)  &seconds& \multicolumn{2}{c}{\num{1.2E7}}\\
\end{tabular}
\caption[Parameters for collider radiation]{Parameters for radiation studies (collider ring) with \SI{14}{\tesla} arc peak field. The number of decays  consider the contribution of both beams.}
\label{rad:tab:ring}
\end{center}
\end{table}

In order to estimate the required shielding thickness for a \SI{10}{\tera\electronvolt} collider, generic shielding studies for the arc magnets were performed with FLUKA~\cite{Calzolari2022IPAC,Lechner2024IPAC}.
The studies considered only muon decay, whereas other source terms (e.g. beam halo losses) still have to be addressed in the future.
Table~\ref{rad:tab:colliderarcdipoles} summarizes the calculated power load and radiation damage in collider ring magnets as a function of the radial absorber thickness (\SI{10}{\tera\electronvolt} collider).
For simplicity, the FLUKA simulation model consisted of a generic string of \SI{16}{\tesla} dipoles, each six meters long; the drift regions between dipoles were assumed to be \SI{20}{\centi\meter} long.

\begin{table}[!h]
\begin{center}
\begin{tabular}{ll|ccc}
         &Unit&  \textbf{20~mm} & \textbf{30~mm} & \textbf{40~mm} \\
        \hline
        Beam aperture  &\si{\milli\meter}& 23.5& 23.5& 23.5\\
        Outer shielding radius  &\si{\milli\meter}& 43.5& 53.5& 63.5\\
        Inner coil aperture  &\si{\milli\meter}& 59& 69& 79\\
        \hline
        Absolute power penetrating tungsten absorber&\si{\watt\per\meter}& 16.2& 7& 3.5\\
        Fractional power penetrating tungsten absorber&\si{\percent}& 3.7& 1.6& 0.8\\
        Peak power density in coils&\si{\milli\watt\per\centi\meter\cubed}& 5.5& 1.8& 0.6\\
        Peak dose in Kapton insulation (1 year)&\si{\mega\gray}& 9.3& 2.9& 1.1\\
        Peak dose in coils (1 year)&\si{\mega\gray}& 7.5& 2.5& 0.9\\
        Peak DPA in coils (1 year)& \num{E-5} DPA & 1.4 & 1.0 & 0.8 \\
\end{tabular}
\caption[Power load and radiation damage in collider]{Power load and radiation damage in collider ring arc magnets (\SI{10}{\tera\electronvolt}) as a function of the radial tungsten absorber thickness. The power penetrating the shielding does not include neutrinos, since they are not relevant for the radiation load to the machine; the percentage values are given with respect to the power carried by decay electrons and positrons. The results include the contribution of both counter-rotating beams.}
\label{rad:tab:colliderarcdipoles}
\end{center}
\end{table}

In the interaction region, which accommodates the final focus magnets and a chicane for background reduction, more radiation is expected to arrive on the machine elements. This is a consequence of the long straight section between the chicane and the chromaticity correction section, which leads to a build-up of decay products. As a consequence, the radial shielding thickness generally needs to be larger than in the arcs in order to remain below critical dose levels. Moreover, the beam size in this section is substantially larger than the one in the arc sections, therefore increasing the aperture requirements.

In Table \ref{rad:tab:focus_geom}, the different IR magnets and the corresponding ionizing dose is reported. Thicker shielding elements are required for the first three dipoles than for the final focus quadrupoles.

In case of 5 years of operation, the dose would remain below 40~MGy in all magnets, which is considered acceptable. However, the dose would become too high for 10 years of operation, exceeding even 70~MGy for one of the final focus quadrupoles (IQF2). Therefore, in case of an extended operational period, even more stringent requirements on the shielding would be required.

\begin{table}[!h]
    \centering
\begin{tabular}{l|cccc}
\textbf{Name} & \textbf{L [m]} & \textbf{Beam aperture (radius) [cm]} & \textbf{Coil aperture [cm]} & \textbf{Peak TID [MGy/y]} \\ \hline
IB2     & 6  & 10& 16.0 & 1.1\\
IB1     & 10 & 10& 16.0 & 2.7\\
IB3     & 6  & 10& 16.0 & 4.3\\
IQF2    & 6  & 10& 14.0 & 6.8\\
IQF2\_1 & 6  & 9.3& 13.3 & 4.0\\
IQD1    & 9  & 10.5& 14.5 & 1.0\\
IQD1\_1 & 9  & 10.5& 14.5 & 3.2\\
IQF1B   & 2  & 6.2& 10.2 & 5.6\\
IQF1A   & 3  & 4.6& 8.6  & 3.2\\
IQF1    & 3  & 3& 7.0  & 3.1\\
    \end{tabular}
    \caption[Final focusing magnets and dose]{Cumulative ionizing dose in final focus quadrupoles and chicane dipoles located in the insertion region (lattice version 0.8).}
    \label{rad:tab:focus_geom}
\end{table}\FloatBarrier
\section{Radiation Protection}
\label{sec:radpro}
The decay of muons in the collider ring produces very energetic neutrinos that have a non-negligible probability to interact far away from the collider in material near to the Earth’s surface, producing secondary particle showers.
The goal is to ensure that this effect does not entail any noticeable addition to natural radiation and that the environmental impact of the muon collider is negligible, i.e. an effective dose of the order of 10~$\mu$Sv/year, similar, for instance, to the impact from the LHC.
For the environmental impact assessment, detailed studies of the expected neutrino and secondary-particle fluxes are being performed with FLUKA.
The latter can be folded with the realistic neutrino source term taking into account the collider lattice to predict the effective dose and to design suitable methods for mitigation and demonstration of compliance. 

The results of the FLUKA simulations are shown in Tables~\ref{tab:effdose_kernel_1.5TeV} and \ref{tab:effdose_kernel_5TeV} for the 1.5~TeV and 5~TeV muon beams. These results are expressed in terms of dose kernel parameters, i.e. peak and lateral width of the effective dose profile at different baseline distances from the muon decay position.

\begin{table}[!h]
    \centering
    \begin{tabular}{rcccc}
         \hline\hline
         \multicolumn{1}{c}{}& \multicolumn{2}{c}{\textbf{$\mu^-$}} & \multicolumn{2}{c}{\textbf{$\mu^+$}} \\ 
         \textbf{Distance} & \textbf{Peak eff. dose [pSv/decay]} & \textbf{$\sigma$ [m]} & \textbf{Peak eff. dose [pSv/decay]} & \textbf{$\sigma$ [m]} \\ \hline 
       $5$~km  & $2.09 \cdot 10^{-7}$ & $0.17$ & $2.19 \cdot 10^{-7}$  & $0.16$ \\
       $10$~km  & $6.57 \cdot 10^{-8}$ & $0.32$ & $6.56 \cdot 10^{-8}$ & $0.32$ \\
       $15$~km  & $3.28 \cdot 10^{-8}$ & $0.47$ & $3.34 \cdot 10^{-8}$ & $0.46$ \\
       $20$~km  & $1.98 \cdot 10^{-8}$ & $0.60$ & $1.99 \cdot 10^{-8}$  & $0.60$ \\
       $40$~km  & $5.42 \cdot 10^{-9}$ & $1.17$ & $5.49 \cdot 10^{-9}$ & $1.17$ \\
       $60$~km  & $2.53 \cdot 10^{-9}$ & $1.71$ & $2.51 \cdot 10^{-9}$ & $1.71$ \\
       $80$~km  & $1.44 \cdot 10^{-9}$ & $2.29$ & $1.42 \cdot 10^{-9}$  & $2.29$ \\
       $100$~km  & $9.20 \cdot 10^{-10}$ & $2.85$ & $9.21 \cdot 10^{-10}$ & $2.84$ \\
       \hline\hline
    \end{tabular}
        \caption[Neutrino radiation at distances from muon decay (1.5 TeV beam)]{Effective dose kernel parameters in [pSv/decay] of neutrino-induced radiation within soil at different baseline distances from the muon decay, for a muon beam energy of $1.5$~TeV. The peak dose per muon decay and the lateral width of the dose profile ($\sigma$) have been derived from Gaussian fits of the FLUKA results. 
        }
    \label{tab:effdose_kernel_1.5TeV}
\end{table}

\begin{table}[!h]
    \centering
    \begin{tabular}{rcccc}
         \hline\hline
         \multicolumn{1}{c}{}& \multicolumn{2}{c}{\textbf{$\mu^-$}} & \multicolumn{2}{c}{\textbf{$\mu^+$}} \\ 
         \textbf{Distance} & \textbf{Peak eff. dose [pSv/decay]} & \textbf{$\sigma$ [m]} & \textbf{Peak eff. dose [pSv/decay]} & \textbf{$\sigma$ [m]} \\ \hline 
       $5$~km  & $1.57 \cdot 10^{-5}$ & $0.05$ & $1.63 \cdot 10^{-5}$  & $0.05$ \\
       $10$~km  & $4.86 \cdot 10^{-6}$ & $0.10$ & $5.38 \cdot 10^{-6}$ & $0.10$ \\
       $15$~km  & $2.54 \cdot 10^{-6}$ & $0.15$ & $2.70 \cdot 10^{-6}$ & $0.14$ \\
       $20$~km  & $1.56 \cdot 10^{-6}$ & $0.19$ & $1.55 \cdot 10^{-6}$  & $0.20$ \\
       $40$~km  & $4.80 \cdot 10^{-7}$ & $0.37$ & $4.62 \cdot 10^{-7}$ & $0.38$ \\
       $60$~km  & $2.33 \cdot 10^{-7}$ & $0.54$ & $2.22 \cdot 10^{-7}$ & $0.55$ \\
       $80$~km  & $1.38 \cdot 10^{-7}$ & $0.71$ & $1.31 \cdot 10^{-7}$  & $0.73$ \\
       $100$~km  & $9.16 \cdot 10^{-8}$ & $0.87$ & $8.63 \cdot 10^{-8}$ & $0.90$ \\
       
         $200$~km& $2.55\cdot 10^{-8}$ & 1.64& $2.30\cdot 10^{-8}$&1.75\\
         $300$~km& $1.11\cdot 10^{-8}$ & 2.52& $1.07\cdot 10^{-8}$ &2.56\\
         \hline\hline
    \end{tabular}
        \caption[Neutrino radiation at distances from muon decay (5 TeV beam)]{Effective dose kernel parameters in [pSv/decay] of neutrino-induced radiation within soil at different baseline distances from the muon decay, for a muon beam energy of $5$~TeV. The peak dose per muon decay and the lateral width of the dose profile ($\sigma$) have been derived from Gaussian fits of the FLUKA results. 
        }
    \label{tab:effdose_kernel_5TeV}
\end{table}

The neutrino flux density arising from the collider ring arcs is expected to be reduced to a negligible level by deforming the muon beam trajectory, achieving a wide-enough angular spread of the neutrinos. 
Wobbling of the muon beam within the beam pipe would be sufficient for 1.5 TeV muon beam energy.
At 5 TeV muon beam energy, the beam line components in the arcs may have to be placed on movers to deform the ring periodically in small steps such that the muon beam direction would change over time. 
%Table~\ref{tab:wobbling_dose_5TeV} presents the effective dose within soil similar to Table~\ref{tab:effdose_kernel_5TeV}, but taking into account the vertical deformation of the beam within $\pm$1 mrad by the movers. It results in a reduction factor of 80-90 of the saturated dose kernels within soil.

\begin{figure}[!h]
    \centering
    \includegraphics[width=0.65\textwidth]{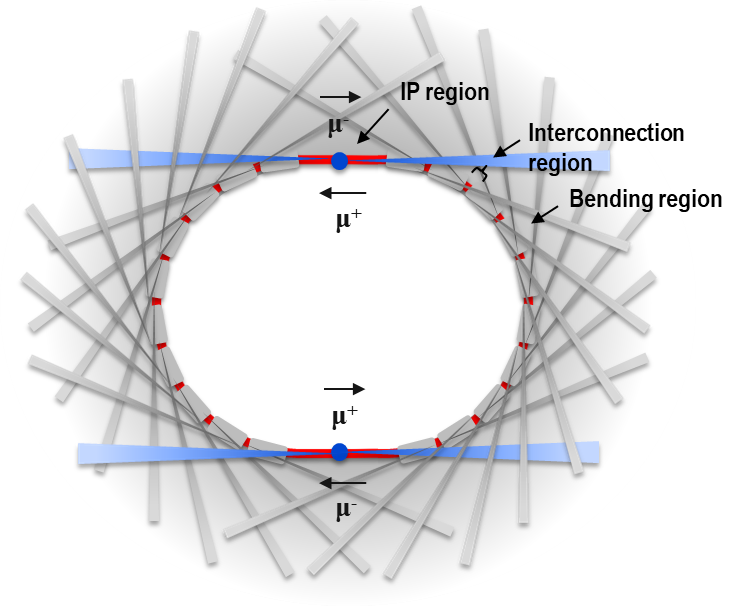}
    \caption{Conceptual view of the collider ring elements contributing to the neutrino fluxes emitted by the collider.}
    \label{fig:coll_elements}
\end{figure}

\begin{figure}[!h]
    \centering
    \includegraphics[width=\textwidth]{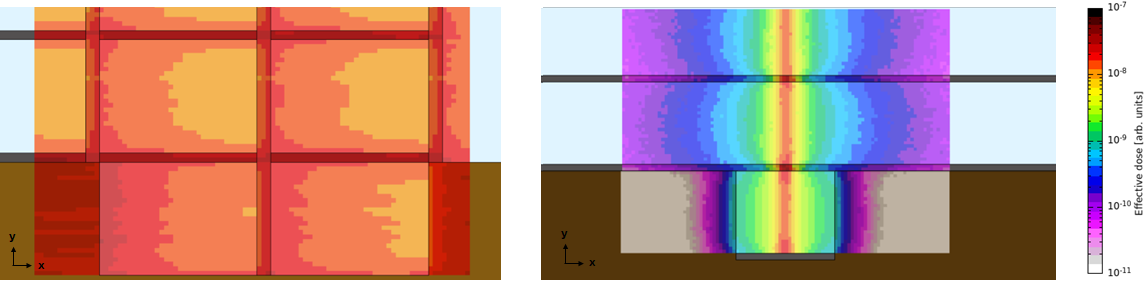}
    \caption{Side (left) and cross-sectional view (right) of the effective dose (in arb. units) for an underground building structure exposed to the neutrino flux from the decay of negative muons in an interconnection region, after the vertical deformation by the movers.}
    \label{fig:cellar_geo}
\end{figure}

\begin{figure}[!h]
    \centering
    \includegraphics[width=\textwidth]{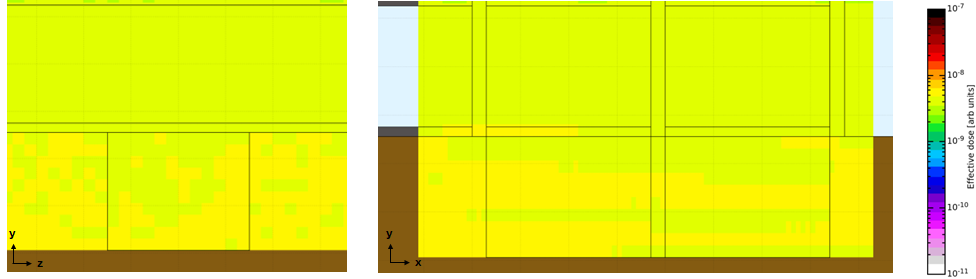}
    \caption{Side (left) and cross-sectional view (right) of the effective dose (in arb. units) for an underground building structure exposed to the neutrino flux from the decay of negative muons within a bending region, after the vertical deformation by the movers.}
    \label{fig:cellar_geo_bends}
\end{figure}

Figure \ref{fig:coll_elements} shows how neutrino fluxes emerge from different elements of the collider lattice. The long straight sections surround the IP regions result in very intense, but localized neutrino fluxes which are directed to unpopulated areas with the careful collider placement.
To estimate the doses from the neutrinos emerging from the interconnection and the bending regions, realistic yet conservative exposure scenarios were investigated, including building structures below and above the ground. The most conservative geometry considered are two consecutive underground rooms that are aligned along the neutrino flux path. The resulting neutrino induced dose distributions for the interconnection and the bending regions are illustrated in Figs~\ref{fig:cellar_geo} and \ref{fig:cellar_geo_bends}, respectively. Assuming
a conservative annual exposure scenario with a 100\% occupancy in the two underground rooms
would lead to a respective effective dose for various relevant distances as given in Tables~\ref{tab:cellar_wobbling_dose_5TeV} and \ref{rad:tab:bend5tev}.

Table \ref{rad:tab:bend5tev} shows the dose contribution from a bending section, for which the collider is approximated as an ideal circle of 1.36 km %1355.65 m 
radius. %The unit comes from the fact that the above numbers have to be only multiplied by decays/meter/year to obtain the annual dose. The bending section length is already embedded in the calculation. 
As can be seen in Fig.~\ref{fig:cellar_geo_bends}, the vertical smearing effect of the movers is overlapped with the horizontal bending of the muon beam resulting in neutrino flux uniformly distributed in the plane perpendicular to the neutrino direction. It was found that the neutrino-induced dose from a bending section is approximately proportional to the square of the muon beam energy.

\begin{table}[!h]
\centering
\begin{tabular}{lcllll}  \hline \hline
 & \multicolumn{2}{c}{$\mu^-$}& \multicolumn{2}{c}{$\mu^+$}&Summed\\
\textbf{Distance [km]} & \textbf{Avg eff. dose [pSv m/decay] }&  & \textbf{Avg eff. dose [pSv m/decay]} &  & \textbf{[pSv m/decay]} \\
\hline
% 5 & 1.19E-08 &  & 1.21E-08 &  & 2.40E-08 \\
% 10 & 3.06E-09 &  & 3.09E-09 &  & 6.15E-09 \\
15 & \num{1.37E-09} &  & \num{1.38E-09} &  & \num{2.75E-09} \\
20 & \num{7.70E-10} &  & \num{7.79E-10} &  & \num{1.55E-09} \\
30 & \num{3.43E-10} &  & \num{3.47E-10} &  & \num{6.90E-10} \\
40 & \num{1.93E-10} &  & \num{1.95E-10} &  & \num{3.88E-10} \\
60 & \num{8.58E-11} &  & \num{8.67E-11} &  & \num{1.73E-10} \\
80 & \num{4.82E-11} &  & \num{4.88E-11} &  & \num{9.70E-11} \\
100 & \num{3.09E-11} &  & \num{3.12E-11} &  & \num{6.21E-11} \\
200 & \num{7.72E-12} &  & \num{7.81E-12} &  & \num{1.55E-11} \\
300 & \num{3.43E-12} &  & \num{3.47E-12} &  & \num{6.90E-12} \\
 \hline \hline
\end{tabular}
\caption[Effective dose for underground structures due to neutrinos from bending magnets (5 TeV w/ movers)]{%Realistic geometry, bending magnets with ideal circle collider - above-ground \& movers case: Neutrino radiation at distances from muon decay (5 TeV beam) }%
Effective dose in [pSv m/decay] of neutrino-induced radiation for an underground building structure at different baseline distances from the muon decay after the vertical deformation by the movers is applied. The muon beam energy is \SI{5}{TeV} and the neutrinos are assumed to emerge from a bending magnet inside the collider arcs, where the mover system is employed. The dose unit comes from the convention where the quoted numbers have to be only multiplied by decays/meter/year to obtain the annual dose. The bending section length is already embedded in the calculation.}
\label{rad:tab:bend5tev}
\end{table}

The dose evaluation for a straight section in an interconnection region is shown in Table~\ref{tab:cellar_wobbling_dose_5TeV}. Assuming \SI{1}{m} of a straight section length, the values for a single beam are comparable with the summed numbers of a bending section shown in Table~\ref{rad:tab:bend5tev}. Due to overlaps, the dose from a bending section always has to take into account a contribution from both beams, while an interconnection straight section generally contributes from a single beam. More studies are needed to understand possible overlaps of neutrino fluxes emerging from the interconnection regions of a realistic collider lattice.

\begin{table}[!h]
\centering

\begin{tabular}{rcc}
\hline \hline
 & \textbf{$\mu^-$} & \textbf{$\mu^+$} \\
\textbf{Distance [km]} &\textbf{ Avg eff. dose [pSv/decay]} &\textbf{ Avg eff. dose [pSv/decay]} \\
\hline
15 & \num{6.55E-09} & \num{6.66E-09} \\
20 & \num{4.74E-09} & \num{4.88E-09} \\
30 & \num{3.04E-09} & \num{3.14E-09} \\
60 & \num{1.32E-09} & \num{1.35E-09} \\
100 & \num{6.55E-10} & \num{6.58E-10} \\
\hline \hline

\end{tabular}
\caption[Effective dose for underground structures due to neutrinos from arc interconnects (5 TeV w/ movers)]{%Neutrino radiation for the double cellar \& movers case at distances from muon decay (5 TeV beam).}
Effective dose in [pSv/decay] of neutrino-induced radiation for an underground building structure at different baseline distances from the muon decay after the vertical deformation by the movers is applied. The muon beam energy is \SI{5}{TeV} and the neutrinos are assumed to emerge from a collider straight section inside the arcs, where the mover system is employed.}
\label{tab:cellar_wobbling_dose_5TeV}
\end{table}

% \begin{table}[!h]
% \centering

% \begin{tabular}{rcc}
% \hline \hline
%  & \textbf{$\mu^-$} & \textbf{$\mu^+$} \\
% \textbf{Distance [km]} & \textbf{Avg eff. dose [pSv/decay]} & \textbf{Avg eff. dose [pSv/decay]} \\
% \hline
% 15 & \num{4.12E-08} & \num{4.10E-08} \\
% 200 & \num{1.15E-08} & \num{1.25E-08} \\
% 300 & \num{6.30E-09} & \num{6.90E-09} \\
% \hline \hline

% \end{tabular}
% \caption{Effective dose of neutrino-induced radiation for the above-ground case with realistic geometry and no movers. Neutrino radiation at distances from muon decay (5 TeV beam)}
% \end{table}

%Instead of considering the saturated effective dose within soil, which is unrealistic for an annual exposure, various more realistic, yet conservative scenarios are under investigation, including exposure in building structures below and above the ground. The most conservative of these is illustrated in Figure~\ref{fig:cellar_geo}, where two consecutive underground rooms are aligned along the neutrino flux path. Assuming a very conservative annual exposure scenario with a 100\% occupancy in the two underground rooms would lead to a effective dose for various relevant distances as given in Table~\ref{tab:cellar_wobbling_dose_5TeV}.
\FloatBarrier
\section{Demonstrators}
\label{demo:sec}

The Muon Cooling Demonstrator Programme will be an essential component of the muon collider R\&D programme. Muon cooling is required in order to deliver the required luminosities but it is a technology that has not been fully proven.

The muon cooling demonstrator will demonstrate
\begin{itemize}
\item Successful integration of cooling equipment.
\item Operation of the cooling equipment with beam.
\item Delivery of required beam physics performance.
\end{itemize}

\begin{figure}[!h]
    \centering
    \includegraphics[width=0.75\textwidth]{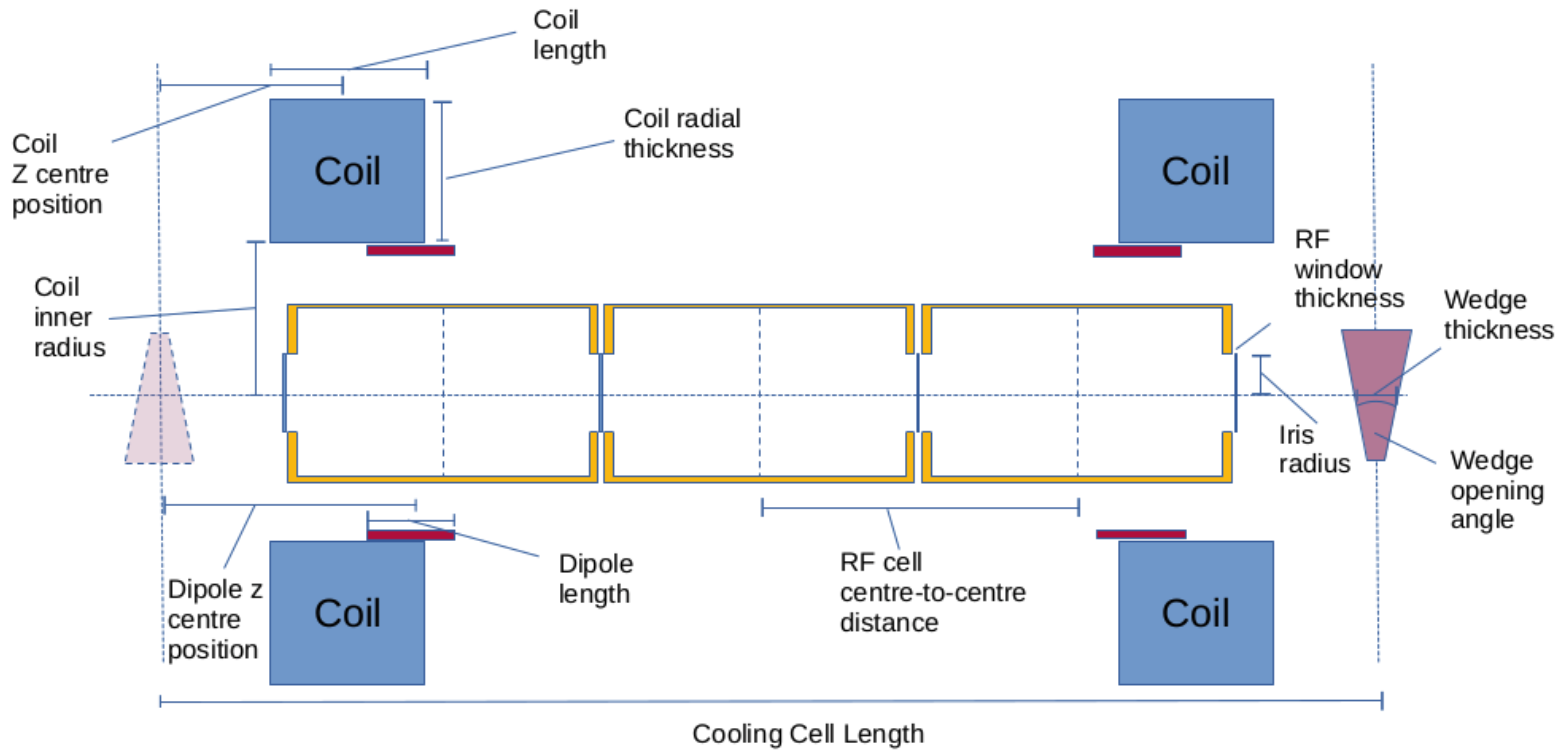}
    \caption{Cooling Cell Schematic showing individual elements.}
    \label{coolcell:fig:schematic}
\end{figure}

Delivery of the demonstration of muon cooling will require a programme of R\&D to understand and mitigate risks surrounding construction of the cooling lattice. The principle issues are:

\begin{enumerate}
\item The cooling cell has RF cavities and solenoids in close proximity. Solenoid fields are known to induce RF breakdown which must be understood in detail.
\item Warm-cold interfaces between adjacent RF cavities and solenoids require careful attention to thermal management.
\item Integration of ancillary equipment such as vacuum, RF power and beam instrumentation may be very challenging to implement in such a compact lattice.
\item Beam instrumentation must enable suitable commissioning of the equipment. For the beam demonstration in particular, where muon rates may be low compared to conventional beams, suitable instrumentation must still be implemented. 
\item The integrated facility must be operable in a routine manner. 
\end{enumerate}
In order to deliver this, a staged R\&D programme is envisaged, with each stage demonstrating the technology more fully:
\begin{enumerate}
\item Several RF test stands will be constructed to understand the limits to RF gradient that can be achieved in the presence of high-field solenoids.
\item A one-cell module will be implemented in order to test the operation of RF cavities in an operational magnetic environment.
\item A multi-cell module will be implemented to demonstrate integration of absorber, RF and magnets.
\item The multi-cell module will be operated with beam in order to demonstrate commissioning and operation of the cooling equipment with beam.
\item A cooling line comprising several cooling modules will be implemented to demonstrate beam physics performance.
\end{enumerate}

The Collaboration has adopted the terminology in Fig. \ref{coolcell:fig:schematic} to designate the elements of a cooling cell. The chosen cooling cell to implement is related to the B5 rectilinear cooling cell. Several important design differences have been implemented compared to B5. The Demonstrator cooling cell parameters are chosen as a compromise between cost and technical challenge. The design will inform subsequent design of the muon collider cooling system.

\begin{figure}[!h]
    \centering
    \includegraphics[width=0.75\textwidth]{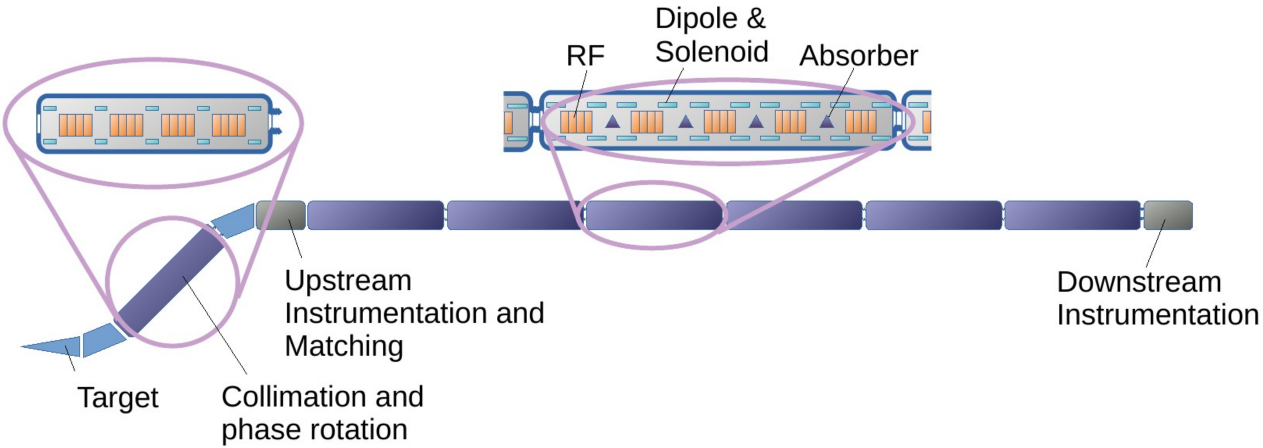}
    \caption{Cooling demonstrator conceptual layout.}
    \label{demo:fig:layoutschematic}
\end{figure}

The cooling channel would be composed of a series of cooling cells grouped into vacuum vessels, as shown in Fig \ref{demo:fig:layoutschematic}. The cooling performance for two different cooling channel lengths is listed in Tab. \ref{demo:tab:coolingperformance}, while the main parameters of the most recent demonstrator cell layout are reported in Tab. \ref{demo:tab:coolingcell}.
\begin{table}[!h]
    \centering
    \begin{tabular}{ll|cc} 
         \textbf{Simulated cooling performance}&  \textbf{Unit}&  \textbf{Start value}&  \textbf{End value (50 m)} 
\\ \hline
         Transverse emittance&  \si{\milli\meter} &   1.85 & 1.46 
\\ 
         Longitudinal emittance&  \si{\milli\meter} &  3.20 &  2.81  \\
         Transmission &  \% &  100 &  95.4  \\
    \end{tabular}
    \caption{Simulated cooling performance.}
    \label{demo:tab:coolingperformance}
\end{table}

%\begin{table}[!h]
%    \centering
%    \begin{tabular}{ll|ccc} 
%         \textbf{Simulated cooling performance}&  \textbf{Unit}&  \textbf{Start value}&  \textbf{End value (32 m)}& \textbf{End value (48 m)} 
%\\ \hline
%         Transverse emittance&  \si{\milli\meter} &  2.37&  1.61& 1.44
%\\ 
%         Longitudinal emittance&  \si{\milli\meter} &  4.99&  3.89& 3.58
%\\ 
%         6D emittance&  \si{\milli\meter\cubed} &  26.21&  9.77& 7.12\\ 
%    \end{tabular}
%    \caption{Simulated cooling performance.}
%    \label{demo:tab:coolingperformance}
%\end{table}

% Rogers:

\begin{table}[!h]
    \centering
    \begin{tabular}{lccc}
        \textbf{Parameter} & \textbf{Unit} & \textbf{Value} \\ \hline
         Cooling Cell Length&  mm& 1000\\
         \multicolumn{3}{c}{\textbf{Beam Physics}}\\ \hline
         Momentum&  MeV/c& 200\\
         Twiss beta function&  mm& 130\\
         Dispersion in X&  mm& -61.5\\
         Dispersion in Y&  mm& -19.7\\
         Beam Pipe Radius&  mm& 81.6\\
         \multicolumn{3}{c}{\textbf{Solenoid Parameters}}\\ 
         &  Unit & Value & Tol\\\hline
         B0&  T& 7 & 0.2\\
         B0.5&  T& 0 & 0.016\\
         B1& T& 1 & 0.02\\
         B2& T&0 & 0.4 \\  \hline
         \multicolumn{4}{c}{\textbf{Coil Geometry}}\\
         \textbf{Parameter} & \textbf{Unit} & \textbf{Coil 1} & \textbf{Coil 2} \\ \hline
         Geometry & -- & \multicolumn{2}{c}{B5-DEMO-MAG-2.4}\\
         Inner Radius& mm&285 & 185\\
         Length& mm&211&63.4\\
         Radial Thickness& mm&76.2&71.7\\
         Z Centre Position& mm&251.8&88.1\\
         Pancake length & mm & 12& 12\\
         Spacer length & mm & 7.9& 13.7\\
         Number pancakes & -- & 11& 3\\
         Current Density& A/mm$^2$& 403.5&632.3\\
         \multicolumn{3}{c}{\textbf{RF Cavity}}\\ \hline
         Centre-to-centre distance& mm&177.5\\
         Gradient E0& MV/m&30\\
         Iris Radius& mm&60\\
         Number of RF Cells& &3\\
         Frequency& GHz&0.704\\
         Synchronous Phase& degree&20\\
         Window Thickness& mm&0.1\\
%         Coupling factor & & 1.2 \\
%         Pulse duration & \si{\micro\second} & 11 \\
%         Shunt impedance (R/Q) & \si{\ohm} & 167 \\
%         Window material & & Beryllium \\ 
%         Total peak RF input power & \si{\mega\watt} & 10 \\
%         Total power dissipation & \si{\mega\watt} & 8.31\\
%         Duty factor (\SI{15.6}{\second} repetition) && \num{6.4E-6}\\
         \multicolumn{3}{c}{\textbf{Wedge}}\\ \hline
         Material& &LiH\\
         Opening Angle& degree&10\\
         Thickness& mm&20\\
         Alignment& &Horizontal\\
         \multicolumn{3}{c}{\textbf{Dipole}}\\ \hline
         Length& mm&100\\
         Polarity& &+ - - +\\
         Field& T&0.2\\
         Z Centre Position& mm&160\\
         Field Direction & &Vertical\\
    \end{tabular}
    \caption{Cooling Cell Table}
    \label{demo:tab:coolingcell}
\end{table}\FloatBarrier
\clearpage
\appendix
\section{Appendix: Top-Level Parameters}
\label{app:top}

Additional parameters relating to the Muon Collider staging options can be found in Table \ref{app:facility_param}.

\begin{table}[!h]
\centering
  \begin{tabular}{|c|c|c||c|c|}   \hline
    Parameter & Symbol & Unit & Stage 1 &Stage 2  \\ \hline
    Centre-of-mass energy & $E_{\mathrm{cm}}$ & TeV & 3 & 10\\
    Target integrated luminosity & $\int{\cal L}_{\mathrm{target}}$ & $\rm ab^{-1}$ & 1 & 10 \\
    Estimated luminosity & ${\cal L}_{\mathrm{estimated}}$ & $10^{34}\rm cm^{-2}s^{-1}$ & 2.1 & 18 \\
    Collider circumference& $C_{\mathrm{coll}}$ & $\rm km$ & 4.5 & 11.4 \\
    Collider arc peak field& $B_{\mathrm{arc}}$ & $\rm T$ & 11 & 14 \\
    Luminosity lifetime & $N_{\mathrm{turn}}$ &turns& 1039 & 1363 \\ 
    \hline
    Muons/bunch & $N$ & $10^{12}$ & 2.2 & 1.8 \\
    Repetition rate & $f_{\mathrm{r}}$ & $\rm Hz$ & 5 & 5\\
    Beam power  & $P_{\mathrm{coll}}$ & $\rm MW$ &5.3  & 14.4 \\
    RMS longitudinal emittance& $\varepsilon_\parallel$ & $\rm eVs$ & 0.025 & 0.025 \\
    Norm.\,RMS transverse emittance& $\varepsilon_\perp$ & \textmu m & 25 & 25 \\
    \hline
    IP bunch length& $\sigma_z $ & $\rm mm$ & 5 & 1.5\\
    IP betafunction& $\beta $ & $\rm mm$ & 5 & 1.5\\
    IP beam size& $\sigma $ & \textmu m & 3 & 0.9 \\
    \hline
    Protons on target/bunch & $N_{\mathrm{p}}$ & $10^{14}$ & 5 & 5 \\
    Proton energy on target  & $E_{\mathrm{p}}$ & $\rm GeV$ & 5 & 5\\
    \hline
  \end{tabular}
\caption[Target parameters for a muon collider for Stage 1 at 3 TeV and Stage 2 at 10 TeV]{Target parameters for a muon collider for Stage 1 at 3 TeV and Stage 2 at 10 TeV. The estimated luminosity refers to the value that can be reached if all target specifications can be reached, including beam-beam effects.}
\label{app:facility_param}
\end{table}

\subsection{Luminosity assumptions}
\label{app:top:lumi}

The luminosity of the muon collider is estimated taking into account several effects.
The beams will perform collisions with the bunch charges $N$ decreasing with time $t$ following
\begin{equation}
N(t)=N_0\exp\left(-\frac{t}{\gamma\tau}\right)
\end{equation}
It should be noted that in the 10 TeV machine after $200\;\rm ms$ still about 15\% of the charge remain, which corresponds to 2\% of the integrated luminosity. One can inject a new bunch without removing the old one. However, for the current luminosity estimate, we assume that the beam is being removed. In contrast, for the radiation load on the detector and arcs as well as the neutrino flux, we assume that the bunch is not being removed.
The hourglass effect reduces the luminosity by a factor 0.76 for round beams with the beta-function and the rms bunch length being equal, as can be easily estimated analytically. The beam-beam forces on the other hand increase the luminosity since the beams focus each other, this requires simulations since the disruption parameter is of the order of 1.
A simple estimate of the combination of both effects is produced by running the muon version of GUINEA-PIG using the target beam parameters. The calculation is performed assuming a longitudinally "round" beam, i.e. all particles are distributed with equal density in the space
\begin{equation}
\left(\frac{\Delta E}{\sigma_E}\right)^2
+\left(\frac{\Delta z}{\sigma_z}\right)^2 \le 2^2
\end{equation}
The beam-beam enhancement factor varies with the bunch charge, and reaches up to 24\% at 10 TeV and full charge. We perform the integration over time and find
\begin{equation}
{\cal L}\approx 1.86\times10^{35}\;\rm cm^{-2}s^{-1}
\end{equation}
The result depends on the actual charge distribution in the bunch, the full model is being developed. However, using other longitudinal profiles with the same RMS bunch length, such as Gaussian distributions or a constant charge profile yield very similar results. We thus use
\begin{equation}
{\cal L}\approx 1.8\times10^{35}\;\rm cm^{-2}s^{-1}
\end{equation}
For the 3 TeV parameters the disruption is higher.

The above luminosity can be roughly estimated analytically using the following assumptions:

\begin{itemize}
  \item One bunch of $\mu^+$ is colliding against one bunch of $\mu^-$ (as for the same total number of particles, it is more efficient to have all the particles in one bunch),
  \item Densities are uncorrelated in the three planes,
  \item Gaussian distributions in the transverse planes,
  \item Same parameters for both bunches,
  \item Round (transverse) beam,
  \item No crossing angle,
  \item No transverse offset,
  \item Ignoring the beam-beam enhancement.
\end{itemize}

\noindent The Muon Collider luminosity formula is typically written as

\begin{equation}
L=\frac{N_0^2 f_0 \gamma}{4 \pi \beta^* \epsilon_\text{n}} F_{\text{HG}}(\beta^* / \sigma_\text{z}) F_{\text{decay}} \quad ,
\label{eq:L_linearised}
\end{equation}

\noindent where $N_0$ is the initial number of muons colliding, $f_0$ is the revolution frequency, $\beta^*$ is the beta function at the interaction point, $\epsilon_\text{n}$ is the normalised transverse emittance and $\sigma_\text{z}$ is the rms longitudinal beam size. Furthermore, $F_{\text{HG}}(\beta^* / \sigma_\text{z})$ describes the usual hourglass effect while $F_{\text{decay}}$ is a new term specific to the muons due to their decay and replenishment. For the first term, assuming that $\beta^*=\sigma_\text{z}$ yields $F_{\text{HG}}(1)~=~0.76$. For the second term, as the muons decay rapidly and new muons arrive with the repetition frequency $f_\text{r}=1/T_\text{r}$, the number of muons needs to be averaged such that

\begin{equation}
<N^2> = \frac{1}{T_\text{r}} \int_{0}^{T_\text{r}}{(N_0e^{-\frac{t}{\gamma \tau_{\upmu0}}})^2dt}= N_0^2 F_{\text{decay}} \quad ,
\end{equation}

\noindent with

\begin{equation}
F_{\text{decay}} = \frac{f_\text{r} \gamma \tau_{\upmu0}}{2}[1-e^{-\frac{2}{f_\text{r} \gamma \tau_{\upmu0}}}] \quad .
\end{equation}

\noindent It is worth reminding that this scheme assumes that there is no muon beam once the new injection arrives, either because muons decayed or because the remaining ones were kicked out. Therefore, the luminosity can be written fully as:

\begin{equation}
L=\frac{N_0^2 f_0 \gamma}{4 \pi \beta^* \epsilon_\text{n}} F_{\text{HG}}(\beta^* / \sigma_\text{z}) \frac{f_\text{r} \gamma \tau_{\upmu0}}{2}[1-e^{-\frac{2}{f_\text{r} \gamma \tau_{\upmu0}}}] \quad .
\end{equation}

\noindent Once applying the aforementioned assumption that $\beta^*=\sigma_\text{z}$,

\begin{equation}
L=\frac{c}{8 \pi^2} \frac{N_0^2 }{\epsilon_\text{n} \epsilon_\text{l}} B_{\text{avg}} \gamma \frac{\sigma_\text{E}}{E} F_{\text{HG}}(1) \frac{f_\text{r} \gamma \tau_{\upmu0}}{2}[1-e^{-\frac{2}{f_\text{r} \gamma \tau_{\upmu0}}}] \quad .
\label{eq:L_full}
\end{equation}

\noindent Here we define $\epsilon_\text{l}[\text{eVs}] = \sigma_\text{t} \frac{\sigma_\text{E}}{E} E$ and $B_{\text{avg}}R=\frac{p}{e}$, where $B_{\text{avg}}$ is the average dipolar magnetic field, $R$ is the average machine radius, $p$ is the muon momentum and $e$ is the elementary charge. Doing the numerical application with the baseline parameters of Table \ref{app:facility_param} ($N_0 = 1.8 \times 10^{12}$, $\epsilon_\text{n} = 25~\upmu\text{m}$, $\epsilon_\text{l} = 0.025~\text{eVs}$, $B_{\text{avg}}=10.5~\text{T}$, $E=5~\text{TeV}$, $\frac{\sigma_\text{E}}{E} = 0.1~\%$, $F_{\text{HG}}(1)=0.76$, $f_\text{r} = 5~\text{Hz}$, $\beta^* = \sigma_\text{z} = 1.5~\text{mm}$), yields $L \approx 18.8\times10^{34}~\text{cm}^{-2}\text{s}^{-1}$. Assuming that one year of run corresponds to $1.2 \times 10^7~\text{s}$ leads to $L_{\text{int}}(1~\text{year}) \approx 2.2~{\text{a}\text{b}^{-1}}$. This means that $10~\text{ab}^{-1}$ can thus be reached in $\sim~5$ years (or $\sim~2.5$ years with two detectors), which should give enough margin for further design and technology studies and a realistic ramp-up of the luminosity.

It is also interesting to normalise the luminosity per beam power, with $P_{\text{1beam}}~[\text{W}] = E~[\text{J}] \times N_0 \times f_\text{r}~[\text{Hz}]$, which yields the following:

\begin{equation}
\frac{L}{P_{\text{1beam}}}=\frac{1}{8 \pi^2 m_{\upmu0} c} \frac{N_0}{\epsilon_\text{n} \epsilon_\text{l}} B_{\text{avg}} \frac{\sigma_\text{E}}{E} F_{\text{HG}}(1) \times F_{\text{energy}} \quad ,
\end{equation}

\noindent where 

\begin{equation}
F_{\text{energy}}=\frac{\gamma \tau_{\upmu0}}{2} [1-e^{-\frac{2}{f_\text{r} \gamma \tau_{\upmu0}}}] \quad .
\end{equation}

\begin{figure}[h]
    \centering
    \includegraphics[width=.45\textwidth]{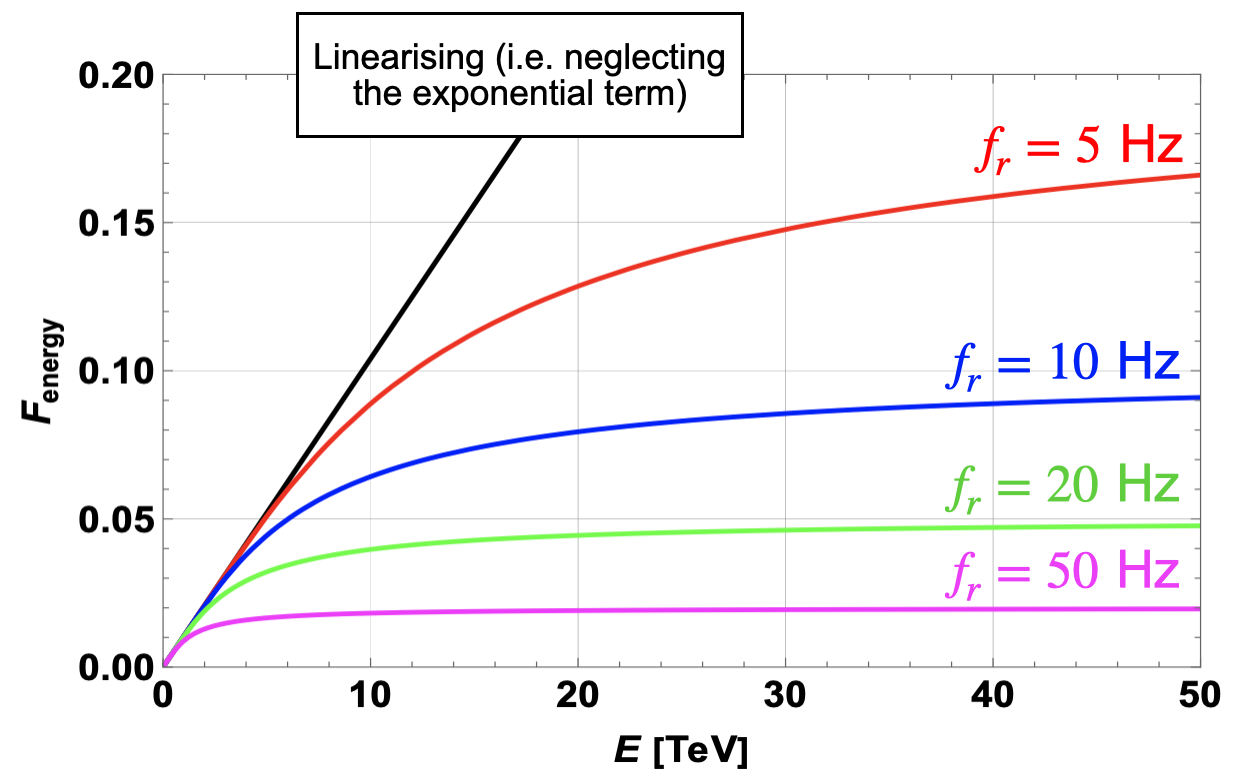}
    \caption{Normalised luminosity per muon beam power as a function of the muon beam energy, assuming that (only) one bunch of $\mu^+$ collides with one bunch of $\mu^-$. This is a worst-case scenario; in reality additional bunches will be stored if $F_{decay}$ becomes significant.}
    \label{fig:LuminosityPerBeamPower}
\end{figure}

\noindent Plotting $F_{\text{energy}}$, the only factor dependant on energy, Fig.~\ref{fig:LuminosityPerBeamPower} is obtained, from which three conclusions can be drawn. First, the derived luminosity formula of Eq. \ref{eq:L_full}, gives the same result as the linearised one in Eq. \ref{eq:L_linearised}(i.e. neglecting the exponential term linked to the muon decay) with the IMCC assumptions ($E=5~\text{TeV}$ and $f_\text{r}=5~\text{Hz}$). Second, these linearised assumptions would however not be true for higher energies and/or higher repetition rates, as the energy factor $F_{\text{energy}}$ converges towards $1/f_\text{r}$. Third, the (linear) luminosity formula can be recovered by using more than one bunch per beam, but then in this case we cannot consider our initial assumption of one bunch of $\mu^+$ colliding against one bunch of $\mu^-$.

Let's consider an integer $n$ and average $N^2$ over $nT_\text{r}$ instead of $T_\text{r}$. This leads to:

\begin{equation}
<N^2> = \frac{1}{nT_\text{r}} \int_{0}^{nT_\text{r}}{(N_0e^{-\frac{t}{\gamma \tau_{\upmu0}}})^2dt}= N_0^2 F_{\text{decay}}(n) \quad ,
\end{equation}

\noindent with

\begin{equation}
F_{\text{decay}}(n) = \frac{f_\text{r} \gamma \tau_{\upmu0}}{2n}[1-e^{-\frac{2n}{f_\text{r} \gamma \tau_{\upmu0}}}] \quad .
\end{equation}

\noindent To conclude: for any energy and repetition rate, one can choose a $n$ such that the exponential term becomes negligible. One then just has to inject the bunches in $n$ different buckets and one recovers the usual luminosity formula (without the exponential term). However, in this case all the beam studies should be performed with these $n$ bunches of $\mu^+$ colliding against $n$ bunches of $\mu^-$. If multiple bunches are stored, a collider design may be implemented with additional interaction points leading to an increase in the physics capability of the facility beyond the baseline assumptions.

\subsection{Muon Decays}
\label{app:top:decay}
Within the main document of Table \ref{top:tab:transmission_estimate}, the total transmission of the muon beams throughout the muon collider complex has been listed. These values include both the losses due to decays and the losses due to beam dynamics effects.
The transmission loss solely due to decays can be safely predicted as it depends only on the energy of the beam, and the length of each system the beam travels through.
This is significant as it dictates the maximum intensity which is physically possible for a perfect system. In addition it allows for an integrated understanding of which systems contribute the most significant decay losses.

\begin{table}[!h]
\centering
\begin{tabular}{l c c c c c c c} \hline \hline
System & Energy In & Energy Out & Lengths & Turns & Transm. & Cumul. Transm. & Total Length \\
& \si{\giga\electronvolt} & \si{\giga\electronvolt} & \si{\meter} & & \% & \% & \si{\meter} \\ \hline
Front End & 0.121 & 0.200 & 150 & 1 & 90.5 & 90.5 & 150 \\
Rectilinear A & 0.200 & 0.162 & 363 & 1 & 80.3 & 72.7 & 513 \\
Rectilinear B & 0.162 & 0.124 & 487 & 1 & 70.6 & 51.3 & 1000 \\
Final Cooling & 0.124 & 0.005 & 100 & 1 & 86.0 & 44.2 & 1100 \\
Pre-Accelerator & 0.005 & 0.250 & 245 & 1 & 77.5 & 34.2 & 1345 \\
LINAC & 0.25 & 1.250 & 500 & 1 & 89.7 & 30.7 & 1845 \\
RLA1 & 1.25 & 5 & 800 & 4.5 & 81.5 & 25.0 & 2645 \\
RLA2 & 5 & 62.5 & 2430 & 4.5 & 92.6 & 23.2 & 5075 \\
RCS1 & 62.5 & 314 & 5990 & 17 & 90.0 & 20.9 & 11065 \\
RCS2 & 314 & 750 & 5990 & 55 & 90.0 & 18.8 & 17055 \\
RCS3 & 750 & 1500 & 10700 & 66 & 90.0 & 16.9 & 27755 \\
RCS4 & 1500 & 5000 & 35000 & 55 & 90.0 & 15.2 & 62755 \\
Collider & 5000 & 5000 & 10000 & 1000 & 72.6 & 11.0 & 72755 \\
\hline \hline
\end{tabular}
\caption{Transmission due to only decays assuming linear change in energy.}
\label{top:tab:decays}
\end{table}

\begin{figure}[!h]
    \centering
    \includegraphics[width=0.9\linewidth]{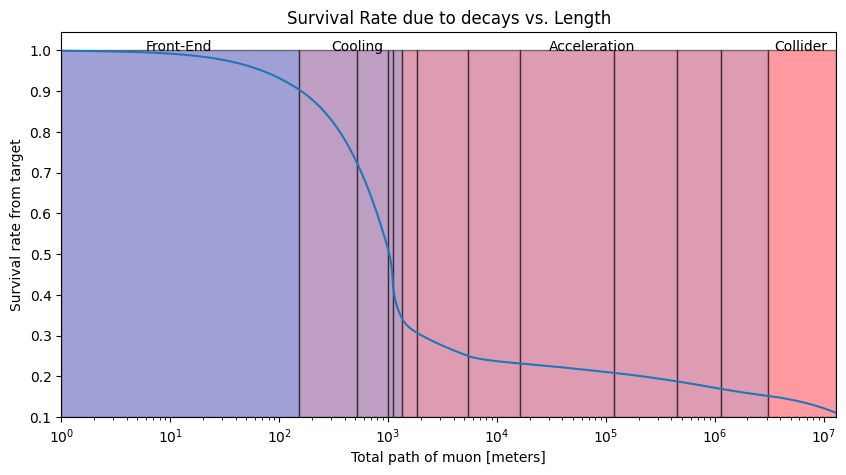}
    \caption{Decrease in survival rate throughout the complex only due to decays, estimated from energy and lengths of each system}
    \label{fig:decays}
\end{figure}

Table \ref{top:tab:decays} shows the transmission for each system only due to decays, which is represented graphically in Figure \ref{fig:decays}.
The total length logarithmically represents the path the muon takes, which is the length of the system multiplied by the number of turns the beam takes.
It is clear that the rate of decay increases most significantly throughout the cooling system. 

 \FloatBarrier
\section{Appendix: Proton Driver}
\label{app:proton}
This section is devoted to the Proton Complex parameters choice. The proton driver of a future Muon Collider is required to deliver a proton-beam of at least 2 MW at a repetition rate of 5 Hz to the pion-production target. The proton-beam energy must be in the multi-GeV range in order to maximize the pion yield. In addition, a particular time structure consisting of a single very short bunch, with a rms pulse length on the order of 2 ns, is needed to allow the muon beam to be captured efficiently in the cooling section. The proton bunch parameters are intimately connected and constrained by beam loading and longitudinal acceptance in the downstream muon accelerator systems and by the acceptance (in time, energy, and power) of the target and pion capture system. The study for the proton complex focused on two different options, the first considers a \SI{5}{\giga\electronvolt} proton beam with a power of \SI{2}{\mega\watt} and the high-level parameters are listed in~\ref{proton:tab:H-main} and~\ref{proton:tab:acc_comp_main}, and the second considers a higher energy and higher power proton beam of \SI{10}{\giga\electronvolt} and \SI{4}{\mega\watt}, and the parameters are listed in Tables~\ref{proton:tab:H-alt} and~\ref{proton:tab:acc_comp_alt}. These two options are equivalent to the luminosity scaling options.

\begin{table}[!h]
\centering
    \caption{H- LINAC parameters for a final energy of 10 GeV.}
\begin{tabular}{ l l l}
Parameters & Unit & main \\
\hline
Final Energy & GeV & 10\\
Repetition Rate & Hz & 5 \\
Max. source pulse length & ms & 5.0\\
Max. source pulse current & mA & 80 \\
Source emittance & mm.mrad & 0.25 \\
Power & MW & 4\\
Linac length & m & 1200\\
RF frequency & MHz & 352, 704 \\
\end{tabular}
\label{proton:tab:H-alt}
\end{table}

\begin{figure}[!h]
  \begin{center}
    \includegraphics[width=0.8\textwidth]{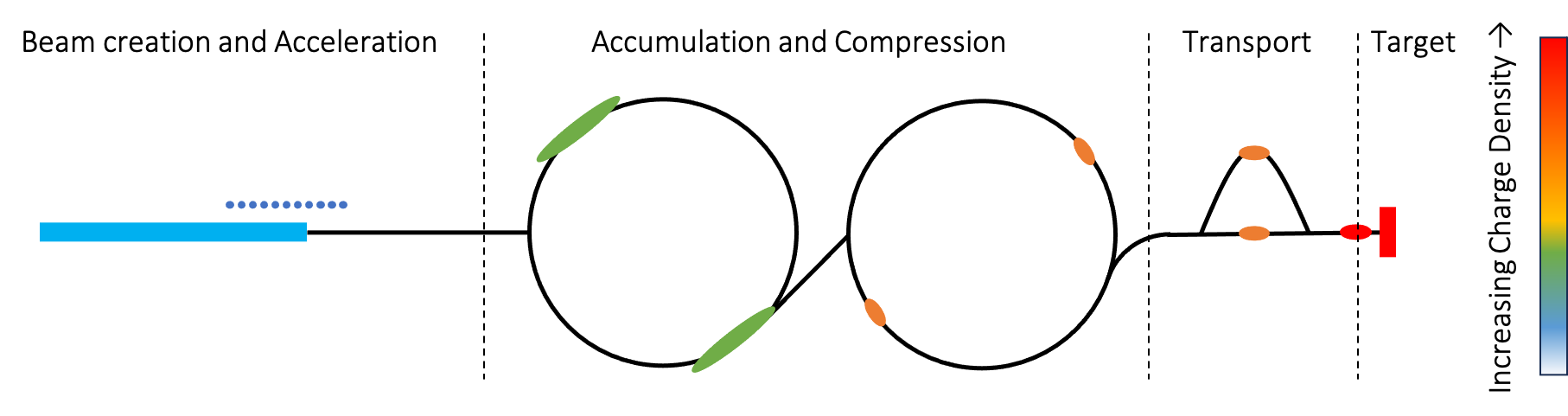}
  \end{center}
  \caption{Schematic of the baseline design for the Proton Complex.}
%\caption{Layout of the main components of the baseline linac\cite{SPLdesign}.}
\label{fig:Baseline}
\end{figure}

Figure~\ref{fig:Baseline} is a schematics of the baseline for the proton complex for both energy options presented. A full energy linac delivers a pulse of H$^-$ to an accumulator ring, after that the pulses are transferred to a compressor ring and rotated longitudinally in order to reach the 2 ns rms bunch length. After the compressor a recombination transport line merge the bunches, which are 2 for now for both cases, and delivers a final single bunch to the target.

\begin{table}[!h]
\centering
    \caption{Alternative Accumulator and Compressor parameters for a 10 GeV beam.}
\begin{tabular}{l l c c}
Parameters & Unit &  Accumulator & Compressor  \\
\hline
Energy & GeV & 10& 10\\
Circumference & m & 300 & 628 \\
Final rms bunch length & ns & 120 & 2 \\
Geo. rms. emit & $\pi$.mm.mrad & 9.5 & 9.5 \\
number of bunches &  & 2 & 2 \\
Number of turns &  & 4167 & 67\\
\hline
RF voltage  & MV & - & 4 \\
RF harmonic &  & - & 2 \\
initial mom. spread & \% & 0.025 & 0.025 \\
final mom. spread & \% & 0.025 & 0.6\\
Protons on target & $10^{14}$ & - & 5 \\
\end{tabular}
\label{proton:tab:acc_comp_alt}
\end{table}

%In the following sections, an overview of the current preliminary parameters for the MuCol proton driver will be outlined and the next steps in the study laid out.

\subsection{High power Linac}
\label{sec:Section111}

The linear accelerator is the first stage of any hadron accelerator complex. The LINAC generates the initial transverse and longitudinal beam emittances and energy spread, defining the beam quality for the next stages of acceleration, accumulation and compression.
%Moreover, the reliability of the linac injector has to be the highest of the entire accelerator complex, any fault of the linac will shut down all other machines. It is also a part of the complex that allows multiple users to be served, it is not by requirement to be a single use machine.
For a project like the Muon Collider, where the repetition rate is low, a high-energy high-power LINAC can be a versatile machine that can serve many other purposes including (and not restricted to) neutrino factories and nuclear science experiments.

%During the study for a Neutrino Factory at CERN, a linac-based solution was adopted, which would be based on the a 5 GeV, high-power version of the Superconducting Proton Linac (SPL), delivering $10^{14}$ protons at the repetition rate of 50 Hz\cite{SPLdesign}. The SPL design still very much actual and can be use as the spring board for the proton driver linac for the a Muon Collider facility at CERN.  In this scenario, the chopped beam from the SPL would be injected into an accumulator ring in order to achieve the required total charge in 1 to 4 bunches which then are transferred to a compressor ring.

\begin{figure}[!h]
  \begin{center}
    \includegraphics[width=0.6\textwidth]{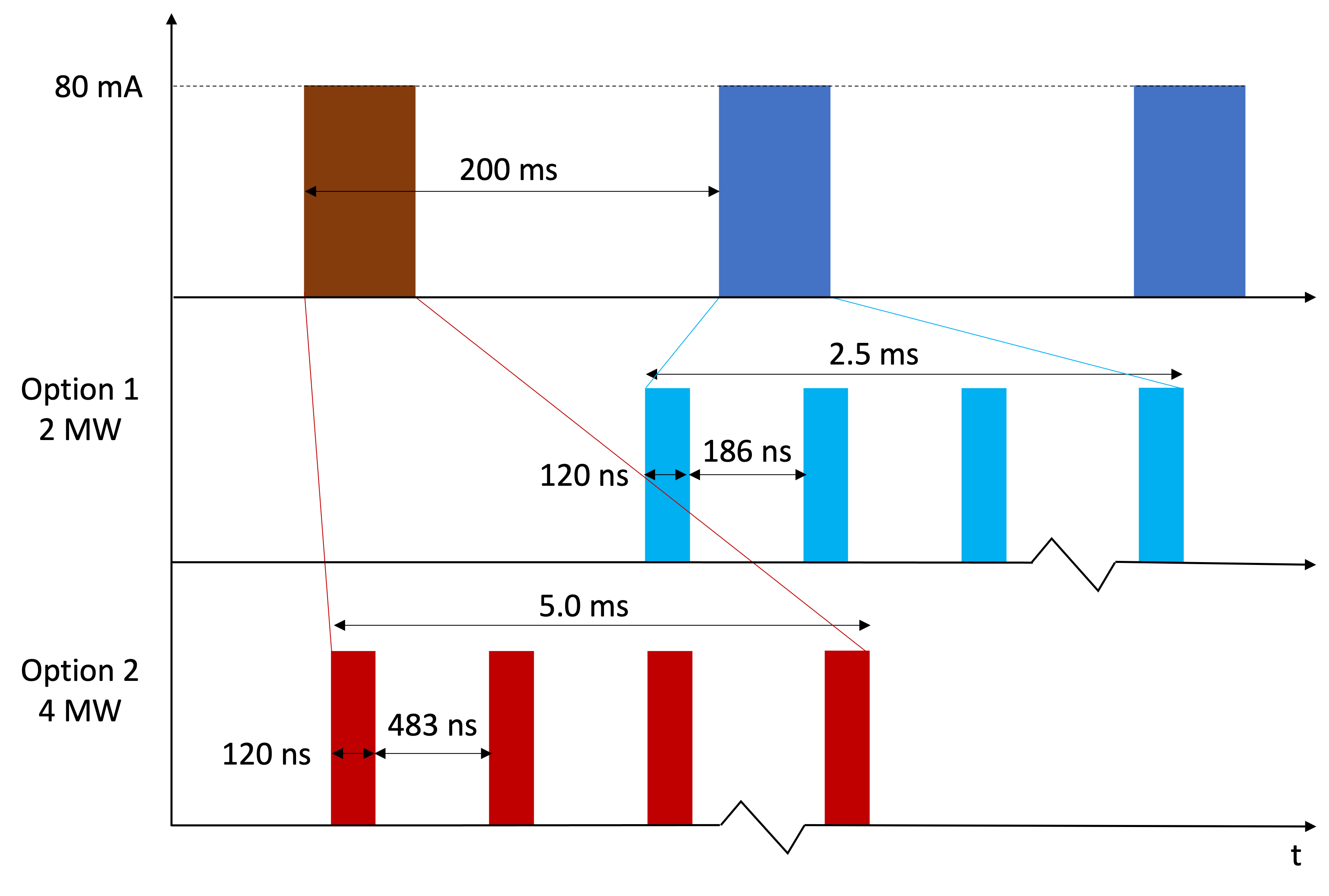}
  \end{center}
  \caption{Possible Chopping scheme for the LINAC considering a double bunch option for both final energies.}
\label{fig:Chopping}
\end{figure}

The main parameters for a Muon Collider LINAC based injector are listed in Tables~\ref{proton:tab:H-main} and~\ref{proton:tab:acc_comp_main} consisting of two options that will drive the final power of the facility. Additional components required for the proton driver includes a H$^{-}$ source for charge-exchange injection, and a low-energy chopper. Fig.~\ref{fig:Chopping} presents a chopping scheme for each of the options where the source current is set at 80 mA and a pulse length from the source pulse length of either 2.5 or 5.0 ms is needed.
   
A study on the losses in the high energy section of a full energy LINAC for both options was carried out~\cite{IPAC24}. The main contributor is black body radiation losses which indicates that the warm sections outside cryomodules and the transfer line to the accumulator ring will have to be cooled to temperatures below 200 K in order to maintain the losses within acceptable levels.

\subsection{Accumulator}

In order to test the stability and possible accumulation schemes for the beam coming from the LINAC we used two lattices developed for the Neutrino Factory at CERN~\cite{Accumulator1,Accumulator2} and based on the parameters listed in Tables~\ref{proton:tab:acc_comp_main} and~\ref{proton:tab:acc_comp_alt}. For both final energies no instabilities are seen throughout the turns needed for accumulation. The total tune spread due to space charge forces is 0.15 for the \SI{5}{\giga\electronvolt}  (\SI{2}{\mega\watt} ) option and less than 0.05 for the \SI{10}{\giga\electronvolt}  (\SI{4}{\mega\watt} ) option.

\subsection{Compressor}

The role of the compressor ring is to transform the 120~ns bunch from the accumulator into a 2~ns bunch, keeping the same bunch intensity. This is done by injecting a longitudinally 'mismatched' bunch into a synchrotron ring with RF cavities.

The current design is inspired by the lattice for short proton bunches for a neutrino factory, which combines a large slippage factor for a fast rotation with a minimized dispersion for a realistic ring aperture~\cite{Compressor}. This is achieved by capping the maximum positive dispersion by adding negative bends. A lattice for the 5 GeV and 10 GeV options exists and simulations including 3D space charge are under way in order to optimized the longitudinal rotation, working points and final emittances.

\subsection{Next Steps}

A new lattice for the linac including cavities with higher geometrical beta ($\beta_g = 1$) we developed and the simulations show a good transport. New calculation on intra-beam stripping of H$^-$ ions for the new lattice was carried out and no major show stopped could be identified, losses are well within the 1 W/m loss budget. 

New simulations for the accumulator for the \SI{5}{\giga\electronvolt} case is also under way where 3D space charge will be used in order to confirm that no major problems are encountered during accumulation.

An alternative compression idea in under study, in this idea there is no need for an accumulator lattice and the current compressor ring is used for final compression. During the simulation for the compression with RF for the \SI{5}{\giga\electronvolt}  case, significant beam emittance blow-up was observed at the end of compression and is under investigation.

%\begin{flushleft}
%\interlinepenalty=10000
%\begin{thebibliography}{99}

%\bibitem{MAP}
%M.A.~Palmer \textit{for the MAP collaboration} ,
%The US Muon Accelerator Program,
%\href{https://doi.org/10.48550/arXiv.1502.03454}
%{doi:10.48550/arXiv.1502.03454}.

%\bibitem{SPLdesign}
%F. Gerigk et al., 
%Conceptual design of the SPL II: A high-power %superconducting $H^{-}$ linac at CERN,
%Tech. Rep. CERN-2006-006, CERN, Geneva, %Switzerland, July, 2006
%{doi:10.5170/CERN-2006-006}.

%\bibitem{IPAC24}
%S. Johannesson \emph{et al.},
%Initial design of a proton complex for the Muon %Collider,
%in \emph{Proc. IPAC'24}, Nashville, TN, May 2024, pp. 2528-2531.
%\url{doi:10.18429/JACoW-IPAC2024-WEPR24}

%\bibitem{Accumulator1}
%M.~Aiba,
%Accumulator Lattice Design For SPL Beam,
%Tech. Rep. CERN-NUFACT-Note-151, CERN, Geneva, Switzerland, February, 2007.

%\bibitem{Accumulator2}
%E.~Benedetto,
%Beam stability in the SPL-Proton Driver accumulator for a Neutrino Factory at CERN,
%Tech. Rep. CERN-NUFACT-Note-156, CERN, Geneva, Switzerland, November, 2009,
%{doi:10.1063/1.3399319}.

%\bibitem{Compressor}
%M.~Aiba,
%Compressor Lattice Design for SPL Beam,
%Tech. Rep. CERN-NUFACT-Note-153, CERN, Geneva, Switzerland, September, 2007.

%\bibitem{Linac4_source}
%E.Sargsyan, 
%LINAC4 H- source studies and future developments,
%IMCC Annual Meeting, 2023, France,
%\href{https://indico.cern.ch/event/1250075/contributions/5356241/}.

%\bibitem{Instabilities}
%R.~Cappi, J.~Gareyte, E.~M\'{e}tral, D.~M\"{o}hl,
%Collective Effects and Final Bunch Rotation in a 2.2 GeV – 44 MHz Proton Accumulator – Compressor for a Neutrino Factory,
%Tech. Rep. CERN/PS 2000-069, CERN, Geneva, Switzerland, September, 2000.

%\end{thebibliography}
%\end{flushleft}
\FloatBarrier
\section{Appendix: Target \& Front-End}
\label{app:target}

\begin{enumerate}
\item The target and solenoid (transverse) capture system, initially with field 15--20 T tapering to 1--2 T, the baseline parameters of which are in Table \ref{target:sec:target-main}. Further information in Section \ref{target:sec:2target} for the baseline design, and Section \ref{target:sec:4target} for the higher-power option;
\item Extraction line for the spent proton beam in Section \ref{front:sec:extraction};
\item Solenoid chicane and proton absorber, in Section \ref{front:sec:chicane};
\item Longitudinal drift;
\item Sequence of RF cavities for bunching, in Section \ref{front:sec:rfbunch};
\item Further sequence of RF cavities for rotating in energy-phase space, also in Section \ref{front:sec:rfbunch}.
\end{enumerate}

\subsection{\SI{2}{\mega\watt} Target and solenoid}
\label{target:sec:2target}

Pions are produced by sending protons onto a graphite target immersed in a strong magnetic field.
Solenoid parameters are listed in Table \ref{mag:tab:developments}.
In the MAP design, resistive magnets (RC1--RC5) were considered, however IMCC is developing a full HTS-based alternative.
The target and target systems are under design, thus no details regarding the expected operation temperature, mechanical response and life-time are listed here. 

Moreover, small discrepancies exist in the components dimensions reported in this chapter, particularly between pion/muon yield studies and design and engineering calculations.
%Both studies are conceptual and have been made in order to probe the sensitivity of different parameters, therefore extensive details of the mechanical design are not included..

Information on the radiation load on the target solenoids is within Section \ref{sec:radshield:target}.

\subsubsection{Production target engineering parameters}
As depicted in Table \ref{target:tab:target_baseline}, the target system is divided into production target, target vessel, target shielding, target shielding vessel, proton beam window and muon beam window. The auxiliary services for cooling of the target and shielding are equally part of the target system and are listed in Table \ref{target:tab:target_baseline-services}.  
For both tables, the main dimensions, key material considerations and important design and integration features are summarized.
%Details about the thermo-mechanical performance of target-systems are out of the scope of this section, and will be published in the interim report.

\begin{table}[!ht]
    \centering
    \resizebox{\textwidth}{!}{
    \begin{tabular}{l|ccc}
     & Material& Box dimensions DxL  
[mm]& Integration \\ \hline \hline
     Production Target& Isostatic Graphite& D30 x L800& Rod supported with transverse CFC \\
 & & &supports attached to cylindrical frame\\ \hline
     Target Vessel& Titanium Grade 5& D346 x L920& Located in the bore of the Target shielding vessel.\\ \hline
     Proton Beam Window& Beryllium& D220 x L0.25& \\ \hline
     Muon Beam Window& Titanium Grade 5& 
D240 x L1& Welded on the vessel\\ \hline
     Target Shielding & Tungsten & D x L2000& Inside Shielding Vessel. Multiple pie-like \\
 & & &blocks stacked together with guiding rods.\\ \hline
     Target Shielding Vessel & Stainless-Steel& D1218 x L2065& Supported by transversal beam \\
 & & &across the cryostat of the solenoid.\\ 
    \end{tabular}
    }
    \caption{Baseline engineering parameters of the carbon target system}
    \label{target:tab:target_baseline}
\end{table}

\begin{table}[!ht]
    \centering
    \resizebox{\textwidth}{!}{
    \begin{tabular}{l|ccccc}
     Cooling& Coolant& Type& Mass Flow& Pressure& Integration \\
 Unit& & & \si{\kilo\gram\per\second}& \si{\bar} &\\ \hline \hline
     Target& Helium& Static / Natural & -& 1& Surrounding target rod \\
 & & convection& & &enclosed by windows and target vessel.\\ \hline
 Target Vessel& Helium& Forced convection& 0.5& 10&Inside double wall target vessel. \\
 & & & & &Routing upstream via the  Solenoid bore.\\ \hline
 Beam window& Helium & Forced convection& 0.005& 1&Double layer window\\
 & & & & &\\ \hline
 Target Shielding& Helium& Forced convection& 0.33& 2&Inside Target shielding Vessel. \\
 & & & & &Routing upstream via the  Solenoid bore.\\ 
    \end{tabular}
    }
    \caption{Baseline engineering parameters of the carbon target auxiliary systems for 2MW.}
    \label{target:tab:target_baseline-services}
\end{table}

The baseline for a 2~MW-class target consists of a solid graphite target. The graphite rod is housed within a double-walled vessel filled with a static helium atmosphere. This helium confinement facilitates the initial stage of heat removal from the graphite rod through natural convection, while raising the sublimation temperature of the graphite when compared to a vacuum environment and providing a non-erosive heat transfer medium. Forced convection cooling is then applied through the vessel's double wall using a 10 bar helium flow.

Titanium is a suitable candidate for the target vessel  due to its low density (reduced interaction with produced pions and muons) and good thermal-shock resistance. However, it is required to use beryllium in proton and muon windows to guarantee a peak power density of approximately $\SI{800}{W/cm^3}$ and yearly DPA around 0.5. On the contrary, adopting titanium would increase these values by an order of magnitude.

The target vessel is surrounded by a helium-cooled, heavy tungsten shield, which reduces power deposition and radiation damage to the solenoid materials to acceptable levels.
For details on the radiation shielding, see Section \ref{sec:radshield}.
The target proximity shielding is housed inside a large stainless-steel vessel, extending from just upstream of the target to around 2 meters downstream. The large size and weight, combined with the need to efficiently extract heat from each tungsten block, resulted in proximity shielding composed of multiple pie-shaped tungsten segments, perforated in specific locations to either guide helium flow or allow for the insertion of longitudinal rods to hold the assembly together. The shielding vessel also hosts a water layer to moderate the neutrons.

Both the cooling and instrumentation routing for the target systems are handled via the upstream side of the assembly.

Downstream of the target and its cooled shielding assembly, the shielding is made of tungsten and has an aperture following the parabolic shape defined in the MAP studies.

\subsubsection{Target radial build}
\label{target:sec:radialbuild}
A preliminary target radial build has been defined and is shown in Table~\ref{target:tab:radialbuild}. This build takes into account a 700~mm inner-radius solenoid coil, the baseline target system dimensions as described in Table~\ref{target:tab:target_baseline}, and the required shielding configuration with a water and Boron-Carbide neutron-absorbing layers (Table~\ref{rad:tab:DPA_coils}). The discrepancy in thickness of tungsten shielding between Table~\ref{target:tab:radialbuild} and Table~\ref{rad:tab:DPA_coils} is explained by the need to integrate other components in the prior as part of the exercise to engineer the entire target-solenoid cryostat. 

\begin{table}[!ht]
    \centering
    \begin{tabular}{l|cccc}
     Component& Material& $r_i$ [mm]& $r_e$ [mm]& ${\Delta}r$ [mm]\\ \hline
     Solenoid coils& HTS& 700& -& -\\
     Insulation& Insulation& 690& 700& 10
\\
     Vacuum& Vacuum& 670& 690& 20
\\
     Thermal shield& Copper \& Water& 651& 670& 19
\\
 Vacuum& Vacuum& 631& 651& 20
\\
 Inner supporting tube& Stainless-steel& 619& 631& 12
\\
 Vacuum& Vacuum& 609& 619& 10
\\
 Outer Target shielding& Tungsten& 599& 609& 10
\\
 Neutron absorber& Boron Carbide& 594& 599& 5
\\
 Target shielding and neutron moderator& Stainless-steel& 589& 594& 5
\\
 & Water& 569& 589& 20
\\
 & Stainless-steel& 564& 569& 5
\\
 & Tungsten& 179& 564& 385
\\
 & Stainless-steel& 174& 179& 5
\\
 Vacuum& Vacuum& 173& 174& 1
\\
 Target vessel& Titanium& 168& 173& 5
\\
 & Helium& 155& 168& 13
\\
 & Titanium& 150& 155& 5
\\
 & Helium& 15& 150& 135
\\
 Target& Graphite& 0& 15&15
\\
    \end{tabular}
    \caption{Target System radial build for a graphite target.}
    \label{target:tab:radialbuild}
\end{table}

\FloatBarrier
\subsection{\SI{4}{\mega\watt} Target and solenoid}
\label{target:sec:4target}
Additional studies on the higher-power \SI{4}{\mega\watt} target option have been performed assuming the initial proton beam of Table \ref{target:sec:target-alt}.
The engineering parameters for such a target are shown in Table \ref{target:tab:target_4MW}, and the cooling in Table \ref{target:tab:target_4MW_cooling}.

\begin{table}[!h]
\centering
\begin{tabular}{l c  |c }
Parameters & Unit & Baseline \\ \hline
Beam power & MW & 4\\
Beam energy & GeV & 10\\
Pulse frequency & Hz & 5\\
Pulse intensity & p+ $10^{14}$ & 5\\
Bunches per pulse &  & 1\\
Pulse length & ns & 2\\
Beam size & mm & 5 or 7.5\\
Impinging angle & \textdegree  & 0\\
\end{tabular}
\caption{Alternative beam from proton driver via carbon target for the \SI{4}{\mega\watt} beam}
\label{target:sec:target-alt}
\end{table}

\begin{table}[!ht]
    \centering
    \resizebox{\textwidth}{!}{
    \begin{tabular}{l|ccc}
     & Material& Box dimensions DxL  
[mm]& Integration \\ \hline \hline
     Production Target& Isostatic Graphite& D30 x L800& Rod supported with transverse CFC \\
 & & &supports attached to cylindrical frame\\ \hline
 Target Forced Convection Inner Vessel& Titanium Grade 5& D42 x L920&Inner vessel to force annular convection. \\
 & & &Suported with the Outer vessel\\ \hline
     Target Vessel& Titanium Grade 5& D346 x L920& Located in the bore of the Target shielding vessel.\\ 
    \end{tabular}
    }
    \caption{Engineering parameters for carbon target for alternative option of 4MW }
    \label{target:tab:target_4MW}
\end{table}

\begin{table}
\centering
\begin{tabular}{ l | l l l l }
Cooling & Coolant & Type & Mass Flow &  Pressure \\
Unit &  &  & \si{\kilo\gram\per\second}& \si{\bar}\\  \hline \hline
Target & Helium & Forced convection & 0.2-0.36&  10 \\
\end{tabular}
    \caption{Baseline engineering parameters of the carbon target auxiliary systems for 4MW.}
    \label{target:tab:target_4MW_cooling}
\end{table}

 \FloatBarrier
\subsection{Front-End}
\label{app:front}
%% ==== START TentativeChapters/Author10-FrontEnd ==== %%
\subsubsection{Chicane and proton absorber}
\label{front:sec:chicane}
The target solenoid is followed by a solenoid chicane which is terminated by a thick beryllium cylinder.
The cylinder absorbs low energy remnant protons which would otherwise irradiate equipment downstream of the chicane.
The concept was initially introduced in \cite{Rogers:2013kna} and initial parameters were defined. 
Further discussion was made in \cite{Stratakis:2014ina}.
In particular, the former study assumed \SI{1.5}{\tesla} solenoid fields, while the MAP and latter study considered \SI{2}{\tesla} solenoid fields in this region.
The latter study also noted that a large proportion of undecayed pions were stopped in the proton absorber which negatively impacted the muon yield.
%In the table below, the parameters from the former study are listed as this came to a definite conclusion for the parameters.

Table \ref{front:tab:chicane} shows the current design parameters for the chicane and the proton absorber.

\begin{table}[h]
    \centering
    \begin{tabular}{l|cc}
     Parameters& Unit& Value\\ \hline
     Chicane bend angle& degree& 15 \\
     Chicane radius of curvature& m& 22 \\
     Proton absorber material& -& Be \\
     Proton absorber thickness& m& 0.1 \\
 Chicane field& T& 1.5 \\
    \end{tabular}
    \caption{Chicane and proton absorber parameters}
    \label{front:tab:chicane}
\end{table}

\subsubsection{Spent proton beam extraction}
\label{front:sec:extraction}
A non-negligible fraction of the primary protons do not have an inelastic nuclear collision in the production target and escape from the graphite rod.
At these energies, the protons are not bent significantly by the chicane and would be lost on the chicane aperture.
In absence of a mitigation strategy, the energy carried by these particles would lead to a high power deposition density in the normal-conducting chicane solenoids. In addition, a high cumulative ionizing dose and displacement damage would be reached within a short operational time. It is therefore necessary to extract the spent protons from the front-end and steer them onto an external beam dump.

Earlier studies explored a possible solution of injecting the proton beam at different angles into the front-end, with an extraction channel envisaged in a gap between the superconducting magnets upstream of the chicane. This concept proved to be unfeasible due to geometrical aspects and the increase of the radiation load to the superconducting coils. As an alternative solution, the spent proton beam could be extracted in the middle of the chicane, by using solenoids with different diameters in order to create a gap for the high-energy protons. Shower simulation studies showed that such an extraction channel in the chicane needs to have a transverse size of a few tens of centimeters, which is challenging for the magnet design. In addition, an internal radiation shielding would be needed to protect the coils from particles, which are still lost in the chicane. The chicane design studies are presently still ongoing. 

%Alternative solutions were explored, . The curvature of the chicane is used to separate protons from lighter particles. The former, weakly impacted by the chicane field, can be extracted with varying chicane aperture providing a gap where the solenoid magnets would be impacted.
%When injecting at different angles, the trajectory is no longer straight, and allows the construction of an extraction channel.
%As shown in Fig. \ref{front:fig:proton_extraction}, the path of the spent proton beam straightens toward the end of the tapering region.
%When adopting this strategy, future magnetic field designs have to be carefully optimized to guarantee the space for the extraction channel.
%% ==== END TentativeChapters/Author10-FrontEnd ==== %%

%\begin{figure}[!ht]
%     \centering
%     \includegraphics[width=0.95\textwidth]{TentativeChapters/Author10-FrontEnd/Figures/TARGET_extraction_channel.pdf}
%     \caption{Trajectory of the beam when injected at different angles. The shallower the angle, the further away the extraction channel has to be put. In the background, the existing magnet configuration is plotted.}
%     \label{front:fig:proton_extraction}
% \end{figure}

\begin{table}[h]
    \centering
    \begin{tabular}{l|cc}
     Parameters& Unit& \\ \hline
     Num. micro bunches& & 21\\
     Longit. emittance& mm& 46\\
     Transv. emittance& um& 17000\\
     Positive muon yield& 1/GeV per p+& 0.024\\
    Negative muon yield& 1/Gev per p+& 0.018\\
    \end{tabular}
    \caption{Outgoing muon beam}
    \label{front:tab:muonbeam}
\end{table}

%% ==== START TentativeChapters/Author10-FrontEnd ==== %%
\subsection{Buncher \& Phase Rotator}
\label{front:sec:rfbunch}
The buncher is comprised of a sequence of RF cavities.
The cavity frequency is chosen to match the distance between nominal RF bunches, so that it varies along the length of the buncher.
The phase is purely bunching.

In the phase rotator, cavities are dephased so that the low energy tail of the beam sees an accelerating gradient and the high energy front of the beam sees a decelerating gradient.

Cavities are placed in a two-cavity LINAC with \SI{0.25}{\meter} separation between adjacent cavity pairs.
Each cavity in the pair is independently phased.
Transversely, the beam is contained in a \SI{2}{\tesla} field.

%The number, frequency and gradient of each cavity is listed for the buncher in table \ref{tab:buncher_rf} and for the rotator in table \ref{tab:rotator_rf}.
%Where 'Number` is listed as 2, this means a single pair of cavities is simulated.
%% ==== END TentativeChapters/Author10-FrontEnd ==== %%
\FloatBarrier
\section{Appendix: Cooling} \label{app:cool}
\subsection{Initial Cooling}
\label{app:initial}

The Helical FOFO Snake (HFOFO) is a design for initial (pre-charge separation) 6D cooling of both signs of muon in a single channel. The HFOFO lattice is composed of alternating-polarity, inclined solenoids, as well as RF cavities and LiH wedge absorbers. Periodic rotations about both the x- and z-axes are applied to the solenoids, as defined by the pitch and roll angles respectively. The effect of these rotations is the generation of a rotating dipole field which enables charge-agnostic focusing. A ``matching section” comprising the first nine solenoids, characterized by unique parameters, is necessary to induce the hallmark helical orbits particles execute in the HFOFO channel. The subsequent portion of the channel, referred to here as the ``steady-state," is built from repeated periods of six units corresponding to six periodic solenoid rotations (where a unit is defined as a set of one solenoid, the RF cavity within it, and the wedge absorber placed after it).
\begin{table}[!h]
\centering
\begin{tabular}{ c | c c c }
Position of unit in lattice & Coil pitch [deg] & Coil roll [deg] & RF gradient [MV/m] \\
\hline
 1 & 0& 0& 20 \\
 2 & 0& 0& 20 \\
 3 & 0.0886& -122.4& 20 \\
 4 &  0.1246& -23.6& 20 \\
 5 & 0.0863& 122.3& 20 \\
 6 & 0.0817& -102.0& 25 \\
 7 &  0.0969& 25.3& 25 \\
 8 &  0.1672& 137.8& 25 \\
 9 & 0.1226& -97.0& 25 \\
 10-end &  0.14& periodic (see additional table) & 25 \\
\end{tabular}
\caption[HFOFO full-channel variable parameters for matching and steady-state sections.]{HFOFO full-channel variable parameters for matching (units 1\textendash 9) and steady-state (units 10\textendash end) sections.}
\label{initialcooling:table1}
\end{table}

Table \ref{initialcooling:table1} contains those parameters which vary along the channel — that is, the rotations of solenoids in the matching and steady-state sections, in addition to the RF gradient. In Table \ref{initialcooling:table2}, a list of parameters which are consistent for the entire channel is given, including the solenoid geometries and further descriptions of the RF system.

\begin{table}[!h]
\centering
\begin{tabular}{ c | c c }
Parameter & Unit & Value \\
\hline
Number of solenoids per period &  & 6 \\
Period length & mm & 4200 \\
Number of periods per channel &  & 30 \\
Coil length & mm & 300 \\
Coil inner radius & mm & 420 \\
Coil outer radius & mm & 600 \\
Spacing between coil centers & mm & 700 \\
RF frequency & MHz & 325 \\
RF length & mm & 249 \\
GH$_2$ density & g/cm$^3$ & 0.014 \\
\end{tabular}
\caption{HFOFO full-channel constant parameters for matching and steady-state sections.}
\label{initialcooling:table2}
\end{table}

The set of six repeated z-rotations (described by roll angles) of solenoids in the steady-state channel are given in Table \ref{initialcooling:table3}, as are the angles of the repeated wedge absorber rotations (about the z-axis).

\begin{table}[!h]
\centering
\begin{tabular}{ c | c c }
Position of unit in period & Periodic coil rolls [deg] & Periodic wedge angles [deg] \\
\hline
1 & 240 & -26.97 \\
2 & 0 & 93.03 \\
3 & 120 & 213 \\
4 & 240 & 333 \\
5 & 0 & 453 \\
6 & 120 & 573 \\
\end{tabular}
\caption{HFOFO periodic parameters.}
\label{initialcooling:table3}
\end{table}

Finally, Table \ref{initialcooling:table4} provides performance results from present G4beamline simulations of HFOFO. The emittances have been calculated using the ICOOL \textit{emitcalc} script. Notably, these simulations use a MAP-era beam file containing only positive muons — though further studies are ongoing to assess the performance and acceptance of the design with more modern input beams. Corrections may be in order to adequately compare the performance to that of other designs.

\begin{table}[!h]
\centering
\begin{tabular}{l | c c c c }
 & $N_{\text{total}}$ & $N_{150<p<350 \text{ MeV}}$ & $\varepsilon_{\perp}$ [m] & $\varepsilon_{\text{L}}$ [m] \\
\hline
Initial & 11452 & 7666 & 0.01604 & 5.748 \\
Final & 5348 & 5139 & 0.003595 & 2.908 \\
 & 47\% transmission & 67\% transmission & Factor of 4.46 & Factor of 1.98 \\

\end{tabular}
\caption[HFOFO performance (with MAP-era $\mu^+$ beam).]{HFOFO performance (with MAP-era $\mu^+$ beam). Emittances calculated with \textit{emitcalc}.}
\label{initialcooling:table4}
\end{table}

 \FloatBarrier
\subsection{Baseline Rectilinear Cooling}
\label{app:6d}
The rectilinear cooling section consists of a number of solenoid magnets with dipole field superimposed.
In the MAP design the dipole field was achieved by means of introducing a tilt in the solenoids but separate dipoles are proposed for this IMCC design. The rectilinear cooling lattice described below is stored in the MuonCollider-WG4 GitHub group, \verb|rectilinear| repository as release (branch) \verb|2024-09-27_release| and described in \cite{zhu2024performance}.
%MAP numbers are reproduced from \cite{Stratakis:2014nna}.

The solenoid field is approximately sinusoidal with a period given by the cell length $L$ so that $B_z(z, r=0) = B_{peak} \sin(2 \pi z/L)$.
Cells in the Rectilinear B lattices are increasingly non-sinusoidal, with a component $B_z(z, r=0) = B_{peak} \sin(4 \pi z/L)$ that gets stronger further down the B lattice.
The peak $B_z$ listed in Table \ref{cool:tab:6d_cell} is the peak field on the axis of the solenoid.
Fields may be higher in the conductor volume.

RF cavities are modelled as perfect cylindrical pillbox cavities operating in TM010 mode.
Several RF cavities are included within each cell.
A thin conductive window electromagnetically seals the RF cavities so that the pillbox model is an adequate approximation to the real cavity field and the cavities can be assumed to be independently phased. 
The RF gradient listed in Table \ref{cool:tab:6d_rf} is the peak gradient.

Updates for the A and B stages of the rectilinear cooling system have been developed, comprising of 10 "B-type" stages, denoted S1 through S10 that yields improved performance over the MAP lattice listed above and has been designed using 352 MHz RF and harmonics.
The performance is summarised in Table  \ref{cool:tab:6d_emit}.

Hardware parameters are described in Table \ref{cool:tab:6d_cell}.
In this lattice, the dipoles were simulated as a magnet independent of the solenoids which were not tilted and the dipole field is listed.

\begin{table}[!h]
    \centering
    \begin{tabular}{l|ccccc}
         &  $\varepsilon_{\rm{T}}$ &  $\varepsilon_{\rm{L}}$ &  $\varepsilon_{\rm{6D}}$ & Stage & Cumulative\\
         &  mm&  mm&  mm$^3$& Transmission & Transmission \%\\ \hline
         Start&      16.96&  45.53&   13500& &  100\\ \hline
         A-Stage 1 &  5.17&  18.31&  492.60& 75.2 & 75.2\\
         A-Stage 2&   2.47&   7.11&   44.03& 84.4 & 63.5\\
         A-Stage 3&   1.56&   3.88&    9.59& 85.6 & 54.3\\
         A-Stage 4&   1.24&   1.74&    2.86& 91.3 & 49.6\\ \hline
         Bunch merge& 5.13&   9.99&   262.5& 78.0 & 38.7\\ \hline
         B-Stage 1&   2.89&   9.09&   76.07& 85.2 & 33.0\\
         B-Stage 2&   1.99&   6.58&   26.68& 89.4 & 29.4\\
         B-Stage 3&   1.27&   4.05&    6.73& 87.5 & 25.8\\
         B-Stage 4&   0.93&   3.16&    2.83& 89.8 & 23.2\\
         B-Stage 5&   0.70&   2.51&    1.32& 89.4 & 20.7\\
         B-Stage 6&   0.48&   2.29&    0.55& 88.4 & 18.2\\
         B-Stage 7&   0.39&   2.06&    0.31& 92.8 & 17.0 \\
         B-Stage 8&   0.26&   1.86&    0.13& 87.9 & 14.9\\
         B-Stage 9&   0.19&   1.72&    0.06& 85.2 & 12.7\\
         B-Stage 10&  0.14&   1.56&    0.03& 87.1 & 11.1\\ 
    \end{tabular}
    \caption[Rectilinear cooling performance]{Rectilinear cooling performance in terms of emittance reduction (transverse, longitudinal and 6D) and transmission per stage.}
    \label{cool:tab:6d_emit}
\end{table}

\begin{table}[!h]
    \centering
    \begin{tabular}{l|ccccccccc}
         &  Cell&  Stage&  Pipe&  Max. $B_z$ & Int.& $\beta_\perp$ & $D_x$ & On-Axis&Wedge\\
         & Length & Length & Radius & On-Axis & $B_y$ & & & Wedge Len.& Angle \\
         &  m&  m&  cm&  T& Tm& cm& mm& cm&deg\\ \hline
         A-Stage 1 &  1.8&  104.4&    28&   2.5&  0.102&   70&   -60& 14.5&  45\\
         A-Stage 2&   1.2&  106.8&    16&   3.7&  0.147&   45&   -57& 10.5&  60\\
         A-Stage 3&   0.8&  64.8&     10&   5.7&  0.154&   30&   -40&   15& 100\\
         A-Stage 4&   0.7&  86.8&      8&   7.2&  0.186&   23&   -30&  6.5&  70\\ \hline
         B-Stage 1&   2.3&  50.6&     23&   3.1&  0.106&   35& -51.8&   37& 110\\
         B-Stage 2&   1.8&  66.6&     19&   3.9&  0.138&   30& -52.4&   28& 120\\
         B-Stage 3&   1.4&  84.0&   12.5&   5.1&  0.144&   20& -40.6&   24& 115\\
         B-Stage 4&   1.2&  66.0&    9.5&   6.6&  0.163&   15& -35.1&   20& 110\\
         B-Stage 5&   0.8&  44.0&      6&   9.1&  0.116&   10& -17.7& 12.5& 120\\
         B-Stage 6&   0.7&  38.5&    4.5&  11.5&  0.087&    6& -10.6&   11& 130\\
         B-Stage 7&   0.7&  28.0&   3.75&    13&  0.088&    5&  -9.8&   10& 130\\
         B-Stage 8&  0.65& 46.15&   2.85&  15.8&  0.073&  3.8&    -7&    7& 140\\
         B-Stage 9&  0.65&  33.8&    2.3&  16.6&  0.069&    3&  -6.1&  7.5& 140\\
         B-Stage 10& 0.63& 29.61&    2.0&  17.2&  0.069&  2.7&  -5.7&  6.8& 140\\ \hline
    \end{tabular}
    \caption[Rectilinear cooling cell hardware]{Rectilinear cooling cell hardware in terms of cell geometry, solenoid fields, dipole fields and wedge geometry}
    \label{cool:tab:6d_cell}
\end{table}

\begin{table}[!h]
    \centering
    \begin{tabular}{l|ccccc}
         &  RF Frequency&  Num. RF&  RF Length&  RF Gradient& RF phase \\
         &  MHz&  &  cm&  MV/m& deg \\ \hline
         A-Stage 1 &  352&  6&  19&   27.4& 18.5 \\
         A-Stage 2&   352&  4&  19&   26.4& 23.2 \\
         A-Stage 3&   704&  5&  9.5&  31.5& 23.7 \\
         A-Stage 4&   704&  4&  9.5&  31.7& 25.7 \\ \hline
         B-Stage 1&   352&  6&  25&   21.2& 29.9 \\
         B-Stage 2&   352&  5&  22&   21.7& 27.2 \\
         B-Stage 3&   352&  4&  19&   24.9& 29.8 \\
         B-Stage 4&   352&  3&  22&   24.3& 31.3 \\
         B-Stage 5&   704&  5&  9.5&  22.5& 24.3 \\
         B-Stage 6&   704&  4&  9.5&  28.2& 22.1 \\
         B-Stage 7&   704&  4&  9.5&  28.5& 18.4 \\
         B-Stage 8&   704&  4&  9.5&  27.1& 14.5 \\
         B-Stage 9&   704&  4&  9.5&  29.7& 11.9 \\
         B-Stage 10&  704&  4&  9.5&  24.9& 12.2 \\ 
    \end{tabular}
    \caption[Rectilinear cooling cell RF parameters.]{Rectilinear cooling cell RF parameters. 0$^o$ phase is bunching mode.}
    \label{cool:tab:6d_rf}
\end{table}

\begin{table}[!h]
    \centering
    \begin{tabular}{l|cc}
         & Beam Size $\sigma_x$ ($\sigma_y$)&Beam Size  $\sigma_x$ ($\sigma_y$)\\
         & Cell Center (max)&Cell Start (min)\\ 
         & mm&mm\\ \hline
         A-Stage 1 & 48.6 (35.4)&38.6 (47.2)\\
         A-Stage 2& 25 (22.1)&23.9 (23.6)\\
         A-Stage 3& 15.6 (15.4)&15.7 (14.6)\\ \hline
         A-Stage 4& 13 (11.9)&12.6 (12)\\
         B-Stage 1& 28.4 (27.5)&23.9 (23.3)\\
         B-Stage 2& 20 (20.3)&19.5 (17.4)\\
         B-Stage 3& 16.4 (16.3)&12.3 (11.2)\\
         B-Stage 4& 13.5 (13.9)&8.9 (7.9)\\
         B-Stage 5& 9.8 (10)&6.2 (5.8)\\
         B-Stage 6& 8.6 (8.4)&3.9 (3.8)\\
         B-Stage 7& 7.7 (7.6)&3.3 (3.2)\\
         B-Stage 8& 5.8 (5.6)&2.3 (2.3)\\
         B-Stage 9& 5.2 (5.1)&1.7 (1.8)\\
         B-Stage 10& 4.7 (4.2)&1.4 (1.4)\\ 
    \end{tabular}
    \caption[Rectilinear cooling cell beam size at the start and center of the beam.]{Rectilinear cooling cell beam size at the start and center of the beam. Horizontally (and vertically).}
    \label{cool:tab:6d_beamsize}
\end{table}

\FloatBarrier
\subsubsection{Low-Stress Rectilinear Cooling}
\label{app:6d:lowstress}
Upon review of the above solenoids, the radial stress was calculated, as shown in Table \ref{mag:tab:rectilinear_magnet}.
In response to this, a low-stress lattice option has been produced, the performance of which is displayed in Table \ref{cool:tab:6d_emit_lowstress}.
The cell details in each stage is listed in Table \ref{cool:tab:6d_cell_lowstress}, and the resulting RF cavity parameters are in Table \ref{cool:tab:6d_rf_lowstress}.

\begin{table}[!h]
\centering

\begin{tabular}{l  |l  l  l c c }
\textbf{VARIANT} & $\varepsilon_{\rm{T}}$ 
  & $\varepsilon_{\rm{L}}$ 
  & $\varepsilon_{\rm{6D}}$ 
  & Stage  & Cumulative \\
\textbf{Low Stress} & mm & mm & mm$^3$ & Transmission  & Transmission \\
\hline
Start & 16.96 & 45.53 & 13500 &  & 100 \\ \hline
A-Stage 1  & 4.977 & 17.83 & 447.3 & 72.6 & 72.6 \\
A-Stage 2 & 2.486 & 7.06 & 44.24 & 82.8 & 60.1 \\
A-Stage 3 & 1.609 & 3.616 & 9.604 & 84.1 & 50.6 \\
A-Stage 4 & 1.247 & 1.74 & 2.863 & 87.4 & 44.2 \\\hline
Bunch merge & 5.13 & 9.99 & 262.5 & 78 & 34.6 \\\hline
B-Stage 1 & 2.892 & 9.239 & 77.77 & 85.3 & 29.5 \\
B-Stage 2 & 2.025 & 6.418 & 26.96 & 90.9 & 26.8 \\
B-Stage 3 & 1.214 & 3.972 & 5.943 & 87.2 & 23.4 \\
B-Stage 4 & 0.8987 & 3.021 & 2.476 & 91.6 & 21.4 \\
B-Stage 5 & 0.6868 & 2.528 & 1.224 & 90 & 19.3 \\
B-Stage 6 & 0.4683 & 2.29 & 0.5099 & 85.3 & 16.5 \\
B-Stage 7 & 0.3642 & 2.035 & 0.2718 & 88.4 & 14.5 \\
B-Stage 8 & 0.2659 & 1.843 & 0.1307 & 84.5 & 12.3 \\
B-Stage 9 & 0.1839 & 1.725 & 0.0586 & 81.4 & 10 \\
B-Stage 10 & 0.1404 & 1.554 & 0.03027 & 82.7 & 8.3 \\

\end{tabular}
\caption[Low-stress rectilinear lattice cooling performance]{New lattice with larger gaps and less solenoid stress. Rectilinear cooling performance in terms of emittance reduction (transverse, longitudinal and 6D) and transmission per stage.}
\label{cool:tab:6d_emit_lowstress}
\end{table}

\begin{table}[!h]
\centering

\begin{tabular}{l|clclccccc}
\textbf{VARIANT}& Cell&  Stage& Pipe&  Max. $B_z$ & Int.& $\beta_\perp$ & $D_x$ & On-Axis& Wedge\\
\textbf{Low Stress}& Length &  Length & Radius &  On-Axis & $B_y$ & & &  Wedge Len.&  Angle \\
 & m&  m& cm&  T& Tm& cm& mm& cm& deg\\
\hline
A-Stage 1 & 1.9&  110.2& 28&  2.5& 0.095& 72& -60& 30& 80\\
A-Stage 2& 1.3&  132.6& 16&  3.6& 0.141& 47& -56& 21.5& 100\\
A-Stage 3& 0.9&  80.1& 10&  5.5& 0.152& 31& -40& 15& 100\\
A-Stage 4& 0.76&  101.08& 8&  6.9& 0.172& 23& -35& 14& 110\\ \hline
B-Stage 1& 2.2&  50.6& 23&  3.3& 0.118& 34& -52& 37& 110\\
B-Stage 2& 1.8&  66.6& 19&  4& 0.144& 28& -52& 28& 120\\
B-Stage 3& 1.5&  90& 12.5&  4.9& 0.144& 19& -41& 24& 115\\
B-Stage 4& 1.25&  68.75& 9.5&  5.9& 0.151& 15& -35& 20& 120\\
B-Stage 5& 0.85&  45.9& 6&  8.8& 0.110& 10& -18& 12.5& 120\\
B-Stage 6& 0.8&  43.2& 4.5&  10.7& 0.080& 6& -10& 11& 130\\
B-Stage 7& 0.8&  32& 3.8&  11.5& 0.078& 5& -9.8& 10& 130\\
B-Stage 8& 0.78&  39& 3&  12.9& 0.064& 4& -7.1& 7& 140\\
B-Stage 9& 0.78&  40.56& 2.3&  13.5& 0.059& 3.5& -6.1& 7.5& 140\\
B-Stage 10& 0.78&  31.98& 2&  14.1& 0.059& 3.1& -5.7& 6.8& 140\\

\end{tabular}
\caption[Low stress rectilinear cooling cell hardware]{New lattice with larger gaps and less solenoid stress. Rectilinear cooling cell hardware in terms of cell geometry, solenoid fields, dipole fields and wedge geometry}
\label{cool:tab:6d_cell_lowstress}
\end{table}

\begin{table}[!h]
\centering

\begin{tabular}{l|ccccc}
\textbf{VARIANT} & rf& Number of& rf  cell& rf gradient& rf phase\\
\textbf{Low Stress} & frequency& rf cells& length& & \\
 & MHz& & cm& MV/m& deg\\
\hline
A-Stage 1  & 352& 6& 20& 25.7& 19.9\\
A-Stage 2 & 352& 4& 20& 26& 23.6\\
A-Stage 3 & 704& 5& 10& 31.6& 22.2\\
A-Stage 4 & 704& 4& 10& 31.6& 23.7\\ \hline
B-Stage 1 & 352& 6& 22& 22.5& 32.8\\
B-Stage 2 & 352& 5& 22& 23.6& 27.1\\
B-Stage 3 & 352& 4& 22& 23.2& 25.5\\
B-Stage 4 & 352& 3& 22& 24.1& 27.9\\
B-Stage 5 & 704& 4& 10& 27& 26.4\\
B-Stage 6 & 704& 4& 8& 31.8& 25.6\\
B-Stage 7 & 704& 4& 8& 31.3& 22.7\\
B-Stage 8 & 704& 4& 8& 25.9& 15.9\\
B-Stage 9 & 704& 4& 8& 23.8& 15.4\\
B-Stage 10 & 704& 4& 8& 24.3& 13.6\\

\end{tabular}
\caption[Low stress rectilinear cooling cell RF parameters]{New lattice with larger gaps and less solenoid stress. Rectilinear cooling cell RF parameters. 0$^o$ phase is bunching mode.}

\label{cool:tab:6d_rf_lowstress}
\end{table}

 \FloatBarrier
\subsection{Final cooling}
\label{app:final}
There are three lattice options for the final cooling in development.
Each correspond to the initial conditions of the 6D cooling lattice before it.
The first assumes MAP parameters of $\varepsilon_T$=\SI{300}{\micro\meter}, $\varepsilon_L=$\SI{1.5}{\milli\meter}. The second takes the beam from the B8 stage of the IMCC rectilinear cooling $\varepsilon_T$=\SI{260}{\micro\meter}, $\varepsilon_L=$\SI{1.8}{\milli\meter}, and the third takes the beam from the B10 stage of $\varepsilon_T$=\SI{140}{\micro\meter}, $\varepsilon_L=$\SI{1.5}{\milli\meter}.

\subsubsection{Final Cooling - from MAP initial conditions} 
\begin{figure}[!h]
    \centering
    \includegraphics[width=1\linewidth]{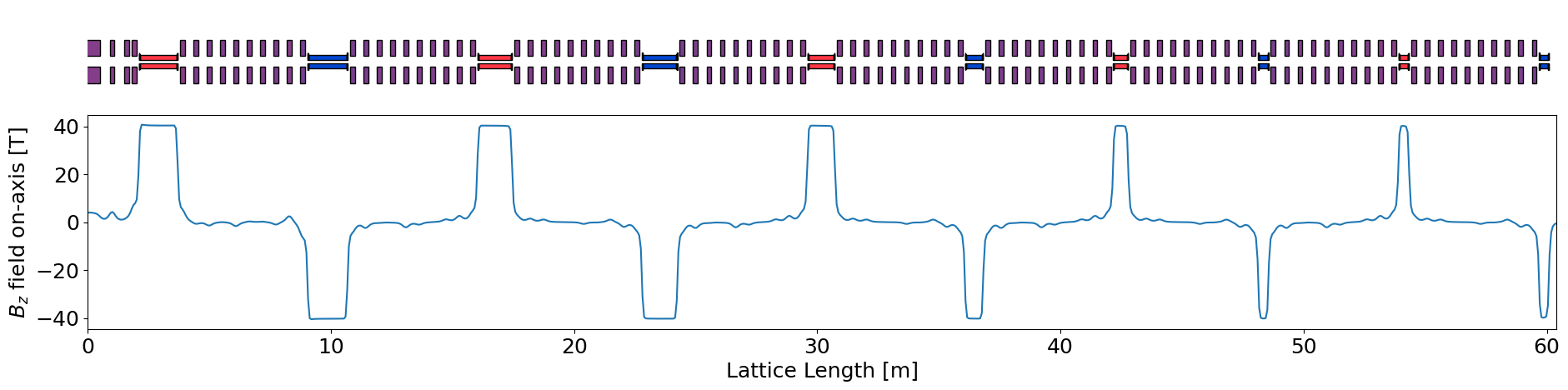}
    \caption{Geometric overview of the Final Cooling lattice from MAP parameters, including Bz field-on-axis. Red for +\SI{40}{\tesla} solenoids, blue for -\SI{40}{\tesla} solenoids, and purple for low-field matching solenoids. Correct relative inner and outer radii.}
    \label{fig:final:MAP}
\end{figure}

The final cooling lattice from MAP initial conditions is made of 10 high field solenoids, which alternate in polarity each cell, represented in Figure \ref{fig:final:MAP}.
10 matching low-field solenoids are placed between the two high-field solenoids, which have space sufficient to fit RF pillboxes, required to reach the kinetic energies and energy spreads referenced in Table \ref{tab:final:MAP}.

The absorbers are modelled as a constant pressure of  \SI{70.8}{\kilo\gram\per\meter\cubed}, which is unrealistic given the beam intensities towards the end of the final cooling lattice~\cite{stechauner_2025_69h4h-rxj29}.  For this reason, the density x length is represented in  Table \ref{tab:final:MAP}.

\begin{table}[!h]
\centering
\begin{tabular}{l|cccccclcc}
Cell No & $\varepsilon_{\rm{T}}$  & $\varepsilon_{\rm{L}}$ & KE& $\sigma KE$& $\rho L(H_2)$& $L_{sol}$ &$B_z$& T (loss)& T (decay)\\
Unit & \si{\milli\meter}& \si{\milli\meter}& \si{MeV}& \si{MeV}& \si{\kilo\gram\per\meter\squared}& \si{\meter} &\si{\tesla}& \%&\%\\
\hline
0 & 0.300& 1.5 &  &  &  &   &&  & 1.00\\
1 & 0.247& 1.88 & 123.49 & 5.42 & 90.4&  1.52 &40& 100.0 & 0.99\\
2 & 0.203& 2.28 & 123.34 & 5.82 & 93.8&  1.57 &-40& 100.0 & 0.96\\
3 & 0.165& 2.83 & 100.41 & 5.25 & 78.2&  1.35 &40& 100.0 & 0.91\\
4 & 0.126& 3.98 & 85.49 & 5.07 & 80.6&  1.38 &-40& 100.0 & 0.85\\
5 & 0.103& 5.13 & 74.48 & 5.66 & 57.8&  1.06 &40& 100.0 & 0.77\\
6 & 0.087& 6.73 & 51.82 & 4.36 & 30.9&  0.68 &-40& 100.0 & 0.67\\
7 & 0.060& 11.82 & 32.01 & 2.40 & 21.4&  0.54 &40& 95.1 & 0.55\\
8 & 0.045& 20.76 & 18.28 & 1.30 & 7.4&  0.34 &-40& 99.7 & 0.43\\
9 & 0.032& 39.25 & 17.94 & 1.25 & 8.4&  0.36 &40& 96.4 & 0.32\\
10 & 0.0222& \textbf{71.72} & 14.70 & 1.21 & 6.4&  0.33 &-40& 89.8 & 0.20\\
\end{tabular}
\caption{Overview of final cooling design from MAP initial conditions}
\label{tab:final:MAP}
\end{table}

 \FloatBarrier
 
 \subsubsection{Final Cooling - from B8}

Updated final cooling lattices were designed based on the output beam emittance from the 6D cooling lattices described in Section \ref{app:6d}. The overall cell layout follows the configuration shown in Figure \ref{fig:final:MAP}, except that the 0°-phase RF cavities used for phase rotation have been removed. Table \ref{tab:final:B8_emit} summarizes the output and cumulative emittances at the end of each stage, with the initial emittance taken from B-Stage 8 in Table \ref{cool:tab:6d_emit}. At the end of the cooling channel, the transverse emittance is reduced to 22.4 $\mu$m, satisfying the luminosity requirement, while the longitudinal emittance reaches 43 mm, which remains below the current requirement of the acceleration system. The main hardware parameters of the absorber, magnet, and RF system are listed in Tables \ref{tab:final:B8_cell} and \ref{tab:final:B8_rf}. The peak magnetic field is kept below 42 T, and the RF frequency gradually decreases along the channel to match the increasing bunch length, as shown in Table \ref{tab:final:B8_longit}.

\begin{table}[!h]
\centering
\begin{tabular}{l|cccc}
Stage  & $\varepsilon_{\rm{T}}$  & $\varepsilon_{\rm{L}}$ & $\varepsilon_{\rm{6D}}$ 
  & Cumulative  \\
 & mm  & mm  & mm$^3$  & transmission \%  \\
\hline
Start & 0.26 & 1.8 & 0.12 & 100 \\
Stage 0 & 0.21 & 2.5 & 0.11 & 99.6 \\
Stage 1 & 0.16 & 5.2 & 0.14 & 90.1 \\
Stage 2 & 0.12 & 8.7 & 0.13 & 79.8 \\
Stage 3 & 0.095 & 10.3 & 0.098 & 72.5 \\
Stage 4 & 0.063 & 15.1 & 0.064 & 65.5 \\
Stage 5 & 0.041 & 22.7 & 0.039 & 55.5 \\
Stage 6 & 0.032 & 32 & 0.035 & 52 \\
Stage 7 & \textbf{0.0224} & \textbf{42.68} & \textbf{0.022} & \textbf{42.7} \\
\end{tabular}
\caption[Short rectilinear final cooling cell performance parameters]{Short rectilinear final cooling cell performance parameters (latest version)}
\label{tab:final:B8_emit}
\end{table}

\begin{table}[!h]
\centering
\begin{tabular}{l|cccl}
Stage & Stage length (m) & Peak on-axis Bz (T) & LH absorber length (m)  &$\rho L(H_2)$ (\si{\kilo\gram\per\meter\squared})\\
\hline
Stage 0 & 2.754 & 33.6 & 0.692  &48.99\\
Stage 1 & 5.195 & -36 & 0.397  &28.11\\
Stage 2 & 5.401 & 35.5 & 0.135  &9.56\\
Stage 3 & 4.268 & -41.8 & 0.053  &3.75\\
Stage 4 & 5.204 & 40.9 & 0.043  &3.04\\
Stage 5 & 6.836 & -41.3 & 0.018  &1.27\\
Stage6 & 5.17 & 38.4 & 0.012  &0.85\\
Stage 7 & 5.565 & -43.4 & 0.014  &0.99\\
\end{tabular}
\caption[Short rectilinear final cooling cell magnet lattice parameters]{Short rectilinear final cooling cell magnet lattice parameters (latest version)}
\label{tab:final:B8_cell}
\end{table}

\begin{table}[!h]
\centering
\begin{tabular}{l|ccccc}
Stage & Frequency & Number of RF cells & Maximum gradient & Phase & RF cell length \\
 & MHz &  & MV/m  & $^\circ$  & m  \\
\hline
stage 0 &  & 0 &  &  &  \\
stage 1 & 142.9 & 5 & 9.2 & 26 & 1.25 \\
stage 2 & 67.3 & 6 & 5 & 18.7 & 1.5 \\
stage 3 & 52.7 & 3 & 4.9 & 50.2 & 0.75 \\
stage 4 & 29.8 & 10 & 1.7 & 15.7 & 2.5 \\
stage 5 & 15.3 & 14 & 1.5 & 23 & 3.5 \\
stage 6 & 10 & 10 & 1.3 & 28.3 & 2.5 \\
stage 7 & 8 & 11 & 1.2 & 40.9 & 2.75 \\

\end{tabular}
\caption[Short rectilinear final cooling cell RF parameters.]{Short rectilinear final cooling cell RF parameters. 0$^o$ phase is bunching mode. (latest version)}
\label{tab:final:B8_rf}
\end{table}

\begin{table}[!h]
\centering
\begin{tabular}{l|ccc}
Stage & Final Pz & Final energy spread & Final $c\sigma\_t$  \\
Units  & \si{\mega\electronvolt\per\clight}  & \si{\mega\electronvolt}  & \si{\clight}  \\
\hline
Start & 135 & 3.9 & 0.04932 \\
Stage 0 & 95.4 & 5.3 & 0.06703 \\
Stage 1 & 65 & 3.7 & 0.2813 \\
Stage 2 & 53 & 2.1 & 0.4926 \\
Stage 3 & 46.2 & 2.4 & 0.6547 \\
Stage 4 & 36.1 & 1.8 & 1.074 \\
Stage 5 & 31 & 1.7 & 1.465 \\
Stage 6 & 30 & 1.7 & 2.492 \\
Stage 7 & 28 & 1.6 & 3.307 \\
\end{tabular}
\caption[Short rectilinear final cooling cell beam longitudinal parameters]{Short rectilinear final cooling cell beam longitudinal parameters (latest version)}
\label{tab:final:B8_longit}
\end{table}

 \FloatBarrier
 
\subsubsection{Final Cooling - From B10}

Another final cooling lattice was also designed based on the output emittance from B-Stage 10. As shown in Table \ref{tab:final:B10_emit}, this design reduces the transverse emittance to 23 $\mu$m, while the longitudinal emittance increases to 22 mm, which is nearly a factor of two smaller than that in Table \ref{tab:final:B8_emit}. This improvement results from the smaller initial transverse emittance, which allows for fewer cooling stages (absorbers) and therefore less beam-length growth caused by passage through the absorbers. The main hardware parameters of the absorber, magnet, and RF system are listed in Tables \ref{tab:final:B10_cell} and \ref{tab:final:B10_rf}, and the corresponding longitudinal beam parameters are given in Table \ref{tab:final:B10_longit}.

\begin{table}[!h]
\centering
\begin{tabular}{l|cccc}
Stage  & $\varepsilon_{\rm{T}}$  & $\varepsilon_{\rm{L}}$ & $\varepsilon_{\rm{6D}}$ 
  & Cumulative  \\
 & mm  & mm  & mm$^3$  & transmission \%  \\
\hline
Start & 0.14 & 1.5 & 0.03 & 100 \\
Stage 0 & 0.12 & 1.9 & 0.03 & 99.5 \\
Stage 1 & 0.08 & 5.2 & 0.034 & 90.6 \\
Stage 2 & 0.053 & 7.7 & 0.023 & 77.9 \\
Stage 3 & 0.041 & 10.9 & 0.019 & 71.9 \\
Stage 4 & 0.029 & 15.7 & 0.014 & 66.8 \\
Stage 5 & \textbf{0.023} & \textbf{22.1} & 0.012 & \textbf{61.4} \\
\end{tabular}
\caption[Long rectilinear final cooling cell performance parameters]{Long rectilinear final cooling cell performance parameters (latest version)}
\label{tab:final:B10_emit}
\end{table}

\begin{table}[!h]
\centering
\begin{tabular}{l|cccl}
Stage & Stage length (m) & Peak on-axis Bz (T) & LH absorber length (m)  &$\rho L(H_2)$ (\si{\kilo\gram\per\meter\squared})\\
\hline
Stage 0 & 2.035 & 40 & 0.183  &12.96\\
Stage 1 & 4.656 & -29.3 & 0.255  &18.05\\
Stage 2 & 4.628 & 39.4 & 0.055  &3.89\\
Stage 3 & 3.89 & -41 & 0.02  &1.42\\
Stage 4 & 4.124 & 39.6 & 0.015  &1.06\\
Stage 5 & 5.068 & -42.7 & 0.0092  &0.65\\

\end{tabular}
\caption[Long rectilinear final cooling cell magnet lattice parameters]{Long rectilinear final cooling cell magnet lattice parameters (latest version)}
\label{tab:final:B10_cell}
\end{table}

\begin{table}[!h]
\centering
\begin{tabular}{l|ccccc}
\hline
Stage & Frequency & Number of RF cells & Maximum gradient & Phase & RF cell length \\
 & MHz &  & MV/m  & $^\circ$  & m  \\
\hline
stage 0 &  & 0 &  &  &  \\
stage 1 & 131.8 & 6 & 6.6 & 14.5 & 1.5 \\
stage 2 & 56.3 & 5 & 4.4 & 31.4 & 1.25 \\
stage 3 & 25.5 & 6 & 2.9 & 17.4 & 1.5 \\
stage 4 & 14.8 & 7 & 1.7 & 45.6 & 1.75 \\
stage 5 & 11.5 & 9 & 1.3 & 41.3 & 2.25 \\
\end{tabular}
\caption[Long rectilinear final cooling cell RF parameters.]{Long rectilinear final cooling cell RF parameters. 0$^o$ phase is bunching mode. (latest version)}
\label{tab:final:B10_rf}
\end{table}

\begin{table}[!h]
\centering
\begin{tabular}{l|ccc}
Stage & Final Pz & Final energy spread & Final $c\sigma\_t$  \\
Units  & \si{\mega\electronvolt\per\clight}  & \si{\mega\electronvolt}  & \si{\clight}  \\
\hline
Start & 95 & 3.35 & 0.04794 \\
Stage 0 & 79.2 & 4.1 & 0.08625 \\
Stage 1 & 46.8 & 2.6 & 0.2489 \\
Stage 2 & 37.1 & 2.1 & 0.6356 \\
Stage 3 & 31.5 & 1.1 & 1.044 \\
Stage 4 & 28.3 & 1.3 & 1.481 \\
Stage 5 & 26.5 & 1.2 & 2.247 \\
\end{tabular}
\caption[Long rectilinear final cooling cell beam longitudinal parameters ]{Long rectilinear final cooling cell beam longitudinal parameters (latest version)}
\label{tab:final:B10_longit}
\end{table}

\FloatBarrier

\subsection{Pre-accelerator}
No pre-accelerator design exists. Table \ref{cool:tab:preacc} gives estimations of design and performance based on induction LINAC technology.

\begin{table}[!h]
    \centering
    \begin{tabular}{cccccc}
           Injection Energy&Extraction Energy&  Pulse Length&  Transmission& Linac Length\\ 
   MeV&MeV& ns& \%&m\\ \hline
           5&250&  15&  86 & 140\\
    \end{tabular}
    \caption[Pre-Accelerator parameters]{Pre-Accelerator (Induction Linac) - see for example RADLAC-1}
    \label{cool:tab:preacc}
\end{table}\FloatBarrier
\section{Appendix: Low Energy Acceleration}
\label{app:lowE}
The low energy acceleration chain brings the muon beams from \SI{250}{\mega\electronvolt} after the pre-accelerator to \SI{62.5}{\giga\electronvolt} for injection into the high energy acceleration chain described in Section \ref{sec:highE}.\\
It is composed of a single-pass superconducting LINAC outlined in Table \ref{low:tab:linac}, followed by two recirculating linear accelerators (RLA), described in Table \ref{low:tab:RLA}.\\
RLA2 has an preliminary optics design. No optics design exists for LINAC and RLA1.
Both RLAs have an assumed racetrack geometry.
The transmission through RLA2 is 92.6\%. The target transmission for LINAC and RLA1 is 90\%, which corresponds to an effective average gradient of $4.1\;\rm MV/m$.

\begin{table}[!h]
    \centering
    \begin{tabular}{l|cc}
         &  CryoModule 1& CryoModule 2\\ \hline
         Initial energy [GeV]&  0.255& --\\
         Final energy [GeV]& -- & 1.25\\
         Frequency [MHz]&  88 / 264& 88 / 264\\
         RF gradient [MV/m]&  5 / 8& 5 / 8\\
         Passes&  1& 1\\
    \end{tabular}
    \caption{Parameters describing the single-pass LINAC that follows the final cooling section.}
    \label{low:tab:linac}
\end{table}

\begin{table}[!h]
    \centering
    \begin{tabular}{l|cclcc}
     & \multicolumn{2}{c}{RLA1} & \multicolumn{2}{c}{RLA2} \\
    \hline
    Initial energy [GeV] & \multicolumn{2}{c}{1.25} & \multicolumn{2}{c}{5} \\
    Final energy [GeV] & \multicolumn{2}{c}{5} & \multicolumn{2}{c}{63} \\
    Energy gain per pass & \multicolumn{2}{c}{0.85} & \multicolumn{2}{c}{13.5} \\
    Frequency [MHz] & 352 & 1056 & 352 & 1056 \\
    No.\~SRF cavities & 40 & 4 & 300 & 40 \\
    RF length [m] & 68 & 3.44 & 510 & 34.4 \\
    RF gradient [MV/m] & 15 & 25 & 15 & 25 \\
    Passes & \multicolumn{2}{c}{4.5} & \multicolumn{2}{c}{4.5} \\
    Linac length [m] & \multicolumn{2}{c}{--} & \multicolumn{2}{c}{2 x 915} \\
    Arc lengths [m]  & \multicolumn{2}{c}{--} & \multicolumn{2}{c}{$\approx$ 8 x 438} \\
    
    \end{tabular}

    \caption{Multi-pass recirculating LINACs}
    \label{low:tab:RLA}
\end{table}

\FloatBarrier \FloatBarrier
\section{Appendix: High Energy Acceleration}
\label{app:highE}
\begin{figure}[ht!]
\centering
{\includegraphics[width=0.9\columnwidth]{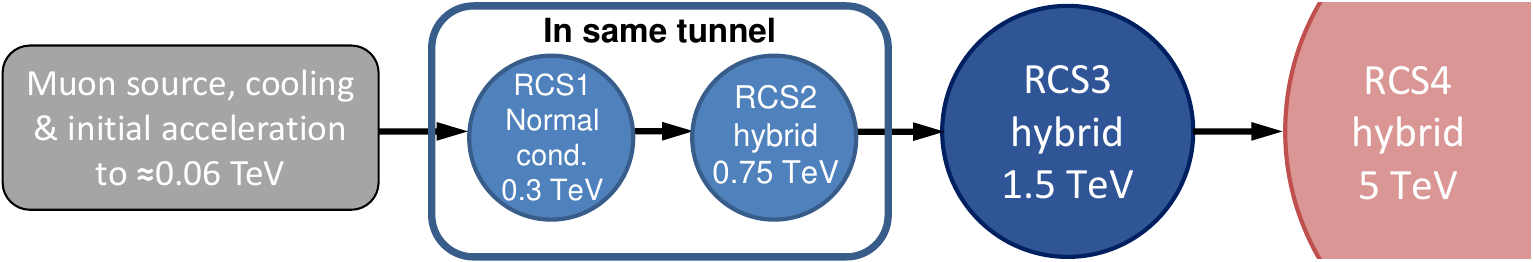}}
\caption{Schematic of the chain of rapid cycling-synchrotrons for the high-energy acceleration complex. From~\cite{batsch:ipac2023}.}
\label{high:fig:rcsoverview}
\vspace*{-1\baselineskip}
\end{figure}

% \subsection{Introduction to the chapter}
As described in \cite{batsch:ipac2023}, an option for the chain of four rapid cycling synchrotrons~(RCS) foresees to accelerate two counter-rotating bunches at a repetition rate of 5\,Hz in stages of \SI{0.30}{\tera\electronvolt} (RCS1), \SI{0.75}{\tera\electronvolt} (RCS2) and \SI{1.5}{\tera\electronvolt} (RCS3) to inject into the \SI{3}{\tera\electronvolt} collider ring, or \SI{5}{\tera\electronvolt} (RCS4), to inject to the \SI{10}{\tera\electronvolt} collider ring.
This scenario is based on the US Muon Acceleration Program (MAP)~\cite{Berg:details,MAP} and applied for a general Greenfield site.
The high-energy stage of the accelerator chain with four RCS is illustrated in Fig.~\ref{high:fig:rcsoverview}.
Corresponding site-specific parameter designs can be found in Section~\ref{app:siteRCS}.
Alternative Fixed Field Accelerator options can be found in Section~\ref{app:FFA}.

The first two RCS share the same tunnel, meaning that they have the same circumference and layout~\cite{Berg:MCpsp}. The bending in the first RCS is provided by normal conducting magnets. The RCS2 to RCS4 are planned as hybrid RCSs where normal conducting magnets cycling from $-B_{\mathrm{nc}}$ to $+B_{\mathrm{nc}}$ are interleaved with strong fixed-field, superconducting magnets.
Within this section, NC magnets are referred to as \textit{pulsed}, and the SC magnets are referred to as \textit{steady}.
This is to reflect the alternative magnet technologies required for the hybrid RCS.
This combination allows for a large energy swing with a high average bending field to minimize the travel distance of the muons and thus their decay losses. 
The absolute value of magnetic field in the normal-conducting dipoles does not exceed \SI{\pm1.8}{T} at injection and extraction for all RCSs to avoid saturation of the magnet yoke.
For the hybrid RCS2 and RCS3, the magnetic field in the SC magnets is \SI{10}{T} to provide a compromise between the magnet filling factor and magnet costs. 
To protect the SC magnets from decay products, the inner aperture of the SC magnets is larger with \SI{10}{T}. Increasing the field to \SI{16}{T} implies higher technological and financial cost without a significant improvement of the machine performance. 
In the case of RCS4 however, the %extraction high-energy and thus 
average magnetic field in the accelerator is assumed to be \SI{16}{T} as a higher magnetic field in the SC magnets helps to reduce the overall circumference and thus the muon decay and RF requirements.
%needs a higher magnetic field in the SC magnets to reduce the overall circumference and therefore the decay and RF requirements. That is why the current magnetic field is \SI{16}{T} in the SC dipoles of the RCS4. 
This requirement may evolve with the optimization of the high-energy chain. 
%Including the cost optimization in the figures of merit may go to a slightly different table.
%The updated tables can be found below in Figs.~\ref{fig:Table1over2_MC-2Tand16T}~-~\ref{fig:Table2over2_MC-2Tand16T}, which are the most up-to-date tables to be used.

The number of synchrotron oscillations per turn is extreme~\cite{batsch:ipac2023}, much larger than the conventional stability limit for stable synchrotron oscillations and phase focusing of $1/\pi$ in a synchrotron with one or few localized RF sections. 
To mitigate resulting beam losses, the RF system must be distributed over the entire RCSs. Tracking simulations on how the number of RF stations influences the longitudinal emittance have been performed. 
For the present design, the minimum number is around 32 RF stations for RCS1 and RCS4, and 24 stations for RCS2 and RCS3~\cite{batsch:ipac2023}. 

It is worth noting that the longitudinal dynamics used values of momentum compaction factor for an RCS lattice design based on FODO cells. 
With a more defined optics design, this number might change and with it the basic parameters of the longitudinal beam dynamics such as the synchrotron tune, bucket area and energy acceptance, which are all a function of the momentum compaction factor.

%\newpage
\subsection*{Parameter tables}
% Table~\ref{high:tab:RCS_key} lists the basic acceleration parameters. For a given survival rate of 90\% per ring, and fixed energy swing, the acceleration time and the average accelerating gradient $G_\text{avg}$, averaged over the full length of the synchrotrons (i.e., also over parts without RF), are fully determined. . Here, a linear increase of the bending field is assumed, even though later, the ramp shape will be only quasi-linear to simplify the magnet powering requirements.
 
Table \ref{high:tab:RCS_key} shows the general RCS parameters, and Table \ref{high:tab:RCS_lattice} specifies lattice parameters. 
The first parameters for the fourth RCS to accelerate to \SI{5}{TeV} are included but may evolve in the near future. 
We assume a survival rate of 90\,\% per ring and linear ramping only considering losses due to muon decay, even though these values are subject to further adjustments to optimize the RF and magnet powering parameters with respect to total costing, ramp shape, bunch matching, and the overall transmission of the entire chain.

\subsection{Site-Based RCS Designs}
\label{app:siteRCS}
Tentative parameter tables to guide future design efforts for the existing site options.
Different assumptions were made for the magnet technologies.
\subsubsection{RCS Layout at CERN}
\label{site:sec:RCS_cern}
The RCS layout on the CERN site is based on the usage of the existing Super Proton Synchrotron (SPS) and Large Hadron Collider (LHC) tunnels to host the high-energy acceleration chain. 
To make a comparison with the greenfield study possible, the same assumptions for the injection energy and the injection bunch population were chosen. 
The survival rate over the whole RCS chain is assumed as $70 \%$, only considering losses due to muon decay, while the individual survival rates of the rings were adjusted to achieve a high extraction energy from the last RCS. 
Table \ref{site:tab:RCS_CERN_key} and \ref{site:tab:RCS_CERN_lattice} for the site-based design correspond to Table \ref{high:tab:RCS_key} and \ref{high:tab:RCS_lattice} for the greenfield design respectively.

\begin{table}[!h]
    \centering
    \begin{tabular}{c|c|ccc} 
         Parameter&  Unit&  RCS SPS&  RCS LHC1&  RCS LHC2\\\hline
         Hybrid RCS&  -&  no&  no&  yes\\
         Repetition rate&  Hz&  5&  5&  5\\
         Circumference& m& 6912& 26659& 26659\\
         Injection energy& GeV& 63& 350& 1600\\
         Extraction energy& GeV& 350& 1600& 3800\\
         Energy ratio&  &  5.6&  4.6&  2.4\\
         Assumed survival rate& &  0.88&  0.86&  0.92\\
         Cumulative survival rate& & 0.88& 0.76&0.7\\
         Acceleration time&  ms&  0.45&  2.6&  4.42\\
         Revolution period&  \textmu s&  23&  88.9&  88.9\\
         Number of turns& & 19& 29& 50\\
         Required energy gain/turn& GeV& 15.1& 43.1& 44\\
         Average accel.~gradient& MV/m& 2.15& 1.62& 1.68\\
         Number of bunches& & 1& 1& 1\\
         Inj.~bunch population& \num{E12}& 2.7& 2.38& 2.04\\
         Ext.~bunch population& \num{E12}& 2.38& 2.04& 1.88\\
         Beam current per bunch& mA& 18.75& 4.29&3.675\\
         Beam power& MW& 803& 523&462\\
         Vert.~norm.~emittance& \textmu m& 25& 25& 25\\
         Horiz.~norm.~emittance& \textmu m&  25&  25&  25\\
         Long.~norm.~emittance&  eVs&  0.025&  0.025&  0.025\\
         Bunch length at injection & ps & 31& 20&14\\ \hline
         Bunch length at ejection &  ps &  20&  14&  10\\
         Straight section length& m & 1033.6& 3989.4& 4003\\
         Length with pulsed dipole magnets& m & 4075& 18630& 12808\\
         Length with steady dipole magnets& m & -& -& 5659\\
         Injection pulsed dipole field& T& 0.32& 0.39& -1.8\\
         Max.~pulsed dipole field& T& 1.8& 1.8& 1.8\\
         Max.~steady dipole field& T& -& -& 10\\
 Ramp rate& T/s& 3280& 541&810\\
 Main RF frequency& GHz& 1.3& 1.3&1.3\\
 Harmonic number& & 29900& 115345&115345\\
    \end{tabular}
    \caption{Key acceleration Parameters for the CERN-site based RCS Acceleration Chain }
    \label{site:tab:RCS_CERN_key}
\end{table}

\begin{table}[!h]
\centering
\begin{tabular}{l c  |c c c }
Parameter & Unit & RCS1& RCS2& RCS3\\
\hline
Fill ratio dipole & \% & 59& 70& 70\\
Cells per arc &  & 22& 43& 30\\
Number of arcs &  & 8& 8& 8\\
Cell length & m & 33.4& 65.9& 94.4\\
Total Arc length & m & 5878.4& 22669.6& 22656\\
Arc Ratio & -  & 0.85& 0.85& 0.85\\
Relative path length difference & \num{E-06} & 0& 0& 1.5\\
Horizontal aperture & mm  & 76.5& 43.9& 76.8\\
Vertical aperture  & mm  & 23.1& 22.9& 21\\
Transition gamma &  & 40.92& 82.62& 59.04\\
Momentum compaction factor & \num{E-04} & 5.973& 1.465& 2.869\\
Horizontal tune (ring) &  & 51.73& 104.96& 70.52\\
Vertical tune (ring) &  & 51.56& 103.86& 69.93\\
Mean horizontal beta & m & 32.95& 72.15& 87.39\\
Mean vertical beta & m & 29.97& 64.32& 81.77\\
Horizontal natural chromaticity (ring) &  & -71.27& -150.52& -98.44\\
Vertical natural chromaticity (ring) &  & -64.63& -145.36& -87.87\\
\end{tabular}
    \caption{Additional Lattice Parameters for the CERN-based RCS Acceleration Chain }
    \label{site:tab:RCS_CERN_lattice}
\end{table}

% Taylor, please decide whether the paragraph below should be moved elsewhere in the document. Many thanks!

To avoid a complete redesign of the RCS optics for the CERN site-based high-energy acceleration chain, eight straight sections were assumed in each of the three RCS. A refined lattice would of course have to take into account that the existing accelerator tunnel of the SPS ring has a six-fold symmetry with only six long straight sections.

\subsubsection{RCS Layout at FNAL}
\label{site:sec:RCS_FNAL}

Tables~\ref{site:tab:RCS_FNAL_key} and \ref{site:tab:RCS_FNAL_add} give tentative parameters for a four ring RCS layout for a Fermilab sited muon collider.
A circumference of \qty{6283}{\meter} is chosen for RCS1 and RCS2 to match that of the existing Tevatron tunnel, while a circumference of \qty{15500}{\meter} is chosen for RCS3 to fit within the Fermilab site boundary.
The circumference of \qty{35437}{\meter} of RCS4 was obtained by optimizing for an extraction energy of \qty{5}{\tera\electronvolt}.
This scenario is described in further detail, along with two other scenarios for the Fermilab RCS chain, in \cite{capobianco-hogan:napac2025-tup082}.

A slightly lower field of \qty{1.75}{\tesla} is assumed for pulsed magnets, while a higher field of \qty{14}{\tesla} is assumed for steady magnets.

\begin{table}[!h]
    \centering
    \begin{tabular}{c|c|cccc} 
         Parameter&  Unit&  RCS1&  RCS2&  RCS3&RCS4\\\hline
         Hybrid RCS&  -&  no&  yes&  yes&yes\\
         Repetition rate&  Hz&  5&  5&  5&5\\
         Circumference& m& 6283& 6283& 15500&35437\\
         Injection energy& GeV& 63.0& 174& 454&1541\\
         Extraction energy& GeV& 174& 454& 1541&5000\\
         Energy ratio&  &  2.756&  2.614&  3.394&3.245\\
         Assumed survival rate& &  0.937&  0.926&  0.907&0.828\\
         Cumulative survival rate& & 0.937& 0.867&0.786&0.651\\
         Acceleration time&  ms&  0.148&  0.468&  1.81&11.5\\
         Revolution period&  \textmu s&  21.0&  21.0&  51.7&118\\
         Number of turns& & 7.04& 22.3& 35.1&97.6\\
         Required energy gain/turn& GeV& 15.7& 12.6& 31.0&35.4\\
         Average accel.~gradient& MV/m& 2.50& 2.00& 2.00&1.00\\
         Number of bunches& & 1& 1& 1&1\\
         Inj.~bunch population& \num{E12}& 2.70& 2.53& 2.34&2.12\\
         Ext.~bunch population& \num{E12}& 2.53& 2.34& 2.12&1.76\\
         Vert.~norm.~emittance& \textmu m& 25.0& 25.0& 25.0&25.0\\
         Horiz.~norm.~emittance& \textmu m&  25.0&  25.0&  25.0&25.0\\
         Long.~norm.~emittance&  eVs&  0.025&  0.025&  0.025&0.025\\
         Straight section length& m & 1478& 1142& 2067&2262\\
         Length with pulsed dipole magnets& m & 2496& 2238& 7811&21748\\
         Length with steady dipole magnets& m & -& 626& 1792&5141\\
         Injection pulsed dipole field& T& 0.635& -1.75& -1.75&-1.75\\
         Max.~pulsed dipole field& T& 1.75& 1.75& 1.75&1.75\\
         Max.~steady dipole field& T& -& 14.0& 14.0&14.0\\
 Ramp rate& T/s& 7553& 7486&1931&303\\
 Main RF frequency& GHz& 1.30& 1.30&1.30&1.30\\
 Harmonic number& & 27246& 27246&67213&153666\\
    \end{tabular}
    \caption{Key acceleration Parameters for the Fermilab-site based RCS Acceleration Chain }
    \label{site:tab:RCS_FNAL_key}
\end{table}

\begin{table}[!h]
\centering
\begin{tabular}{l c  |c c c l}
Parameter & Unit & RCS1& RCS2& RCS3&RCS4\\
\hline
Fill ratio dipole & \% & 33.1& 34.2& 51.6&72.3\\
Cells per arc &  & 12& 8& 12&42\\
Number of arcs &  & 12& 12& 12&6\\
Cell length & m & 30.8& 47.6& 86.1&126\\
Transition gamma &  & 38.2& 24.7& 35.7&60.4\\
Momentum compaction factor & \num{E-04} & 6.84& 16.44& 7.85&2.74\\
\end{tabular}
    \caption{Additional Lattice Parameters for the Fermilab-based RCS Acceleration Chain }
    \label{site:tab:RCS_FNAL_add}
\end{table}\FloatBarrier
\subsection{Fixed Field Acceleration}
\label{app:FFA}

Milestone 17 \cite{ffa:MS17} outlines the possibility of using vertical-excursion Fixed-Field Accelerator (vFFA) \cite{ffa:ffatheory} rings as alternatives to one or more of the RCS rings, giving example parameters for RCS1 and RCS4 greenfield equivalents. These would provide the possibility for acceleration unconstrained by magnet ramp rates, removing issues for power conversion and storage, and enabling the construction of rings with full-superconducting magnet technology (thereby enhancing power efficiency). The relative isochronicity of the vFFA concept mitigates the need for frequency cycling in the RF systems, and enables the possibility of on-crest acceleration for an increased acceleration efficiency at a given RF voltage. Table \ref{tab:ffa} lists the design parameters of these FFA rings alongside key parameters for comparison to the RCS equivalents. 

However, the use of FFA arcs implies a closed orbit that moves as a function of energy. This increases requirements for element apertures. Milestone 17 presents a scheme for the implementation of dispersion suppressors to reduce the impact of this upon the RF systems. Depending on the specific execution of these schemes, further optimisation of the parameters in the included table could be possible to reduce peak fields and reduce the size of the machine by separating RF requirements from arc design requirements. 

\begin{table}[!h]
\centering
\caption{FFA alternative tentative parameters}
\label{tab:ffa}
\begin{tabular}{ l c c | c  c }
Design Parameter & Symbol & Unit & vFFA1 & vFFA4 \\
\hline
Orbit radius at centre of F-magnet & r\_0 & m & 953 & 5570 \\
F-magnet bending angle  & $\theta\_F$ & rad. & 0.01033 & 0.00786 \\
Number of cells & N\_c &  & 790 & 1000 \\
F-magnet half-opening angle & $\beta\_F$ & rad. & 0.0015640 & 0.0019700 \\
D-magnet half-opening angle & $\beta\_D$ & rad. & 0.0011800 & 0.0011660 \\
F-magnet orbit inclination & $\gamma\_F$ & rad. & 0.252 & -0.492 \\
vFFA normalised field index & m & 1/m & 31.94 & 12.13 \\ \hline 
Comparison Parameter &  &  &  &  \\ \hline
Circumference [m] &  & m & 5990 & 35000 \\
Injection Energy [TeV] &  & TeV & 0.06 & 1.5 \\
Extraction Energy [TeV] &  & TeV & 0.3 & 5 \\
Ramp Rate [T/s] &  & T/s & 0 & 0 \\
Vertical Excursion [m] &  & m & 0.048 & 0.099 \\
Relative path length difference &  &  & 0 & 0 \\
Peak Dipole Field On Orbit [T] &  & T & 6.93 & 13.59 \\
Peak Dipole Field (Good Field Region) [T] &  & T & 12.52 & 29.04 \\
Drift length [m] &  & m & 1.18 & 1.03 \\
Tune &  &  & (0.382, 0.079) & (0.460, 0.057) \\

\end{tabular}

\end{table}

 \FloatBarrier
\section{Appendix: Machine-Detector Interface}
\label{app:MDI}
The beam-induced background arising from muon decay poses a significant challenge for the physics performance of a multi-TeV muon collider.
The machine-detector interface relies on massive absorbers in close proximity to the interaction point (IP) to reduce the number of secondary particles reaching the detector.
This section describes the geometrical features of the shielding and quantifies the flux of secondary background particles.
In addition, the ionizing dose and displacement damage in different parts of the detector are presented. 

\subsection{Nozzle geometry and material composition}
\label{mdi:sec:nozzle}
The innermost part of the machine-detector interface consists of a nozzle-like shielding, which defines the inner detector envelope.
The nozzle extends from the last magnet ($L^*=6$~m) to almost the IP and must be made of a high-$Z$ and high density material to shield efficiently the electromagnetic showers induced by the decay electrons and positrons.
All studies carried out so far were based on the slightly modified nozzle geometry than the one developed within the Muon Accelerator Program (MAP) \cite{Mokhov2011,Mokhov2012}.
Although the MAP nozzle was optimized for a center-of-mass energy of 1.5~TeV, it has been used as a starting point for the first 10~TeV studies (see, for example, Refs.~\cite{Calzolari2022,Calzolari2023}). 

\begin{figure}[!h]
    \centering
    \includegraphics[width=0.6\linewidth]{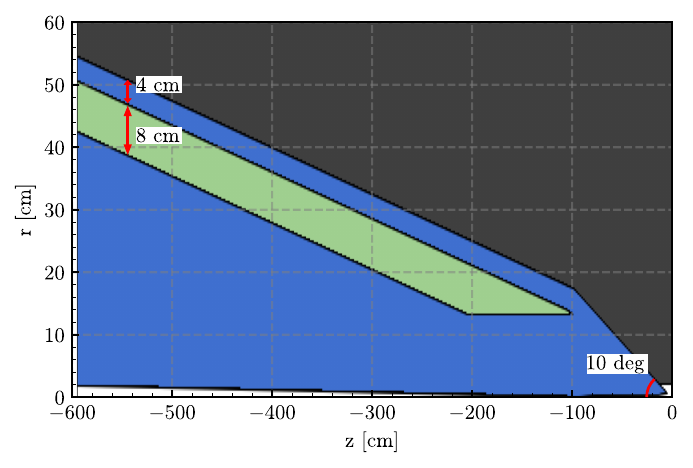}
    \caption{Left nozzle geometry dimensions. The blue layer is made of INERMET180 (registered trademark), a heavy tungsten alloy, while the green one is composed of borated polyethylene.}
    \label{mdi:fig:nozzle_geom}
\end{figure}

\begin{table}[!h]
    \centering
    \begin{tabular}{l|l}
         z [cm]& r [cm]\\ \hline
         \multicolumn{2}{l}{\textbf{Outer surface of nozzle}}\\
         595& 55\\
         100& 17.57\\
         6& 1\\ 
         \multicolumn{2}{l}{\textbf{Outer surface of the borated polyethylene layer}}\\
         595& 51\\
         100& 13.57\\
         \multicolumn{2}{l}{\textbf{Inner surface of the borated polyethylene layer}}\\
         595& 43\\
         204.49&13.47\\
         100&13.47\\
         \multicolumn{2}{l}{\textbf{Inner aperture of the nozzle}}\\
         595&1.78\\
         100&0.3\\
         15&0.6\\
         6&1\\
    \end{tabular}
    \caption{Nozzle Dimensions}
    \label{mdi:tab:nozzle_geom}
\end{table}

Figure~\ref{mdi:fig:nozzle_geom} illustrates the modified MAP nozzle geometry in the $z-r$ plane, where $z$ is the beam axis and $r$ is the radial coordinate.
The nozzle is assumed to have azimuthal symmetry around the $z$-axis.
The figure shows only the nozzle on the left side of the IP; the second nozzle has the same shape but is mirrored with respect to the interaction point.
The nozzle is assumed to consist mainly of INERMET180 (registered trademark), a tungsten-based alloy (blue color), with a layer of borated polyethylene on the outer surface (green color) to thermalize and absorb neutrons before they reach the detector.
Using a tungsten alloy (instead of pure tungsten) is required to allow the manufacture of such shielding elements, however such a choice reduces slightly the shielding effectiveness of the nozzle due to the lower material density.
The beam pipe connecting the two opposite nozzles is made of beryllium, with an internal radius of \SI{2.3}{\centi\meter} and a thickness of \SI{1}{\milli\meter}.

The nozzle tip is located at a distance of \SI{6}{\centi\meter} from the IP. The inner aperture of the nozzle features three different angles, with an aperture bottleneck at \SI{100}{\centi\meter} from the IP. In the region between \SI{100}{\centi\meter} and the first magnet at \SI{600}{\centi\meter}, the inner nozzle surface increases and is defined by the required beam clearance to avoid direct halo losses on the aperture.
The outer surface of the nozzle follows a conical shape, with two different angles.
Near the interaction point, the inclination amounts to 10 degrees, which determines the angular acceptance of the detector.
All the space outside the nozzle and the central beam pipe can be occupied by the detector.
The present setup is of conceptual nature, without yet considering engineering aspects or a possible support structure for the nozzle.

Table \ref{mdi:tab:nozzle_geom} summarizes the coordinates of the inner aperture and outer surface of the nozzle, respectively.
Table \ref{mdi:tab:nozzle_elements} provides the material components of the nozzle.

\begin{table}[!h]
    \centering
    \begin{tabular}{cccc} \hline
         Component&  Density [g/cm3]&  Element& Atomic Fraction (mass fraction if negative)\\ \hline
         EM Shower Absorber&  18&  W& -0.95\\
         &  &  Ni& -0.035\\
         &  &  Cu& -0.015\\ \hline
         Neutron Absorber&  0.918&  H& 0.5\\
         &  &  C& 0.25\\
         &  &  B& 0.25\\ \hline
    \end{tabular}
    \caption{Material composition of nozzle}
    \label{mdi:tab:nozzle_elements}
\end{table}

\subsection{Beam-induced background}
\label{mdi:sec:bib}
%% ==== START TentativeChapters/Author24-MDI ==== %%
The number of background particles entering the detector per bunch crossing depends on the nozzle geometry, the nozzle material composition and the interaction region layout. Table~\ref{tab:MDI_bib_particles} summarizes the number of secondary electrons, positrons, photons and neutrons reaching the detector in a 10~TeV muon collider. The numbers were obtained with FLUKA Monte Carlo simulations, considering the nozzle introduced in the previous section. The bunch intensity was assumed to be 1.8$\times$10$^{12}$ muons. Only secondary particles with energies above a given threshold value were considered (see Table~\ref{tab:MDI_particle_thresholds}).

\begin{table}[h]
\begin{center}
\caption{Particle production and transport thresholds assumed in the background simulations.}
\label{tab:MDI_particle_thresholds}
\begin{tabular}{lc}
\textbf{Particle type} & \textbf{Threshold} \\
\hline
Electrons, positrons and photons & 100 keV \\
Hadrons and muons & 100 keV \\
Neutrons & 0.01 meV \\
\end{tabular}
\end{center}
\end{table}

The number of background particles presented in this section includes only the contribution from muon decay, which is expected to be the dominant source of beam-induced background. Other background sources can include muon halo losses on the aperture and incoherent electron-positron pair production.
%% ==== END TentativeChapters/Author24-MDI ==== %%

\subsection{Ionizing dose and displacement damage in detector}
\label{mdi:sec:dose}
%% ==== START TentativeChapters/Author24-MDI ==== %%
To evaluate the cumulative radiation damage in detector equipment, two quantities have been considered: the total ionizing dose and the 1~MeV neutron-equivalent fluence in Silicon. The former is a measure for the radiation damage in organic materials and compounds, while the latter is related to the displacement damage. %Figures~\ref{fig:MDI_radiation_dose} and \ref{fig:MDI_radiation_1MeVneq} show the spatial distribution of the two quantities in a CLIC-like detector (10~TeV), assuming a MAP-like nozzle. The figures include the vertex detector, the inner and outer tracker, as well as the electromagnetic calorimeter. 

%% ==== END TentativeChapters/Author24-MDI ==== %%
\FloatBarrier
\section{Appendix: Magnets}
\label{app:mag}
Table \ref{mag:tab:developments} provides a summary of the magnet parameters for the study so far.

%\subsection{Magnet Needs and Challenges}
%\label{mag:sec:needs}
The short muon lifetime (2.2 \textmu s at rest) and production of bright muon beams results in a unique set of demands for magnet technologies, including large-bore high-field solenoids, dipoles and quadrupoles, compact ultra-high-field solenoids, and very fast-ramping dipoles.
%The technological overlap with other fields of magnet science such as fusion is a strong motivator for present and future research and development.
Activities within the scope of the IMCC has led to the most advanced set of main magnet conceptual designs and performance parameters.
These parameters are an evolution of previous studies, in particular the U.S. Muon Accelerator Program (MAP) \cite{palmer2015}, extending the performance space by considering recent advances in magnet technology.
%In particular, work carried out so far has a strong focus on High Temperature Superconducting (HTS). The driving reasons for this are the higher field reaches possible, considerations of efficient cryogenic operation, and potentially limited future helium inventory.

%We have taken MAP results as the starting point to identify the main challenges and technology options, and rank priorities.
This section will primarily consider the design challenges of the HTS 6D cooling solenoids, and the aperture-field developments of the collider dipoles and quadrupoles.

\subsection{Cooling Solenoids}
\label{mag:sec:cool}
The overview of the cooling system parameters are in Section \ref{cool:sec}, which factors in our evolving understanding of acceptable solenoid parameter limits. We are presently performing analysis and optimization on this latest configuration.
The 6D cooling section is crucial for producing a high-brightness muon beam, necessary for achieving the required luminosity at the interaction point. In this section the particles are cooled in the 6D phase space (position and momentum), the beam is focused and the bunch size is manipulated through the ionization cooling process.
%To the first order, the final emittance of the muon beam is inversely proportional to the strength of the final cooling solenoids.
%The design study from MAP was based on a \SI{30}{\tesla} final cooling solenoid, and demonstrated that an emittance roughly a factor of two greater than the transverse emittance goal can be achieved \cite{Stratakis2015}.
%Other studies \cite{palmer2011muon} show that fields in the range of \SI{50}{\tesla} improve the final emittance requirements and offer further gains in beam brightness.
%To improve upon these results, we are considering an HTS final cooling solenoid with the potential to reach an excess of \SI{40}{\tesla}. 

\subsubsection{Baseline 6D Cooling solenoids}
\label{mag:sec:6d_cool}

\begin{table}[!h]
\begin{center}
\begin{tabularx}{\linewidth}{lXX|XXXXXX}
\hline\hline
Cell & $E_\text{Mag}$ & $e_\text{Mag}$ & Coil & \textbf{$J_E$} & $B_\text{peak}$   & $\sigma_\text{Hoop}$ (Max.)  &$\sigma_\text{Radial}$ (Min.) &$\sigma_\text{Radial}$ (Max.)  \\
& (MJ) & (MJ/m$^3$) & & (A/mm$^2$) & (T)  & (MPa) & (MPa) & (MPa) \\
\hline\hline
A1  & 5.4   & 21     & A1-1   & 57.6  & 5.2   & 42   & -8  & 0 \\
A2  & 22.1  & 106.1  & A2-1   & 149.5 & 11.6  & 194 & -48 & 0 \\
A3  & 5.0   & 49.5   & A3-1   & 131.5 & 10.1  & 121 & -25 & 0 \\
A4  & 8.0   & 92.3   & A4-1   & 193.2 & 13.8  & 225 & -51 & 1 \\
B1  & 9.1   & 49.8   & B1-1   & 96.9  & 7.7   & 104 & -24 & 0 \\
B2  & 15.6  & 64.2   & B2-1   & 102.1 & 9.2   & 131 & -32 & 0 \\
B3  & 36.9  & 105.9  & B3-1   & 127.9 & 12.9  & 208 & -57 & 0 \\
B4  & 32.2  & 78.9  & \textbf{B4-1}   & 103.0  & 10.6  & 281 & -1  & \textbf{24} \\
B4  & 32.2  & 78.9  & B4-2   & 110.9  & 9.9  & 132 & -49  & 1 \\
B5  & 17.3  & 88.9   & B5-1   & 179.6 & 14.7  & 295 & -2  & 17 \\
B5  &       &        & B5-2   & 154.0 & 14.7  & 212 & -57 & 1 \\
B6  & 8.3   & 96.6   & \textbf{B6-1}   & 214.4 & 15.3  & \textbf{339} & -5  & 18 \\
B6  &       &        & B6-2   & 211.5 & 12.0  & 214 & -6  & 6 \\
B6  &       &        & B6-3   & 212.7 & 12.4  & 162 & -46 & 0 \\
B7  & 8.2   & 87.7   & \textbf{B7-1}   & 183.3 & 14.7  & 264 &  0  & \textbf{25} \\
B7  &       &        & B7-2   & 153.9 & 11.1  & 175 & -4  & 10 \\
B7  &       &        & B7-3   & 210.3 & 13.2  & 180 & -45 & 1 \\
B8  & 8.8   & 92.1   & \textbf{B8-1}   & 193.7 & 16.5  & 270 & -6  & \textbf{38} \\
B8  &       &        & \textbf{B8-2}   & 202.1 & 15.4  & 270 & -6  & \textbf{29} \\
B8  &       &        & B8-3   & 212.8 & 13.2  & 187 & -50 & 0 \\
B9  & 7.5   & 76.5   & \textbf{B9-1}   & 256.4 & 17.2  & 281 &  0  & \textbf{37} \\
B9  &       &        & B9-2   & 88.4  & 10.0  & 95  & -2  & 12 \\
B9  &       &        & B9-3   & 204.9 & 13.2  & 184 & -46 & 0 \\
B10 & 5.0   & 68.6   & \textbf{B10-1}  & 326.8 & 19.2  & \textbf{378} &  0  & \textbf{49} \\
B10 &       &        & B10-2  & 146.1 & 11.1  & 105 & -4  & 13 \\
B10 &       &        & B10-3  & 207.8 & 12.5  & 158 & -43 & 1 \\
\end{tabularx}
\end{center}
\caption[Solenoid types in the latest 6D cooling optics]{Table of various parameters for 14 cell types and 26 unique solenoid types in the latest 6D cooling optics \cite{zhu2024performance}. Values correspond to solenoids operating in their respective cells within a lattice. In bold, the parameters exceeding the considered design limits. The reported parameters will vary depending on the solenoid operational conditions (e.g., stand-alone or single cell operation).}
\label{mag:tab:rectilinear_magnet}
\end{table}

In the current configuration, a total of 3054 solenoids are spread over a 0.85 km distance. There are 14 unique cell types, and 26 unique solenoid types.
During the beam dynamics studies, we integrated a magnet design guide to constrain allowable magnet geometries and current densities based on key solenoid parameters (stresses $\sigma$, stored magnetic energy $e_m$, critical current density $J_c$).
The limits are evaluated considering tape characteristics based on industrial production (Fujikura FESC-SCH ReBCO tape) \cite{Fujikura}.
The parameters and limits implemented (considering stand-alone, single solenoid operation) are: hoop stress, $\sigma_{\theta}< 300$ MPa; radial tensile stress, $\sigma_{r}< 20$ MPa; and stored magnetic energy density, $e_m< 150$ MJ/m$^3$. The limits are identified from average single HTS tape characteristics with an adequate safety margin, ensuring a conservative approach given the considered homogenized coil representation.
Additionally, a limit on the maximum current density was considered. The $J_E$ values were compared to the critical current density $J_c$ values from the measurements reported in \cite{fujita2019flux}, aiming at 2.5 K margin for HTS operating at 20 K.
%This yielded a final optics with assumed solenoid geometries near or within allowed design limits, which can be further optimized.
We report in Tab. \ref{mag:tab:rectilinear_magnet} the main parameters of each cooling cell type and unique solenoid type.
These values are computed based on the lattice design within Section \ref{app:6d}.
%Values are computed assuming each cell is nested within a lattice of neighboring cells of the same type. 

Observing Tab. \ref{mag:tab:rectilinear_magnet}, we find some solenoids exceed allowed design limits, primarily in terms of large hoop stresses (B6-1, B10-1) and tensile radial stresses (B4-1, B7-1, B8-1, B8-2, B9-1, B10-1).
The most concerning solenoid is B10-1, with a hoop stress of 378 MPa, tensile radial stress of 49 MPa, and peak field on the coil of 19.2 T, exceeding its $J_c$ by 114\%.
Critical to the solenoid configuration identified for the latest 6D cooling optics is the gap distance to the beam pipe and to the RF cavities. Following integration studies on the 6D cooling cell demonstrator, we found that larger gaps are needed to integrate the solenoids with RF cavities and absorbers within each cooling cell, making this solution not feasible from the point of view of cell integration. Therefore, another iteration of the design parameters is expected. The proposed solenoid configuration for the latest optic is thus a first step in the definition of an integrated design, combining the beam optics requirements with a more comprehensive engineering design of the cooling cell solenoids. Further optimization will be needed, starting from the initial set of solenoids and integrating the inputs from the WP8 cell integration studies.
% Importantly though, most of the solenoids are within or near the allowed design limits demonstrating the success of the iteration of design parameters with beam optics to produce an initial set of solenoids.

\subsubsection{Low Stress 6D Cooling solenoids}
\label{mag:sec:6d_cool_lowstress}

\begin{table}[!h]
\centering

\begin{tabular}{ccc|cccccc}
\hline\hline
Cell & $E_\text{Mag}$ & $e_\text{Mag}$ & Coil & \textbf{$J_E$} & $B_\text{peak}$   & $\sigma_\text{Hoop}$ (Max.)  &$\sigma_\text{Radial}$ (Min.) &$\sigma_\text{Radial}$ (Max.)  \\
& (MJ) & (MJ/m$^3$) & & (A/mm$^2$) & (T)  & (MPa) & (MPa) & (MPa) \\
\hline\hline
A1 & 9.8 & 32.6 & A1-1 & 68.6 & 6.4 & 66.1 & -16.7 & 0.2 \\
A2 & 38 & 72.7 & A2-1 & 94.7 & 10.7 & 149.2 & -40 & 0.7 \\
A3 & 9.1 & 87.5 & A3-1 & 168.5 & 11.9 & 188.4 & -41.3 & 1.5 \\
A4 & 13.1 & 83.9 & A4-1 & 164.6 & 14 & 218.5 & -49.6 & 5.6 \\
B1 & 13.1 & 13.8 & B1-1 & 32.1 & 5 & 33.5 & -5.5 & 0.1 \\
B2 & 25.4 & 31.3 & B2-1 & 52.3 & 7.5 & 71.3 & -14 & 0.2 \\
B3 & 32.9 & 44.1 & B3-1 & 84.6 & 8.2 & 154.3 & -2.3 & 5.2 \\
B3 &  &  & B3-2 & 67.3 & 9.3 & 84.9 & -22.3 & 0.3 \\
B4 & 47.5 & 88 & B4-1 & 115.2 & 9.2 & 231.5 & -3.2 & 16.4 \\
B4 &  &  & B4-2 & 110.1 & 12.3 & 176 & -52.7 & 1 \\
B5 & 11.2 & 51 & B5-1 & 141.6 & 12.2 & 220.1 & -7 & 10.8 \\
B5 &  &  & B5-2 & 113.4 & 12 & 112.6 & -32.9 & 3.5 \\
B6 & 14.8 & 66.7 & B6-1 & 183.4 & 13.8 & 276.3 & -42.4 & 14.3 \\
B6 &  &  & B6-2 & 132.8 & 12.2 & 184.1 & -161.3 & 7.9 \\
B6 &  &  & B6-3 & 136.3 & 10.8 & 106.5 & -43.1 & 1.1 \\
B7 & 10.8 & 47.8 & B7-1 & 220.2 & 14.5 & 294.3 & -26.4 & 13.8 \\
B7 &  &  & B7-2 & 113 & 9.4 & 127.8 & -95.3 & 5.4 \\
B7 &  &  & B7-3 & 116.2 & 10.2 & 81.5 & -26.4 & 1 \\
B8 & 6.1 & 27.1 & \textbf{B8-1} & 221.6 & 15.3 & 293.3 & -13.8 & \textbf{23.7} \\
B8 &  &  & B8-2 & 115.2 & 6.1 & 114.8 & -17.4 & 6.6 \\
B8 &  &  & B8-3 & 78.6 & 6 & 23.2 & -13.8 & 0.3 \\
B9 & 15 & 48.9 & \textbf{B9-1} & 223 & 15.7 & 301.4 & -64.4 & \textbf{28.8} \\
B9 &  &  & B9-2 & 107.3 & 7.7 & 192.3 & -20.5 & 17.4 \\
B9 &  &  & B9-3 & 106.6 & 10.3 & 74.8 & -64.9 & 0.7 \\
B10 & 7.2 & 21.2 & \textbf{B10-1} & 254.6 & 16.5 & 333.6 & -19.8 & \textbf{30.9} \\
B10 &  &  & B10-2 & 106.7 & 6.3 & 117.5 & -18.9 & 8.8 \\
B10 &  &  & B10-3 & 65.6 & 8.3 & 30.2 & -19.9 & 1.3 \\

\end{tabular}
\caption[Solenoid types in low stress 6D cooling optics]{New lattice with larger gaps and reduced solenoid stress (low stress variant). In bold, the parameters exceeding the considered design limits.}
\label{mag:tab:rectilinear_magnet_low_stress}
\end{table}

The rectilinear cooling complex has an alternative lattice in Section \ref{app:6d:lowstress}, with the aim of reducing the coil stresses and enlarge the gap distance between the solenoids and the other integrated cell systems (beam pipe, RF cavity, absorbers), in response to the investigation of Section \ref{mag:sec:6d_cool}.
The recalculated stresses are reported in Table \ref{mag:tab:rectilinear_magnet_low_stress}. In lattice operation, only three solenoids showed values exceeding the positive radial stress limit (B8-1, B9-1, B10-1), with B9-1 and B10-1 exceeding the hoop stress limit. This is expected, since these three magnets exhibit also the highest peak field values on coils (over 15 T) in a nested coil configuration. 
An improvement on the feasibility of the 6D cooling magnet configuration has been made in this alternative layout, with higher radial and axial gaps separating the different cell systems. A coil optimization is needed to further increase the gaps, considering the inputs from the cooling cell demonstrator study, aiming also to lower the stresses in the last three B-type cells. 
This variant is not yet considered part of the baseline design due to its lower performance. Hence, further iterations in the design parameters for this lattice version are expected.

\subsection{Collider Magnets}
\label{app:mag:collider}
Section \ref{col:sec} presents the collider parameters, including the radiation shielding requirements from muon decay. To achieve a compact ring while allowing sufficient shielding, the ARC and Interaction Region (IR) magnets must feature \textbf{high magnetic fields and large apertures}.

The main arc magnets are \textbf{combined-function magnets} (dipole/quadrupole and dipole/sextupole) designed for magnetic fields up to \textbf{16 T and 160 mm aperture}, though this exceeds current technological limits and requires further optimization. The \textbf{IR quadrupoles} are expected to reach magnetic fields up to \textbf{20 T} and apertures up to \textbf{200 mm}.

\begin{figure}[h!]
    \centering
    \begin{minipage}{0.45\textwidth}
        \centering
        \includegraphics[width=\textwidth]{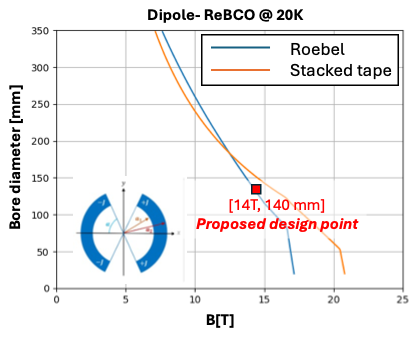}
    \end{minipage}
    \begin{minipage}{0.45\textwidth}
        \centering
        \includegraphics[width=\textwidth]{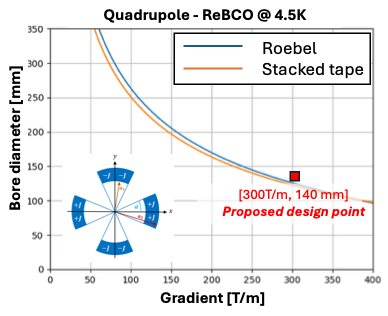}
    \end{minipage}

    \vspace{0.3cm} % Adjust spacing between rows

    \begin{minipage}{0.45\textwidth}
        \centering
        \includegraphics[width=\textwidth]{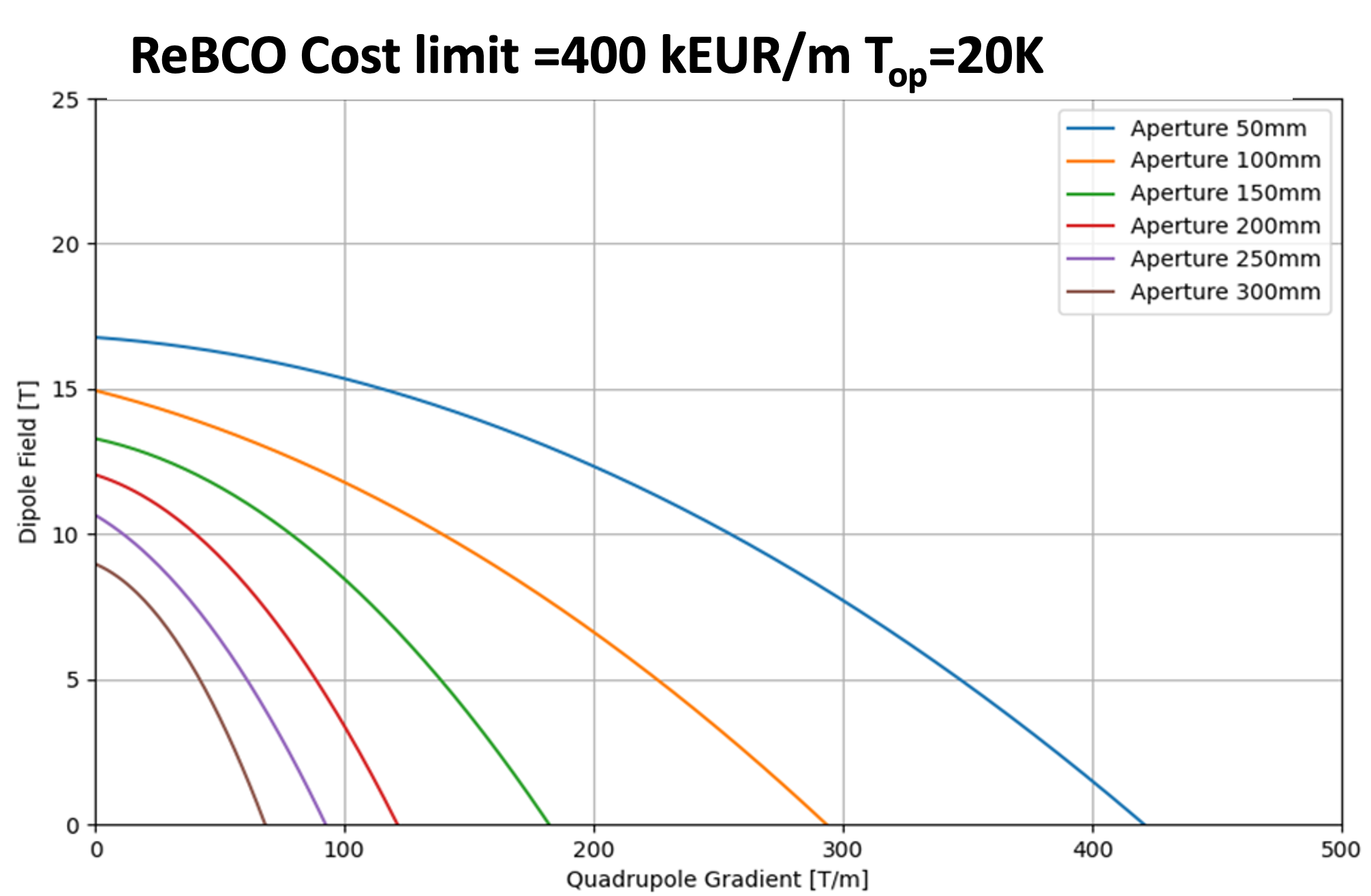}
    \end{minipage}
    \begin{minipage}{0.45\textwidth}
        \centering
        \includegraphics[width=\textwidth]{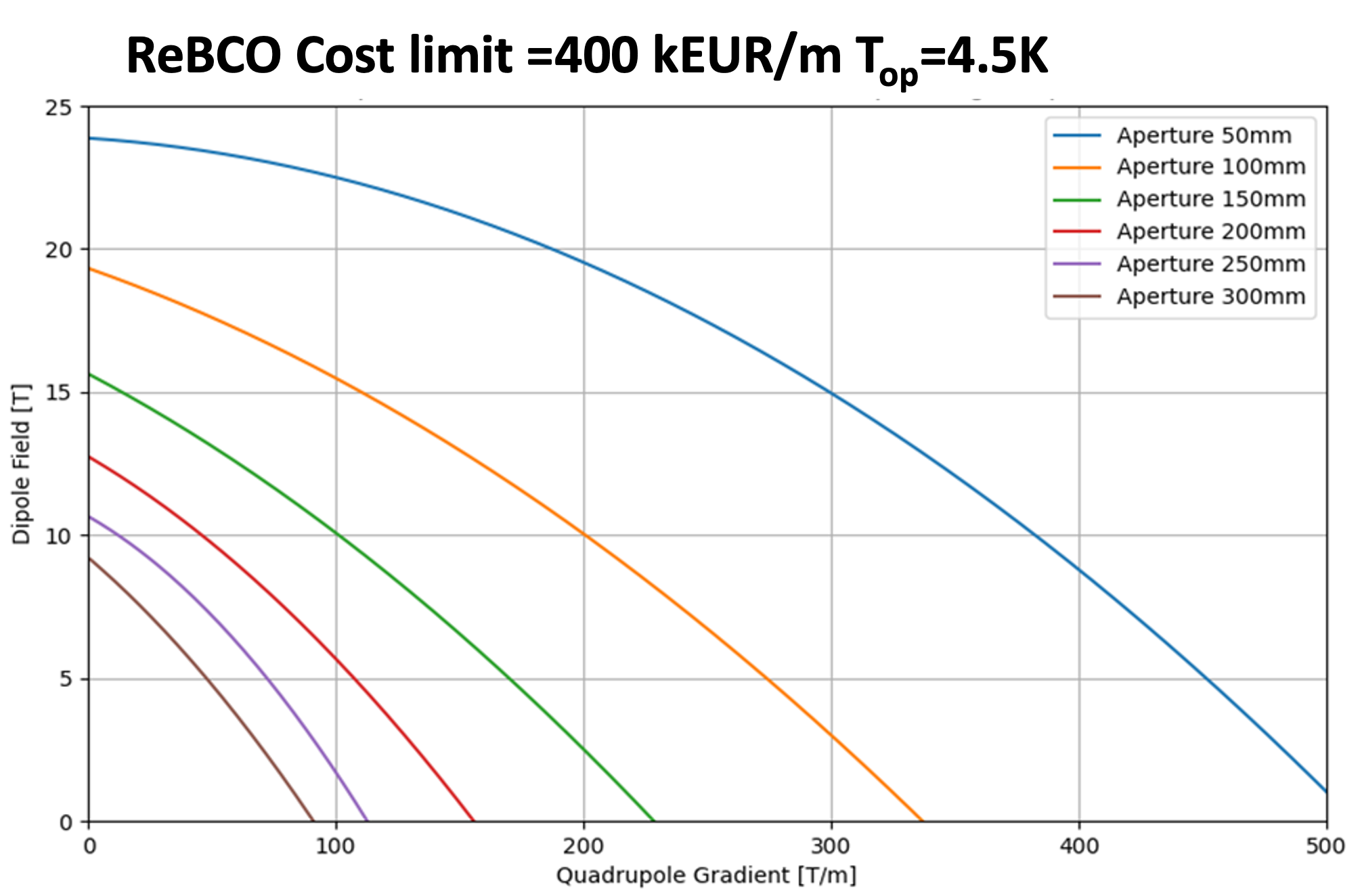}
    \end{minipage}

    \caption{Performance upper-limit plots for ReBCO-based magnets. \textit{Top left}: A–B plots for dipoles at 20 K. \textit{Top right}: A–G plots for quadrupoles at 4.5 K. \textit{Bottom}: B–G plots for combined-function magnets at 20 K (left) and 4.5 K (right).}
    \label{mag:fig:ColliderAB}
\end{figure}

Analytical evaluations based on \textbf{sector-coil geometries} were used to generate \textbf{performance limit plots} -- A–B (aperture vs. field) for dipoles, A–G (aperture vs. gradient) for quadrupoles, and B–G (field vs. gradient) for combined-function magnets—considering \textbf{NbTi, Nb$_3$Sn, and ReBCO} superconductors.
\begin{itemize}
    \item \textbf{NbTi (1.9 K)}: inadequate due to low margins and high energy deposition.
    \item \textbf{Nb$_3$Sn (4.5 K)}: viable up to 14 T, suitable for $\sim$ 3 TeV machines but insufficient for some of the magnet performances required for the 10 TeV.
    \item \textbf{ReBCO}: best-performing option with high fields (10–20 K operation), presently considered as baseline, though R\&D is needed to address \textbf{cost and quench protection} challenges.
\end{itemize}

If ReBCO costs are reduced by a factor 3–4, feasible designs could span \textbf{14–16 T} with \textbf{100–140 mm apertures} at operating temperature in the range 4.5 K - 20 K. For IR quadrupoles, operation at 4.5 K could enable up to \textbf{300 T/m gradients} with apertures up to \textbf{140 mm}.

A \textbf{semi-analytic design tool} has supported fast iteration with beam dynamics, cryogenics, and energy deposition studies. The resulting performance limit plots for \textbf{dipoles, quadrupoles, and combined-function magnets} are shown below in Table \ref{mag:fig:ColliderAB}.

\FloatBarrier
\section{Appendix: Radiofrequency Cavities}
\label{app:rf}

\subsection{RF systems for rectilinear cooling}
\label{rf:sec:cool}
The preliminary RF cavity design for each stage of the rectilinear cooling channel was developed based on the shape presented in \cite{Barbagallo:2024zak} following the beam dynamics specification in Table~\ref{cool:tab:6d_rf}. 
The other geometrical parameters characterizing the cavity shape are chosen to maximize the shunt impedance ($R/Q\cdot Q_{0}$) and reduce surface losses ($P_{\mathrm{diss}}$) on the windows and cavity walls.
The peak surface electric field ($E_{\mathrm{peak}}$) is also minimized to avoid RF breakdown risk. The RF cavity frequency ($f_{0}$), the cavity length ($L_{\mathrm{cav}}$), and the nominal RF gradient along the cavity axis ($E_{\mathrm{nom}}$) for the studied RF cavities are reported in Table~\ref{cool:tab:6d_rf}. Table~\ref{rf:tab:cool_cavity} summarizes the relevant RF figures of merit computed for the operating frequencies of the studied cavities. Most of the power is dissipated in the cavity walls. 

\begin{table}[!h]
    \centering
    \begin{tabular}{c|cccccccc}
          & $Q_0$& $t_f$& $DF$& R/Q&  ${P_{\mathrm{diss}}}$ & $\frac{{P_{\mathrm{diss,Be}}}}{{P_{\mathrm{diss}}}}$ & $E_{\mathrm{peak,Cu}}$&$E_{\mathrm{peak,Be}}$\\
         & $10^4$& \textmu s& $10^{-4}$& $\Omega$ & MW/cavity & - & MV/m & MV/m\\ \hline
         Stage A1& 3.06& 31.203& 1.17& 171.73& 4.25&0.377& 11.72&27.383\\
         Stage A2& 3.14& 32.087& 1.21& 149.68& 4.34&0.085& 23.249&26.511\\
         Stage A3& 2.20& 11.248& 0.43& 160.36& 2.06&0.201& 20.802&31.507\\
         Stage A4&  2.22&  11.345&  0.43&  150.21&  2.21& 0.085& 27.873&31.829\\
         Stage B1&  3.91&  39.954&  1.51&  183.70&  2.678& 0.23& 12.162&21.25\\
         Stage B2&  3.56&  36.323&  1.37&  170.47&  2.807& 0.164& 15.251&21.757\\
         Stage B3&  3.15&  32.148&  1.21&  141.27&  4.07& 0.031& 26.175&24.429\\
         Stage B4&  3.59&  36.71&  1.38&  154.02&  3.92& 0.009& 27.732&22.823\\
         Stage B5&  2.23&  11.366&  0.43&  140.85&  1.18& 0.026& 24.116&22.027\\
         Stage B6&  2.22&  11.36&  0.43&  137.40&  1.89& 0.007& 33.288&26.514\\
         Stage B7&  2.22&  11.354&  0.43&  136.87&  1.97& \num{3.08E-03} & 35.25&25.981\\
         Stage B8&  2.22&  11.347&  0.43&  137.39&  1.79& \num{8.32E-04} & 34.67&22.885\\
         Stage B9&  2.22&  11.344&  0.43&  138.11&  2.14& \num{3.16E-04} & 38.528&23.268\\
         Stage B10&  2.22&  11.342&  0.43&  139.06&  1.51& \num{1.56E-04} & 32.522&18.341\\
    \end{tabular}
    \caption{RF figures of merit for the RF cavities in the rectilinear cooling channel}
    \label{rf:tab:cool_cavity}
\end{table}

 The filling time $t_{\mathrm{f}}$, which is the time required to fill the cavity to the nominal voltage~$V_{\mathrm{nom}} = E_{\mathrm{nom}} L_{\mathrm{cav}}$, is given by:

\begin{equation}
 t_{\mathrm{f}} \approx \frac{2Q_{\mathrm{L}}}{\omega_{0}} \ln\left(\frac{2\beta_{\mathrm{c}}}{\beta_{\mathrm{c}} - 1}\right),
\label{eq:filling_time}
\end{equation}

\noindent
where $Q_{\mathrm{L}}=Q_{0}/(1+\beta_{\mathrm{c}})$, with $Q_{0}$ being the intrinsic quality factor, $\beta_{\mathrm{c}}$ the coupling factor, and $\omega_{0}$ is the angular frequency of the cavity's operating mode.
The beam duty factor ($DF$) can be calculated as the ratio between the average power and the peak dissipated power: 

\begin{equation}
     DF = \frac{P_{\mathrm{ave}}}{P_{\mathrm{diss}}} = \frac{\int_{0}^{\infty}P(t)\mathrm{d}t \cdot f_{\mathrm{b}}}{V_{\mathrm{acc}}^{2}/(R/Q \cdot Q_{0})},
\label{eq:duty_factor} 
\end{equation}
\noindent
where $P(t)$ is the time-dependent power calculated from the cavity voltage profile and $V_{\mathrm{acc}} = TTF\cdot V_{\mathrm{nom}}$ the accelerating cavity voltage, with $TTF$ being the Transit-Time factor, given by:

\begin{equation}
TTF = \frac{\int_{z_{\mathrm{min}}}^{z_{\mathrm{max}}}{E}_ze^{j k z} \, dz}{\int_{z_{\mathrm{min}}}^{z_{\mathrm{max}}}{E}_z \, dz},
\label{eq:transit_time_factor} 
\end{equation}

\noindent
where $k=\omega_{0}/(\beta c)$ is the wave number with $c$ being the speed of light in a vacuum and $\beta$ the relativistic velocity factor. The geometric shunt impedance, $R/Q$, is calculated, considering the TTF as:

\begin{equation} \left(\frac{R}{Q}\right)=\frac{| V_{z}(0,0)|^{2}}{\omega_{0}U_{0}}TTF^2,
\label{Geometric shunt impedance}
\end{equation}

\noindent
where $U=\omega_{0}$ is the energy stored in the cavity. 

\begin{comment}
The peak dissipated power is computed as:

\begin{equation} 
P_{\mathrm{diss}}=\frac{V_{\mathrm{acc}}^{2}}{(R/Q)\cdot Q_{0}}.
\label{Peak dissipated power}
\end{equation}

\noindent

\end{comment}

Table~\ref{rf:tab:cool_power} reports the power requirements for each stage of the cooling channel. The peak input RF power is given by:

\begin{equation} 
P_{\mathrm{g}}=P_{\mathrm{diss}}\beta_{\mathrm{c}}.
\label{Peak input RF power}
\end{equation}

The duty factor of the RF power source ($DF_{\mathrm{g}}$) is given as the ratio between the average power of the generator and the peak input RF power. 

\begin{equation}
     DF_{\mathrm{g}} = \frac{P_{\mathrm{ave,g}}}{P_{\mathrm{g}}}  = \frac{P_{\mathrm{g}} t_{\mathrm{f}} \cdot f_{\mathrm{b}}}{P_{\mathrm{g}}},
\label{eq:duty_factor_generator} 
\end{equation}

The total plug power for the RF systems was calculated considering the generator ($\eta_{\mathrm{G}}$) and modulator ($\eta_{\mathrm{M}}$) efficiencies reported in Table~\ref{rf:tab:cool_const} as:

\begin{equation}
     P_{\mathrm{g,ave,tot}} = \frac{N_{\mathrm{cav}}P_{\mathrm{ave,g}}}{\eta_{\mathrm{G}}\eta_{\mathrm{M}}} ,
\label{eq:total_plug_power} 
\end{equation}
\noindent
where $N_{\mathrm{cav}}$ is the total number of cavities for each stage.

\begin{table}[!h]
    \centering
    \begin{tabular}{ccc|c}
         Parameters&  Symbol&  Unit&  Value\\ \hline
         Coupling factor&  $\beta_{\text{c}}$&  -&  1.2\\
         Bunch repetition frequency&  $f_\text{b}$&  Hz&  5\\
         Generator efficiency&  $\eta_\text{G}$&  -&  0.7\\
         Modulator efficiency&  $\eta_\text{M}$&  -&  0.9\\
    \end{tabular}
    \caption{RF parameters for the rectilinear cooling channel}
    \label{rf:tab:cool_const}
\end{table}

 For the RF frequency, cavity length and nominal RF gradient of the rectilinear cooling RF system, please refer to Table \ref{cool:tab:6d_rf}.
Table \ref{rf:tab:cool_dyn}
 displays in addition the RF cavity window radius, window thickness and the relativistic beta of the muon beam at each stage. 
\begin{table}[!h]
    \centering
    \begin{tabular}{l|ccc}
         & Window&  Window&Relativistic\\
          & radius &  thickness &$\beta$ \\
      & mm&  $\mu$m&- \\ \hline
         Stage A1 & 240&  120&0.923\\
         Stage A2 & 160&  70&0.894\\
         Stage A3 & 100&  45&0.894\\
         Stage A4 &  80&   40&0.901\\
         Stage B1 &  210&   100&0.886\\
         Stage B2 &  190&   80&0.885\\
         Stage B3 &  125&   50&0.887\\
         Stage B4 &  95&   45&0.886\\
         Stage B5 &  60&   30&0.889\\
         Stage B6 &  45&   20&0.888\\
         Stage B7 &  38&   20&0.887\\
         Stage B8 &  28&   20&0.884\\
         Stage B9 &  23&   10&0.881\\
         Stage B10&  20&   10&0.884\\
    \end{tabular}
    \caption{Beam dynamics specifications for the RF cavities in the rectilinear cooling channel}
    \label{rf:tab:cool_dyn}
\end{table}

\begin{table}[!h]
    \centering
    \begin{tabular}{l|cccccc}
          & $P_g$& $DF_g$& $N_{cav}$& $P_{g,tot}$& $P_{g,av}$&$P_{plug,tot}$\\
         & MW/cavity& $10^{-4}$& -& MW& kW&kW\\ \hline
         Stage A1& 5.094& 1.560& 348& 1772.7& 277.09&439.83\\
         Stage A2& 5.21& 1.610& 356& 1854.9& 297.87&472.82\\
         Stage A3& 2.468& 0.567& 405& 999.4& 56.70&90.00\\
         Stage A4&  2.655&  0.573&  496&  1317.1&  75.41& 119.70\\
         Stage B1&  3.214&  2.077&  132&  424.2&  88.1& 139.843\\
         Stage B2&  3.368&  2.097&  185&  623.1&  130.682& 207.432\\
         Stage B3&  4.882&  1.611&  240&  1171.6&  188.801& 299.684\\
         Stage B4&  4.701&  1.843&  165&  775.673&  142.945& 226.897\\
         Stage B5&  1.419&  0.573&  275&  390.1&  22.37& 35.51\\
         Stage B6&  2.262&  0.71&  220&  497.7&  35.35& 56.11\\
         Stage B7&  2.363&  0.613&  160&  378&  23.17& 36.783\\
         Stage B8&  2.143&  0.615&  284&  608.5&  37.449& 59.443\\
         Stage B9&  2.573&  0.571&  208&  535.1&  30.556& 48.517\\
 Stage B10& 1.806& 0.572& 188& 339.8& 19.434&30.848\\
    \end{tabular}
    \caption{RF power requirements in the rectilinear cooling channel}
    \label{rf:tab:cool_power}
\end{table}

\FloatBarrier

\subsection{RF systems for low-energy acceleration}
\label{rf:sec:low}
In the low-energy acceleration, only the design of RLA2 is being considered for the computation of RF parameters. 
The baseline cavity geometry is chosen to be the LEP2 cavity. A summary of the assumed parameters can be found in Table \ref{rf:tab:LEP2_cavity_parameters}. 
For the calculation of the losses in the power generation, the parameters of the ILC-powering system were used (Table \ref{rf:tab:ILC_power_consumption}). The resulting powering parameters for the RLA2 cavities can be found in table \ref{rf:tab:low}.

\begin{table}[h!]
    \centering
    \begin{tabular}{lcc|c|c}
Parameter & Symbol & Unit & Value & Value \\ 
& & & linearizer & accelerator\\
\hline
Fundamental mode RF frequency	& $f_\text{RF}$	& MHz & 352 & 1056 \\
Accelerating gradient & $G_\text{acc}$ & MV/m & 15 & 25 \\
Geometric shunt impedance & $R/Q$ & $\Omega$ & 247.25 & 360.72\\
Active length & $l_\text{active}$ & m & 1.686 & 0.845\\
Total length & $l_\text{total}$ & m & 1.851 & 1.011\\
Number of cells & - & - & 4 & 6\\
$E_\text{peak} / E_\text{acc}$ & - & - & 2.4 &  2.4\\
$B_\text{peak} / E_\text{acc}$ & - & mT/(MV/m) & 3.9 & 3.9\\
Iris aperture (inner/end cell) & - & mm & 286/241 & 94/80\\
Cavity quality factor & $Q_0$ & - & $\geq 1\times10^{10}$ & $\geq 1\times10^{10}$\\
Cell-to-cell coupling & $k_\text{cc}$ & \% & 1.51 & 1.62\\
    \end{tabular}
    \caption[Parameters of the LEP2 cavity]{Parameters of the LEP2 cavity from \cite{rf:LEP2_cavity}} 
    \label{rf:tab:LEP2_cavity_parameters}
\end{table}

\begin{table}[h!]
    \centering
    \begin{tabular}{lc|cc}
         Parameter &  Unit &  RLA2 acc& RLA2 lin\\ \hline
         Synchronous phase &  \textdegree &  95& 275\\
         Frequency&  MHz&  352& 1056\\
%         Number of cavity cells&  -&  4& 6\\
%         Active cavity length&  mm&  1686& 845\\
%         Total cavity length&  mm&  1851& 1010\\
         Number of bunches/species &  - &  \multicolumn{2}{c}{1} \\
         Combined beam current ($\mu^+$, $\mu^-$) &  mA&  \multicolumn{2}{c}{134} \\ \hline
         Total RF voltage &  GV&  15.2& 1.69\\
         Total number of cavities &  - &  600& 80\\
         Total number of cryomodules & -  & 200&16\\
         Total RF section length & m& 1110.6&80.8\\
         External Q-factor & \num{E6}& 0.38&0.21\\
         Cavity detuning for beam loading comp. & kHz& 0.04&0.21\\ \hline
         Beam acceleration time & \textmu s& \multicolumn{2}{c}{35.5}\\
         Cavity filling time & \textmu s& 344&65\\
         RF pulse length & ms& 0.38&0.1\\
         RF duty factor & \%& 0.19&0.05\\
         Peak cavity power & kW& 3425&2965\\
         Average RF power & MW& 5.16&0.16\\
    \end{tabular}
    \caption[RF parameters for the low-energy acceleration chain]{RF parameters for the low-energy acceleration chain. For the synchronous phase, \SI{90}{\degree} is defined as being on-crest}
    \label{rf:tab:low}
\end{table}

\begin{table}[h!]
    \centering
    \begin{tabular}{lc|c}
Parameter & Unit & Value \\ 
\hline
Max. klystron power & MW & 10 \\
Klystron efficiency & \% & 65 \\
Additional power requirement & \% & $\sim$32\\
Klystron repetition rate &  Hz & 5 \\
Klystron frequency & MHz & 1300\\
RF pulse length & ms & 1.65 \\
RF duty factor & \% & 0.83 \\
\end{tabular}
    \caption[ILC RF-power parameters]{ILC RF-power parameters \cite{rf:ILC-tdr} in the Distributed Klystron Scheme (DKS). The additional power requirement includes low-level RF overhead as well as distribution losses. }
    \label{rf:tab:ILC_power_consumption}
\end{table}

\FloatBarrier

\subsection{RF systems for high-energy acceleration}
\label{rf:sec:high}
A first approximation of the power requirements for the RCS chain has been performed using the ILC cavities, cryomodules, and powering infrastructures \cite{rf:ILC-tdr} as a baseline, the results of which can be found in Table~\ref{rf:tab:high}. 

\begin{table}[h!]
\begin{center}
\begin{tabular}{lcccccc}
Parameter & Unit & {\textbf{RCS1}} & {\textbf{RCS2}} & {\textbf{RCS3}} & {\textbf{RCS4}} & {\textbf{All}}\\
\midrule
Synchronous phase & \textdegree & 148 & 153 & 134 & 118 & - \\
Number of bunches/species & - & 1 & 1 & 1 & 1 & - \\
Combined beam current ($\mu^+$ and $\mu^-$) & mA & 43.3 & 39.0 & 19.6 & 5.4 & - \\
Total RF voltage & GV & 27.6 & 17.5 & 15.7 & 72.7 & 133.0 \\
Total number of cavities & - & 865 & 548 & 492 & 2275 & 4180 \\
Total number of cryomodules & -  & 97 & 61 & 55 & 253 & 466 \\
Total RF section length & m & 1079 & 684 & 614 & 2838 & 5214 \\
\midrule
External Q-factor & $10^{6}$& 1.29 & 1.76 & 1.84 & 4.34 & - \\
Cavity detuning for beam loading comp. & kHz& -1.04 & -0.90 & -0.54 & -0.19 & - \\
Max. detuning due to orbit length change & kHz&0 & 6.63 & 1.17 & 1.3 & - \\
\midrule
Beam acceleration time & ms & 0.34 & 1.1 & 2.37 & 6.37 & - \\
Cavity filling time & ms & 0.26 & 0.43 & 0.45 & 1.06 & - \\
RF pulse length & ms & 0.60 & 1.53 & 2.82 & 7.43 & - \\
RF duty factor & \%& 0.30 & 0.76 & 1.41 & 3.72 & - \\
\midrule
Total number of klystrons & - & 109 & 50 & 41 & 91 & 291 \\
Cavities per klystron & - & 8 & 11 & 12 & 25 & - \\
Peak cavity power & kW & 855 & 634 & 598 & 294 & - \\
Total peak RF power & MW & 739 & 348 & 294 & 668 & - \\
Peak RF power to beam & MW & 634 & 310 & 222 & 347 & - \\
\midrule
Average cavity power & kW & 2.57 & 4.85 & 8.44 & 10.9 & - \\
Average RF power to cavity during cycle & MW & 2.22 & 2.66 & 4.15 & 24.8 & 33.8 \\
Average wall plug power for RF system & MW & 4.22 & 5.11 & 7.79 & 42.2 & 59.3 \\
HOM power losses per cavity per bunch & kW & 23.0 & 22.2 & 12.2 & 3.79 & - \\
Average HOM power per cavity & W & 78 & 244 & 280 & 242 & - \\
\end{tabular}
\end{center}
\caption[RF parameters for the RCS chain.]{RF parameters for the RCS chain. The average RF power uses the RF pulse length as a reference within the cycle, assuming a \SI{5}{Hz} repetition rate. The wall plug power includes an additional power input requirement of $\sim32\,\%$ above the cavity input power as well as a klystron efficiency of $65\,\%$, both according to the ILC DKS scheme \cite{rf:ILC-tdr}. The number of cryomodules is based on the assumption of the integration of 9 cavities into one cryomodule. The total RF section length only takes the total length of the cavities, but not the additional space for the cryomodules or interconnects into account. The synchronous phase is defined as \SI{90}{\degree} being on-crest.}
\label{rf:tab:high}
\end{table}

The parameters of the ILC cavity can be found in Table \ref{rf:tab:TESLA_cavity_parameters}. To calculate the losses, parameters from the ILC DKS powering scheme are used (Table \ref{rf:tab:ILC_power_consumption}).
While these parameters are used for initial beam dynamics and power requirements studies, other frequencies and cavities are under investigation for muon acceleration. 
The power requirements do not consider cryogenic losses and the impact of the detuning, which is necessary due to the orbit change during the acceleration. 
The calculated parameters assume a linear ramp of the magnet system.
In the accelerator, a harmonic magnet ramp is foreseen, which will require additional cavities. 

\begin{table}[h!]
    \centering
    \caption[Parameters of the TESLA cavity]{Parameters of the TESLA cavity from \cite{rf:ILC-tdr} and \cite{rf:TESLA_wakefields}.}
    \label{rf:tab:TESLA_cavity_parameters}
    \begin{tabular}{lcc|c}
Parameter & Symbol & Unit & Value \\ 
\hline
Fundamental mode RF frequency	& $f_\text{RF}$	& MHz	&1300 \\
Accelerating gradient & $G_\text{acc}$ & MV/m & 30 \\
Geometric shunt impedance & $R/Q$ & $\Omega$ & 518\\
Geometry factor & $G$ & $\Omega$ & 271\\
Active length & $l_\text{active}$ & m & 1.065\\
Total length & $l_\text{total}$ & m & 1.247 \\
Number of cells & - & - & 9\\
$E_\text{peak} / E_\text{acc}$ & - & - & 2.0 \\
$B_\text{peak} / E_\text{acc}$ & - & mT/(MV/m) & 4.26 \\
Iris aperture (inner/end cell) & - & mm & 70/78 \\
Cavity quality factor & $Q_0$ & - & $\geq 1\times10^{10}$ \\
Longitudinal loss factor ($\sigma_z = 1mm$) & $k_{||}$ & V/pC & 11.05\\
Cell-to-cell coupling & $k_\text{cc}$ & \% & 1.87 \\
\end{tabular}
\end{table}

In comparison to last year's parameter report, the synchronous phases were adjusted to minimise the bucket area differences at the transition between the accelerators. To calculate the power and coupling parameters, a detuning of $\Delta\omega_m = \Delta\omega_\mathrm{opt}/\sin{\varPhi_s}$ was assumed \cite{rf:thiele_ipac2025-mops046}. 

%\begin{thebibliography}{99}
%
%\bibitem{rf:LEP2_cavity}
%C.Arnaud \textit{et al.},
%STATUS REPORT ON S I JPERCONDUCTING N B CAVITIES FOR LEP, in \emph{Proc.\ Fourth Workshop on RF Superconductivity, KWK, Tsukuba, Japan, 14--18 Aug.\ 1989}, pp.\ 19--35,
%\href{https://accelconf.web.cern.ch/SRF89/papers/srf89a02.pdf}{https://accelconf.web.cern.ch/SRF89/papers/srf89a02.pdf}.
%\bibitem{rf:TESLA_cavity}
%D.~Proch, The TESLA cavity: Design considerations and RF properties, in \emph{Proc.\ Sixth Workshop on RF Superconductivity, Newport News, United States, 4--8 Oct.\ 1993}, pp.\ 382--397,
%\href{https://accelconf.web.cern.ch/SRF93/papers/srf93g01.pdf}{https://accelconf.web.cern.ch/SRF93/papers/srf93g01.pdf}.
%
%\bibitem{rf:ILC-tdr}
%C. Adolphsen \textit{et al.}, 
%The International Linear Collider Technical Design Report - Volume 3.II: Accelerator Baseline Design,
%\href{https://doi.org/10.48550/arXiv.1306.6328}    
%{doi.org:10.48550/arXiv.1306.6328}   
%\bibitem{ch12:TESLA_wakefields}
%E. Plawski, 
%The wakefields and loss factors in superconducting accelerating cavities for TESLA collider,
%\href{https://doi.org/10.1109/PAC.1999.792319}   
%{doi.org:10.1109/PAC.1999.792319}  
%
%\end{thebibliography}

\subsubsection{RF system for the RCS layout at CERN}
\label{site:sec:RCS_RF_CERN}
The design of the RF system for the RCS at CERN is in Table \ref{site:tab:rf_rcs_cern} and is based on the same assumptions as the RF system for the greenfield study. The assumptions are presented in \ref{rf:sec:high}.

\begin{table}[h!]
\begin{center}
\begin{tabular}{lccccc}
& & {\textbf{RCS1}} & {\textbf{RC2}} & {\textbf{RCS3}} & \\
Parameter & Unit & {\textbf{SPS}} & {\textbf{LHC}} & {\textbf{LHC}} & {\textbf{All}}\\
\midrule
Synchronous phase & \textdegree & 140 & 117 & 135 & - \\
Number of bunches/species & - & 1 & 1 & 1 & - \\
Combined beam current ($\mu^+$ and $\mu^-$) & mA & 37.5 & 8.56 & 7.36 & - \\
Total RF voltage & GV & 23.1 & 48.4 & 62.7 & 134.0 \\
Total number of cavities & - & 724 & 1514 & 1964 & 4202 \\
Total number of cryomodules & -  & 81 & 169 & 219 & 469 \\
Total RF section length & m & 903 & 1888 & 2449 & 5241 \\
\midrule
External Q-factor & $10^{6}$& 1.14 & 2.58 & 5.06 & - \\
Cavity detuning for beam loading comp. & kHz& -0.97 & -0.31 & -0.20 & - \\
Max. detuning due to orbit length change & kHz&0 & 0 & 1.95 & - \\
\midrule
Beam acceleration time & ms & 0.45 & 2.58 & 4.41 & - \\
Cavity filling time & ms & 0.22 & 0.32 & 1.24 & - \\
RF pulse length & ms & 0.66 & 2.9 & 5.65 & - \\
RF duty factor & \%& 0.33 & 1.45 & 2.82 & - \\
\midrule
Total number of klystrons & - & 104 & 109 & 58 & 271 \\
Cavities per klystron & - & 7 & 14 & 34 & - \\
Peak cavity power & kW & 961 & 511 & 217 & - \\
Total peak RF power & MW & 695 & 773 & 427 & - \\
Peak RF power to beam & MW & 557 & 368 & 326 & - \\
\midrule
Average cavity power & kW & 3.19 & 7.4 & 6.15 & - \\
Average RF power to beam during cycle& MW & 2.31 & 11.2 & 12.1 & 25.6 \\
Average wall plug power for RF system & MW & 4.34 & 18.0 & 22.7 & 45.0 \\
HOM power losses per cavity per bunch & kW & 21.7 & 5.78 & 5.28 & - \\
Average HOM power per cavity & W & 98 & 150 & 233 & - \\
\end{tabular}
\end{center}
\caption[RF Parameters for the CERN-based RCS Acceleration Chain.]{RF Parameters for the CERN-based RCS acceleration chain. For the synchronous phase, \SI{90}{\degree} is defined as being on-crest. All other assumptions are discussed in \ref{rf:sec:high}. }
\label{site:tab:rf_rcs_cern}
\end{table}
\FloatBarrier
\section{Appendix: Power Converters}
\label{power:sec}
%% ==== START TentativeChapters/Author14-PowerConverters ==== %%
\subsection{Resistive magnets equivalent circuital model}
\label{power:sec:magnet_circuit}
At the present state, we are considering that all the resistive magnet length is occupied by dipole magnets. This approach is conservative for the power converters because dipole magnets have the largest energy density. The remainder of this chapter focuses exclusively on the resistive pulsed dipole magnets. Based on the preliminary magnet designs, two representative configurations are used for sizing the power converters across all accelerator scenarios:

\begin{itemize}
\item \textbf{Dipole 1:} \quad $L_{\text{mag}} = 95~\mu\text{H/m}$,\quad $R_{\text{mag}} = 0.93~\text{m}\Omega/\text{m}$ 
\item \textbf{Dipole 2:} \quad $L_{\text{mag}} = 95~\mu\text{H/m}$,\quad $R_{\text{mag}} = 0.41~\text{m}\Omega/\text{m}$
\end{itemize}

Using these parameters, the corresponding peak voltages and powers required from the power converters are computed. The results are reported in Table~\ref{rf:tab:PC_table1_CERN_Scenario} for the CERN scenario and Table~\ref{rf:tab:PC_table2_GF_Scenario} for the Green Field scenario.
\begin{table}[!h]
\begin{center}
\caption{Important dimensioning values for the power converters of the RCS, CERN scenario}
\label{rf:tab:PC_table1_CERN_Scenario}
        \begin{tabular}{l|ccc}
        \hline\hline
             & RCS & RCS & RCS\\
              & SPS &  LHC1 & LHC2 \\ \hline
             Length NC magnets [m]  & 4103 & 18650 & 12940\\
             Max B [T] & 1.8 & 1.8 & 1.8\\
             Gap dimensions [mm] & 100 x 30 & 100 x 30 & 100 x 30\\
             Acceleration time [ms] & 0.45 & 2.60 & 4.42\\
             Capacitor Energy [MJ] & 43 & 164 & 71\\
             Magnetic Energy [MJ] & 26 & 96 & 56\\
             PC Inductive pk Voltage [MV] & 10.0 & 7.9 & 6.4\\
             PC Resistive pk Voltage [MV] & 0.029 & 0.133 & 0.092\\
             Resistive / Inductive ratio [\%] & 0.29 & 1.7 & 1.4\\
             PC pk current[kA] & 12 & 12 & 12\\
             PC pk Power [GW] & 110 & 87 & 70\\
             duty cycle [\%] & $\approx 0.45$ & $\approx 2.8$ & $\approx 4.42$\\
             \hline\hline
             Pulse 2 Pulse repeatability @+- 2 sigma [ppm] & >=100& >=100&>=100\\
            Control accuracy [ppm] & >=100& >=100 &>=100\\
\end{tabular}
\end{center}        
\end{table}
\begin{table}[!h]
\begin{center}
\caption{Important dimensioning values for the power converters of the RCS, Green Field scenario}    
\label{rf:tab:PC_table2_GF_Scenario}
        \begin{tabular}{l|cccc}
        \hline\hline
             & RCS1 & RCS2 & RCS3 & RCS4\\ 
              &  &  & & \\ \hline
             Length NC magnets [m]  & 3654 & 2539 & 4366 & 20376\\
             Max B [T] & 1.8 & 1.8 & 1.8 & 1.8\\
             Gap dimensions [mm] & 100 x 30 & 100 x 30 & 100 x 30 & 100 x 30\\
             Acceleration time [ms] & 0.34 & 1.10 & 2.37 & 6.37\\
             Capacitor Energy [MJ] & 51 & 17 & 25 & 112\\
             Magnetic Energy [MJ] & 31 & 13 & 20 & 88\\
             PC Inductive pk Voltage [MV] & 11.8 & 5.0 & 4.0 & 7.0\\
             PC Resistive pk Voltage [MV] & 0.026 & 0.018 & 0.031 & 0.146\\
             Resistive / Inductive ratio [\%] & 0.22 & 0.36 & 0.77 & 2.08\\
             PC pk current[kA] & 12 & 12 & 12 & 12\\
             pk Power [GW] & 130 & 55 & 44 & 77\\
             duty cycle [\%] & $\approx 0.34$ & $\approx 1.1$ & $\approx 2.37$ & $\approx 6.37$\\
             \hline\hline
            Pulse to pulse repeatability @+- 2 sigma [ppm] &>=100& >=100&>=100 &>=100\\
            Control accuracy [ppm] & >=100& >=100&>=100 &>=100\\
        \end{tabular}
\end{center}
\end{table} 
\FloatBarrier
\subsection{Partition of the total power into different sectors}
\label{power:sec:sectors}
Given the extremely high voltage and power levels required across the full accelerator, it becomes necessary to divide the system into many sub-converters, referred to as Power Electronics cells (PE cells).
In addition to the scale of the electrical power, an important challenge lies in achieving accurate current control across all PE cells—particularly in systems where the cells operate independently, as in the LHC sector model. For example, in the full-wave resonant topology, implementing current control would require each PE cell to be equipped with a fast, high-power active filter. This adds substantial complexity and cost, especially when regulation must occur within less than 1~ms.

An alternative approach, inspired by the CERN SPS configuration, is to connect all PE cells in series within a single circuit. This guarantees the same current through each cell, simplifying control to only ensuring repeatability from one pulse to the next.
\begin{figure}[h]
  \centering
  \includegraphics[width=\linewidth]{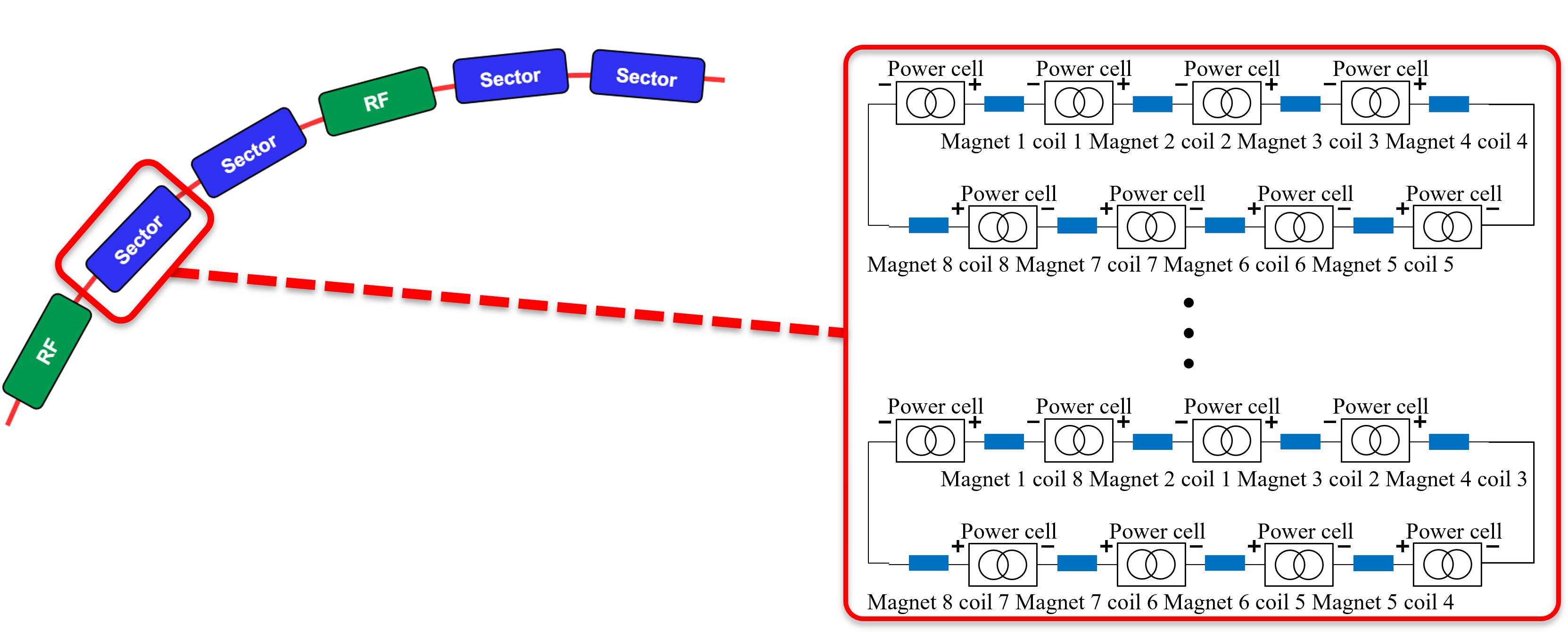}
    \caption{subdivision of he the total power in sectors (LHC style)}
    \label{fig:PC_Sectors}
\end{figure}
\begin{figure}
    \centering
    \includegraphics[width=\linewidth]{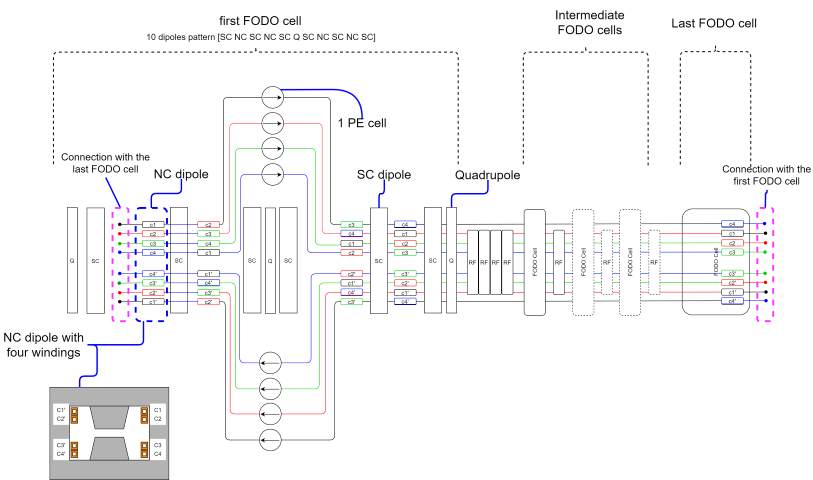}% <- your other file
    \caption{subdivision of he the total power in series (SPS style)}
    \label{fig:PC_Serier}
\end{figure}

\subsection{Power cell topologies}
\label{power:sec:topologies}
\begin{figure}[h]
  \begin{center}
    \includegraphics[width=0.8\textwidth]{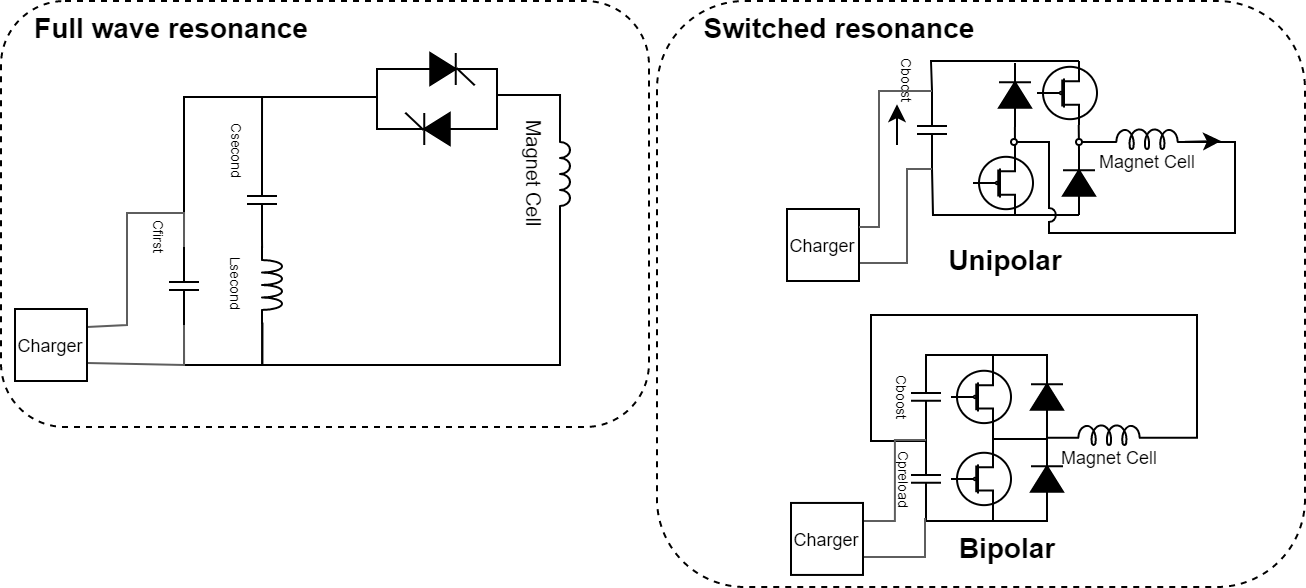}
  \end{center}
\caption{resonating circuits: full wave resonance (left), switched resonance (right)}
\label{fig:PC_ResonatingCircuits}
\end{figure}

To meet the high power and voltage demands of the RCS, the power converters operate using pulsed resonant circuits. Two main types are considered: the full-wave resonance circuit and the switched resonance circuit. These configurations are illustrated in Figure \ref{fig:PC_ResonatingCircuits}. 
In this method, a natural resonance is triggered by connecting one or more pre-charged capacitor bank to the magnets, thus initiating an RLC resonance. This process can be repeated at a frequency of five Hz, aligning with the desired repetition rate. The circuits represented in the figure are identified as "Full wave resonance" (left) and "Switched resonance" (right).
Both topologies rely on pre-charging one or more capacitors to an initial voltage, followed by activating a switch to discharge the energy into the load. As the load is almost purely inductive, the capacitors are nearly fully recharged at the end of the pulse. The switch is then opened and remains off until the next pulse cycle begins. The pulse typically lasts a few percent of the total repetition period, which is approximately 200~ms. 

The full wave resonance can be composed by either one, two or more parallel branches resonating with the magnets. One switch per branch ignite the resonance startig from pre-loaded capacitors and returning to the initial value (minus the losses of the system) at the end of the oscillation. 

\begin{figure}[h]
  \begin{center}
    \includegraphics[width=0.9\textwidth]{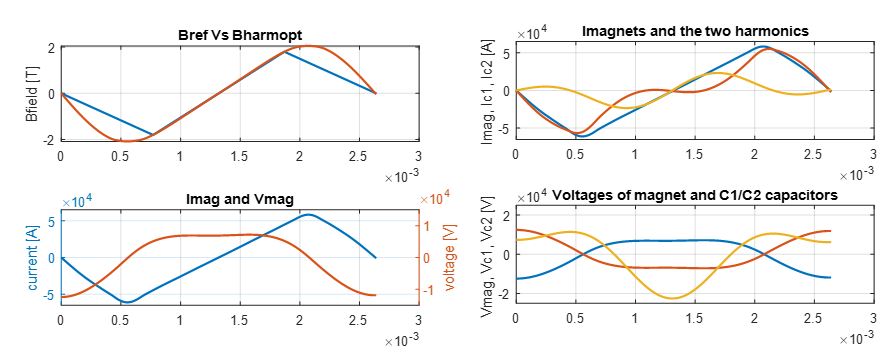}
  \end{center}
\caption{full wave resonance example}
\label{fig:PC_fullwaveexample}
\end{figure}

Fig.~\ref{fig:PC_fullwaveexample} show an example of a full wave resonance with two harmonics, the fundamental and the second. Because of the additional harmonics the total installed capacitive and inductive energy, is much higher that the energy required by the magnet at peak flux density. In addition the discharge of the capacitors is bipolar which poses significant overdimensioning constraints to the capacitors

\begin{figure}[h]
  \begin{center}
    \includegraphics[width=0.6\textwidth]{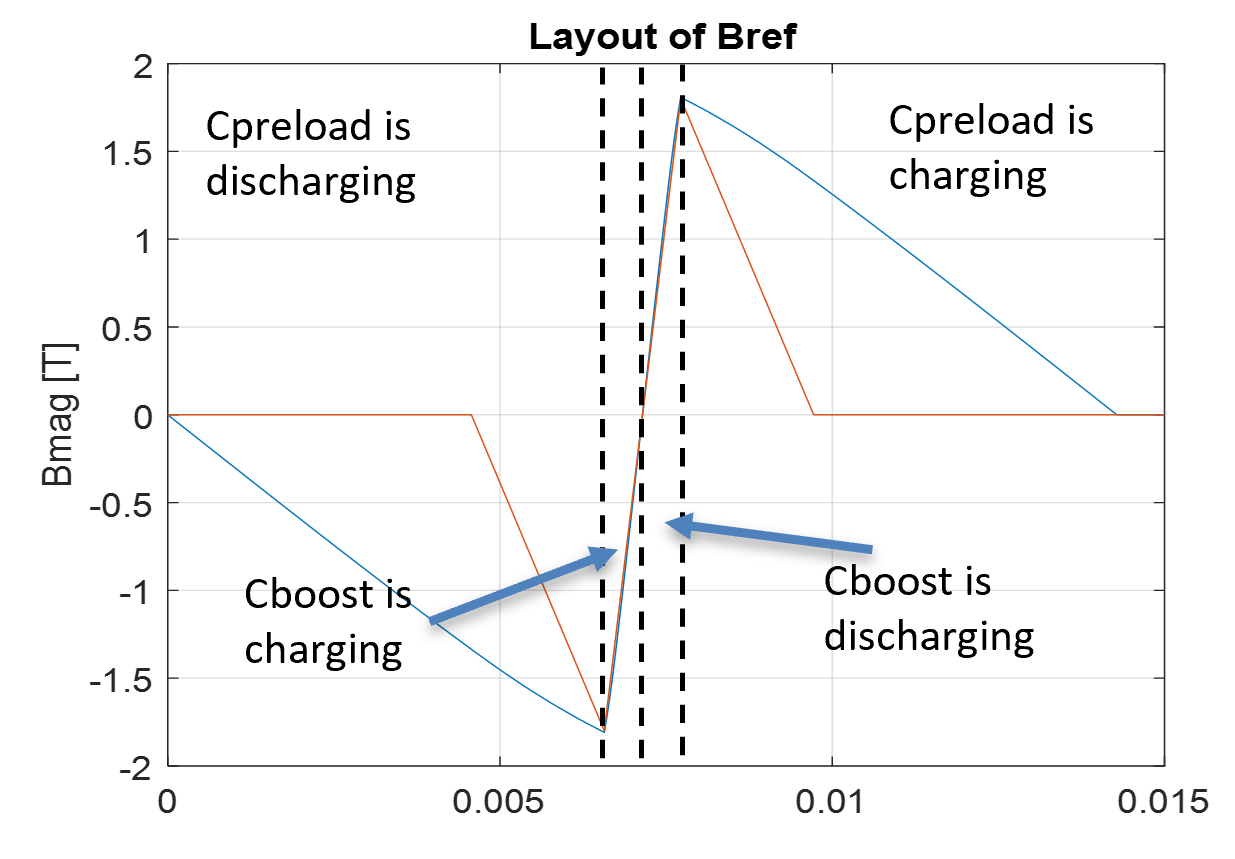}
  \end{center}
\caption{Switched resonance principle}
\label{fig:PC_SwitchedParts}
\end{figure}

In the switched resonance circuit, a distinct approach is employed. This method leverages two distinct simple resonances for different segments of the resonating wave, as depicted in Fig.~\ref{fig:PC_SwitchedParts}. To initialize the pulse, the preload capacitor (see Fig.~\ref{fig:PC_ResonatingCircuits}, right panel) is engaged until the current attains the target negative value. Subsequently, the current pathway is altered by toggling the switches S1 and S2, bringing the boost capacitors into operation.

\begin{figure}[H]
  \begin{center}
    \includegraphics[width=0.9\textwidth]{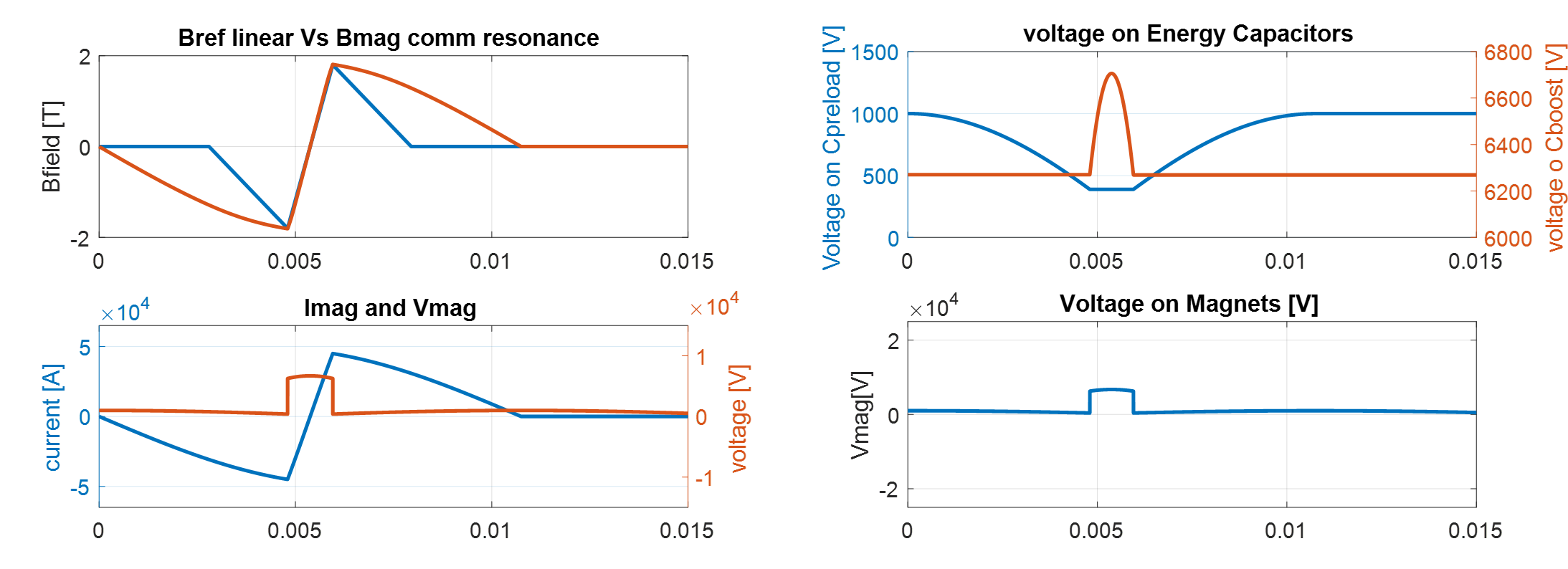}
  \end{center}
\caption{Switched resonance example}
\label{fig:PC_Switchedexample}
\end{figure}

An example of transient with the switched resonance circuit is shown in Fig. \ref{fig:PC_Switchedexample}.

\subsection{The control problem}
Tables \ref{rf:tab:PC_table1_CERN_Scenario} and \ref{rf:tab:PC_table2_GF_Scenario} show some parameters related to the quality of control. In particular they refer to:
\noindent\textbf{\underline{Pulse-to-pulse repeatability @ $\pm 2\sigma$}:}\\
This means that on 95\% of the pulses, the maximum absolute difference between the current at any time instant of any pulse and the average of them is less than 100\,ppm.

\noindent\textbf{\underline{Control accuracy:}}\\
This means that if we have two separate circuits and we need the currents to be the same in both of them, the control will not be able to make it more accurately than that.
In both cases the tables report a target value rather than a limit. We still don't know the statistical parameters of the charger and the jitter of the IGBT+driver; therefore it is difficult to say if we can fit in. Simulations with educated guess values, show, nevertheless, that the reported values would represent a limit with respect to what it is possibly achievable.
%% ==== END TentativeChapters/Author14-PowerConverters ==== %%
\FloatBarrier
\section{Appendix: Impedance}
\label{app:imp}

Transverse HOMs generated by the TESLA cavities would be the main impedance source for the RCS chain.
Table~\ref{impedance:tab:HOMS_t} details the shunt impedance, quality factor and resonance frequency of the HOMs considered.
Transverse coherent stability simulations were performed to evaluate the impact of the RF cavities and vacuum chambers.
To mitigate the instabilities, a transverse damper system can be used to damp the transverse centroid motion of the bunches, and/or chromaticity can be introduced with sextupoles.
Parametric scans were performed to find if those are needed and, if necessary, the chromaticity $Q^{\prime}$ required.
The chromaticity was scanned from $Q^{\prime}=-20$ to $Q^{\prime}=+20$, and the transverse damper from a 4-turn to a 100-turn damping time, with an additional case without damper.

\begin{table}[!h]
    \centering
    \begin{tabular}{cccc}
         Frequency $f_{res}$ &  $\frac{R_s}{Q}$ &  Q factor & Shunt impedance $R_s$\\
         \si{\giga\hertz} & [\si{\kilo\ohm\per\meter}] & [\num{1E4}] & [\si{\mega\ohm\per\meter}]\\ \hline
         1.659&    0.10&    \num{31.4}    &   32.61\\
         1.705&    1.05&    \num{1.35}    &   14.16\\
         1.706&    1.21&    \num{1.34}    &   16.27\\
         1.728&    0.97&    \num{0.0413}  &   0.4\\
         1.729&    0.45&    \num{0.0381}  &   0.17\\
         1.736&    1.25&    \num{0.0516}  &   0.64\\
         1.737&    0.95&    \num{0.0574}  &   0.54\\
         1.761&    0.35&    \num{0.583}   &   2.04\\
         1.762&    0.28&    \num{0.621}   &   1.72\\
         1.788&    0.16&    \num{0.867}   &   1.43\\
         1.789&    0.18&    \num{0.890}   &   1.61\\
         1.798&    0.11&    \num{1.23}    &   1.29\\
         1.799&    0.10&    \num{1.21}    &   1.27\\
         1.865&    0.79&    \num{3.91}    &   30.87\\
         1.865&    0.83&    \num{4.12}    &   34.07\\
         1.874&    1.09&    \num{3.88}    &   42.32\\
         1.874&    1.07&    \num{4.39}    &   47.14\\
         1.88&     0.22&    \num{4.23}    &   9.38\\
         1.88&     0.24&    \num{5.15}    &   12.21\\
         2.561&    0.13&    \num{0.0620}  &   0.08\\
         2.561&    0.12&    \num{0.0527}  &   0.07\\
         2.577&    2.05&    \num{0.364}   &   7.46\\
    \end{tabular}
    \caption{HOMs from TESLA cavity, complete table, for a single cavity.}
    \label{impedance:tab:HOMS_t}
\end{table}

Tracking simulations were performed using Xsuite~\cite{impedance:bib:xsuite} and PyHEADTAIL~\cite{bib:pyheadtail}.
The bunch motion is simulated through the complete RCS chain.
Muon decay is not included in these simulations, therefore the bunch intensity remains constant through the chain, equal to the intensity of \num{2.7e12}~muons per bunch at injection in RCS 1.
Results showed that a positive chromaticity of $Q^{\prime} = +20$ is needed in the accelerators to stabilize the beams and leave enough margin for some initial transverse offset of the bunches, and a 20-turn transverse damper also helps stabilize the beams~\cite{impedance:bib:amorim_rcs_imcc_2024, bib:amorim_tesla_vs_low_loss}.

\begin{table}[!h]
\centering

\begin{tabular}{l c c c c}
& RCS1& RCS2& RCS3&RCS4\\
\hline
Number of cavities & 700  & 380& 540&3000\\

\end{tabular}
\caption{RCS impedance model assumption for number of TESLA cavities}
\end{table}

\subsection{Impedance model for the 10 TeV collider ring}
\label{app:imp:coll}

In the \SI{10}{\tera\eV} collider ring, the main impedance source would be the resistive-wall contribution from the magnets' vacuum chamber.
To protect the superconducting magnet coils from muon decay induced heating and radiation damage, a tungsten shield is proposed to be the inserted in the magnet cold bore as detailed in Section~\ref{sec:radshield} and described in Ref.~\cite{bib:shielding_requirements}.

Previous parametric studies performed with Xsuite and PyHEADTAIL showed that a minimum chamber radius of \SI{13}{\milli\metre}, together with a copper coating on the inner diameter are required to ensure coherent transverse beam stability.
The current dipole magnet radial build detailed in Table \ref{col:tab:arcs} foresees a \SI{23.5}{\milli\metre} inner radius, with a \SI{10}{\micro\metre} copper coating.
The vacuum chamber properties used for the impedance model computation are summarized in Table~\ref{impedance:tab:collider}.

A particularity of the collider ring is its isochronous operation (i.e. with $\eta\approx0$)~\cite{impedance:bib:ng_quasi_isochronous_buckets}, obtained with the flexible momentum compaction cells described in Section~\ref{col:sec}.
This is to avoid the large RF voltage that would be needed to bunch beams with very short length and large energy spread.
However this freezes the synchrotron motion of the particles within the bunch and can lead to beam breakup instabilities such as those encountered in Linacs~\cite{impedance:bib:kim_transverse_instability}.

Transverse coherent beam stability simulations were performed with Xsuite and PyHEADTAIL, including the effect of muon beam decay~\cite{impedance:bib:amorim_coll10tev_imcc_2024}.
The beam parameters used for these simulations are summarized in Table~\ref{impedance:tab:collider_param}.
With a chromaticity of $Q^{\prime} = 0$, the beam becomes unstable over its lifetime in the collider, leading to large transverse emittance growth~\cite{impedance:bib:amorim_coll10tev_imcc_2024}.
A slightly positive chromaticity of $Q^{\prime} = +2$ is needed to introduce a betatron frequency spread that helps stabilize the beam.

% Assuming that the innermost surface seen by the beam is the tungsten shield, impedance models were created for a range a radius between \SI{10}{\milli\metre} and \SI{40}{\milli\metre}.
% A scan on the copper coating thickness was also performed, from \SI{0.1}{\micro\metre} to \SI{100}{\micro\metre}.
% The impedance models were generated using the ImpedanceWake2D code~\cite{bib:iw2d_website}.
% Simulation parameters are reported in Table~\ref{impedance:tab:collider}.

\begin{table}[!ht]
    \centering
    \begin{tabular}{cc|c}
         Parameter&  Unit& Value\\ \hline
         Circumference &   \si{\meter}& 10 000\\
         Beam energy &  \si{\tera\electronvolt} & 5\\
         Bunch intensity at injection &  muons/bunch & \num{1.80E+12}\\
         1$\sigma$ bunch length& \si{\milli\meter} & 1.5\\
         Longitudinal emittance $\epsilon_l = \sigma_z \sigma_E$ &  \si{\mega\electronvolt\meter} & 7.5\\
         Transverse normalized emittance&  \si{\micro\meter\radian} & 25\\
         Momentum compaction factor&  & 0\\
         Total RF voltage & \si{\mega\volt} & 0\\
    \end{tabular}
    \caption{\SI{10}{\tera\electronvolt} collider machine and beam parameters.}
    \label{impedance:tab:collider_param}
\end{table}

\FloatBarrier
\section{Appendix: Demonstrators}
\label{app:demo}

\subsection{CTF3 Building background and current use}

The TT7 option has been extensively studied in 2024, the results of which are shown in Table \ref{demo:tab:TT7}. Civil engineering studies reveal this option to be more expensive and complex than initially expected. for this reason we launched a new study to explore the suitability of reusing the CTF3 building to host the demonstrator facility. The reason not to consider it as a first instance was the fact that at the moment there is not an existing extraction system in the CERN PS that could send beam towards CTF3, in contrast to TT7 where the simple installation of a dipole in a transfer line would have provided an option with less impact on operating machines. 

\begin{table}[!h]
\centering
\begin{tabular}{l lcc}
\hline \hline
\textbf{Area} & \textbf{Parameter Name} & \textbf{Value} & \textbf{Unit} \\
\hline
Proton Beam & Beam Energy & 14 & GeV \\
 & Protons/pulse & \num{1E+13} & Protons \\
 & Pulse rep. rate & 0.064 & Hz (15.6s) \\
 & Avg beam power & 1.5 & kW \\
 & Avg beam power (target assumption) & 5 & kW \\
 & RMS pulse length & 7.65 & ns \\
\hline
Proton transfer line & Number of dipoles & 4 (H), 4 (V) & - \\
 & Number of quadrupoles & 7 & - \\
 & Number of correctors & 5 & - \\
\hline
Target & Proton Beam Energy & 14 & GeV \\
 & Proton Beam RMS size & 2 & mm \\
 & Target Material & Graphite & - \\
 & Target Length & 90 & cm \\
 & Target Radius & 0.6 & cm \\
 & Horn Current & 220 & kA \\
 & Ltot & 200 & cm \\
 & Target Pion Momentum Range & 210 -- 330 & MeV/c \\
 & Target Pion $\epsilon_T$ acceptance & 2 & mm rad \\
 & Simulated Pion Yield per POT & \num{7.90E-04} & - \\
 & Remote handling & YES & - \\
\hline
Decay channel & Decay Channel Lattice & 3 quad triplets & - \\
\& magnetic chicane & Decay Channel Length & 9.5 & m \\
 & Pion Momentum (Nominal) & 270 & MeV/c \\
 & Pion Momentum Acceptance & ±50\% & \% \\
 & Target Muon Momentum Range & 190 -- 210 & MeV/c \\
 & Target Muon $\epsilon_T$ acceptance & 2 & mm·rad \\
 & Chicane Type & 3-bend & - \\
 & Dispersion at Chicane En & -0.4 & m \\
 & $\beta$ at BPS Injection & 3 & m \\
\hline
Beam preparation system & Number of RF cavities & 16 & - \\
 & RF peak gradient & 15 & MV/m \\
 & RF phase & 0 & degrees \\
 & RF frequency & 704 & MHz \\
 & Dipole field & 0.67 & T \\
 & Dipole length & 1.04 & m \\
\hline \hline

\end{tabular}
\caption{TT7 Beamline parameters}
\label{demo:tab:TT7}
\end{table}

The CTF3 building hosted the LIL (Linear Injector of LEP) machine and was later dedicated to the experimental activities around the CLIC study for linear colliders. Although there is no extraction to it, CTF3 has already many characteristics that are needed for the Demonstrator, namely sufficient length that would be sufficient not only for the facility but also for eventual future extensions, a Klystron gallery and all the infrastructure and services necessary to operate such a facility. 

Moreover, as former building for the injector of LEP, it has a connection to the PS tunnel and therefore no major civil engineering works will be needed to reconnect it to the PS. Only a well-shielded target area shall have to be created, with therefore the hope that costs and efforts can be mostly concentrated on the components of interest.
The beamline parameters of a muon cooling demonstrator at the CTF3 facility are displayed in Table \ref{demo:tab:ctf3_beamline}.

Today CTF3 hosts the CLEAR facility, whose continuation is fully compatible with the new facility. 

\begin{table}[!h]
\centering
\begin{tabular}{llcc}
\hline \hline
\textbf{Area} & \textbf{Parameter Name} & \textbf{Value} & \textbf{Unit} \\
\hline
Proton Beam & Beam Energy & 14 & GeV \\
 & Protons/pulse & \num{1E+13} & Protons \\
 & Pulse rep. rate & 0.064 & Hz (15.6s) \\
 & Avg beam power & 1.5 & kW \\
 & Avg beam power (target assumption) & 5 & kW \\
 & RMS pulse length & 7.65 & ns \\
\hline
Extraction & Extraction dipole (New, after PS septa) & 1 & - \\
 & Bumper magnets & 4 & - \\
 & Septa & 2 & - \\
 & KFA71 Kicker & 1 & - \\
\hline
Proton transfer line & Number of dipoles & 1 & - \\
 & Number of quadrupoles & 5 & - \\
 & Number of correctors & TBD & - \\
\hline
Target & Proton Beam Energy & 14 & GeV \\
 & Proton Beam RMS size & 2 & mm \\
 & Target Material & Graphite & - \\
 & Target Length & 90 & cm \\
 & Target Radius & 0.6 & cm \\
 & Horn Current & 220 & kA \\
 & Ltot & 200 & cm \\
 & Target Pion Momentum Range & 210 -- 330 & MeV/c \\
 & Target Pion $\epsilon_T$ acceptance & 2 & mm rad \\
 & Simulated Pion Yield per POT & \num{7.90E-04} & - \\
 & Remote handling & YES & - \\
\hline
Decay channel & Decay Channel Lattice & 3 quad triplets & - \\
\& magnetic chicane & Decay Channel Length & 9.5 & m \\
 & Pion Momentum (Nominal) & 270 & MeV/c \\
 & Pion Momentum Acceptance & ±50\% & \% \\
 & Target Muon Momentum Range & 190 -- 210 & MeV/c \\
 & Target Muon $\epsilon_T$ acceptance & 2 & mm·rad \\
 & Chicane Type & 2-bend & - \\
 & Dispersion at Chicane End & 0 & m \\
 & $\beta$ at BPS Injection  & <1 & m \\
\hline
Beam preparation system & Number of RF cavities & 16 &  \\
 & RF peak gradient & 15 & MV/m \\
 & RF phase & 0 & degrees \\
 & RF frequency & 704 & MHz \\
 & Dipole field & 0.67 & T \\
 & Dipole length & 1.04 & m \\
\hline
Matching section & TBD & - & - \\
\hline \hline

\end{tabular}
\caption{CTF3 Beamline parameters}
\label{demo:tab:ctf3_beamline}
\end{table}

\subsection{Proposed demonstrator layout at CTF3:}

\begin{figure}[!h]
    \centering
    \includegraphics[width=1\linewidth]{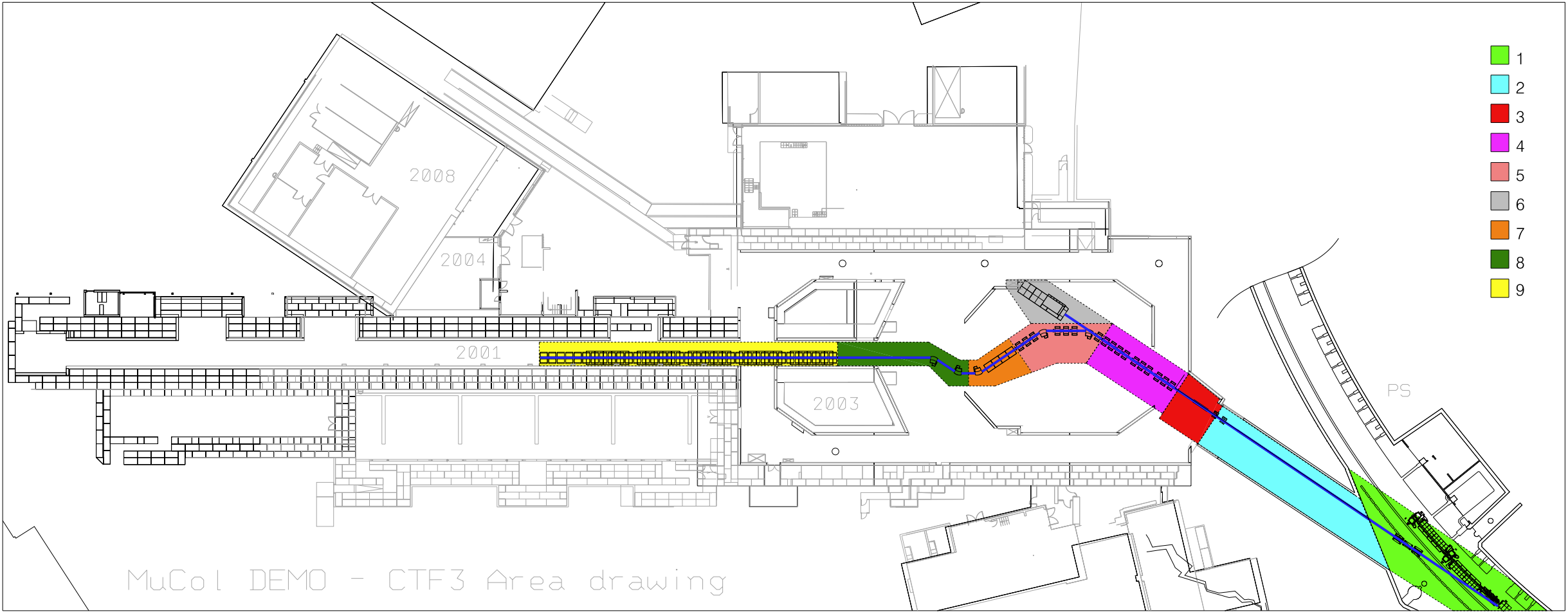}
    \caption{CTF3 Area Definition: Proton extraction (light green), Proton Transfer line (cyan), Target area (red), pion decay channel (magenta), magnetic chicane (peach), proton and pion dump (grey), beam preparation system (orange), matching section (dark green), cooling cell channel (yellow).}
    \label{demo:fig:CTF3_area}
\end{figure}

\begin{table}[!h]
\centering
\begin{tabular}{llcc}
\hline \hline
\textbf{Area} & \textbf{Parameter Name} & \textbf{Value} & \textbf{Unit} \\
\hline
Proton transfer line & Gallery length & 22.3 & m \\
 & Gallery width & 5.3 & m \\
 & Gallery height & 3 & m \\
\hline
Target area & Gallery length & 5 & m \\
 & Gallery width & 11 & m \\
 & Gallery height & 3 & m \\
\hline
Pion Decay channel & Gallery length & 9 & m \\
 & Gallery width & 3 & m \\
 & Gallery height & 5 & m \\
\hline
Magnetic chicane & Gallery length & 11 & m \\
 & Gallery width & 3 & m \\
\hline
Beam Preparation system & Gallery length & 5 & m \\
 & Gallery width & 3 & m \\
\hline
Matching section & Section length & 14 & m \\
\hline
Cooling channel & Channel length & 36 & m \\
 & Channel width & 3 & m \\
 & Channel height & 2.5 & m \\
 & Number of cooling cells & 30 & - \\
\hline
Klystron gallery & Gallery length & 41.5 & m \\
 & Gallery width & 6.5 & m \\
 & Gallery height & 5 & m \\
 & Number of klystrons & 15 & - \\
 \hline \hline 
\end{tabular}
\caption{CTF3 Site-specific parameters}
\label{demo:tab:ctf3_gallery}
\end{table}

\begin{itemize}
    \item Extraction. Protons would be extracted from the PS and transported to CTF3 by reopening and adapting the historic link.
    \item Floor usage. Depicted in Figure \ref{demo:fig:CTF3_area}:
    \begin{itemize}
        \item The lower floor hosts the proton transfer line, target station, pion decay channel, magnetic chicane, beam-preparation system, matching section and the cooling-cell gallery.
        \item The upper floor (gallery) hosts klystrons, with waveguide penetrations to the cooling cell gallery—reusing existing infrastructure instead of building a new surface hall. Details of the CTF3 klystrons are in Table \ref{demo:tab:ctf3_klystron}.
        \item Target area. Located around the former combiner-ring centre; the layout permits the construction of a target area with limited works and independent access relative to CLEAR.
    \end{itemize}
\end{itemize}

\begin{table}[!h]
\centering
\begin{tabular}{llcc}
\hline \hline
\textbf{Area} & \textbf{Parameter Name} & \textbf{Value} & \textbf{Unit} \\
\hline
Klystron & RF Frequency & 704.4 & MHz \\
 & Modulator & Scandinova K200 &  \\
 & Klystron Voltage & 125 & kV \\
 & Klystron Current & 242 & A \\
 & Input power & 1320 & W \\
 & klystron Efficiency & 0.75 &  \\
 & Modulator output power & 30.25 & MW \\
 & Klystron Output Power & 22.69 & MW \\
 & RF system rep. rate & 5 & Hz \\
 & RF pulse length & 15 & us \\
 & High voltage pulse length & 17 & us \\
 & RF power average & 1701.56 & W \\
 & Modulator efficiency & 0.9 &  \\
 & modulator power consumption & 2856.94 & W \\
 & klystron solenoid power consumption & 5000 & W \\
 & Total average power comsumption & 7856.94 & W \\
\hline \hline
 \end{tabular}
\caption{CTF3 Klystron parameters}
\label{demo:tab:ctf3_klystron}
\end{table}
    
\subsubsection{Work underway - Scope of current studies}

\begin{itemize}
    \item Beamline \& optics. End-to-end lattice from PS extraction to the target; decay channel, momentum-selection chicane and BPS matched to the cooling section (baseline “B5-like” cell).
    \item Integration \& access. 3D integration of the tunnel straight and gallery; installation/maintenance scenarios; co-existence planning with CLEAR.
    \item Assembly/disassembly. Removal of remaining CLIC demonstrator hardware where needed; definition of transport paths and lifting means.
    \item Civil engineering. Reopening and adapting the PS link; localized works for the target area, access enlargements and main patio improvement; no enlargement of the main gallery and no new surface building are foreseen.
    \item Services. Reuse of surrounding power, cooling-water and ventilation with targeted upgrades; RF plant staged in the gallery.
    \item Radiation protection. RP modelling of the target/shielding and an operational zoning scheme that does not interfere with CLEAR. 
\end{itemize}

\subsubsection{How CTF3 compares to TT7}

\begin{itemize}
    \item What TT7 would need: To host the same demonstrator, TT7 would require tunnel enlargement ($\approx$+3 m width and $+$1–1.4 m height over tens of meters) and a new surface building (klystrons and services), with access road modifications.
    \item What CTF3 avoids: CTF3 already provides the straight tunnel and the klystron gallery; only local, low impact works (PS link, access improvements, patio/penetrations, target area) are foreseen. Installation can be scheduled outside the accelerator access chain, with no apparent interference with CLEAR. Table \ref{demo:tab:ctf3_gallery} displays the general CTF3 area dimensions.
\end{itemize}

\FloatBarrier

\subsection{Net assessment}

CTF3 is simpler and more cost-efficient in every major aspect except one: the new PS extraction/transfer to CTF3, which is the principal project challenge and integration work.\FloatBarrier
\section{Appendix: CERN Civil Engineering}
The Collider Complex is displayed in Figure~\ref{fig_Muon_Collider_Complex}, presented below. An injector complex has been designed and implemented, initiating at LINAC~4 and ultimately, injecting into a new 10\,km Collider Ring from the LHC. The LINAC 4, SPL and ARCR (Accumulator Ring Compressor Ring) are aiming to be equivalent to the Proton Driver as described in Section~\ref{sec:proton}. The entirety of the complex's surface works would be constructed on CERN land across both the Meyrin and Pr\'{e}vessin sites, minimising territorial and environmental impacts.

\begin{table}[!h]
\centering
\begin{tabular}{lcl} \hline\hline
 \textbf{Structure} & \textbf{Length (m)} & \textbf{Cross Section} \\
\hline
 \multicolumn{3}{l}{\textit{LINAC 4}} \\
 \multicolumn{3}{l}{SPL} \\
 SPL to SPS Transfer & 650 & \diameter 4m \\
 \multicolumn{3}{l}{SPS sector} \\
 Transfers to Pr\'{e}vessin & 930 & \diameter 4m \& Single Width Tunnel \\
 ARCR & 628 & \diameter 200m Ring, Double Width Tunnel \\
 \hline
 Target & 50 & 50m x 30m \\
\hline
 Cooling & 1000 & Double Width Tunnel w/ Surface Structure \\ \hline
 Cooling to SC LINAC Transfer & 100 & Single Width Tunnel \\
 SC LINAC & 200 & Double Width Tunnel w/ Surface Structure \\
 SC LINAC to RLA 1 Transfer & 110 & Single Width Tunnel \\
 Racetrack (RLA 1) & 700 & Single Width Tunnel w/ Surface Structure \\
 RLA 1 to RLA 2 Transfer & 600 & Single Width Tunnel\\
 Racetrack (RLA 2) & 2300 & Single Width Tunnel w/ Surface Structure \\
 RLA 2 to SPS Transfer Lines (2: $\mu^+, \mu^-$) & 1010 & \diameter 4m  \\
\hline
 \multicolumn{3}{l}{\textit{SPS}} \\
 SPS to LHC Transfer Lines (\textit{TI12}) & 536 & \diameter 3.5m  \\
 SPS to LHC Transfer Lines (\textit{TI18}) & 258 & \diameter 3.5m  \\
\hline
 \multicolumn{3}{l}{\textit{LHC}} \\
 LHC to Collider Ring Transfer & 4012 & \diameter 4m  \\
\hline
 MUON Collider Ring & 10000 & \diameter 5.5m  \\
\hline\hline

\end{tabular}
\caption[Muon Collider Sequence at CERN]{Muon Collider Sequence at CERN. (\textit{Italics} shows existing tunnels). "Single Width" refers to a (5m x 4m) tunnel, whereas "Double width" refers to a (8m x 4m) tunnel.}
\end{table}

\begin{figure}[h!]
\centering
\includegraphics[width=\textwidth]{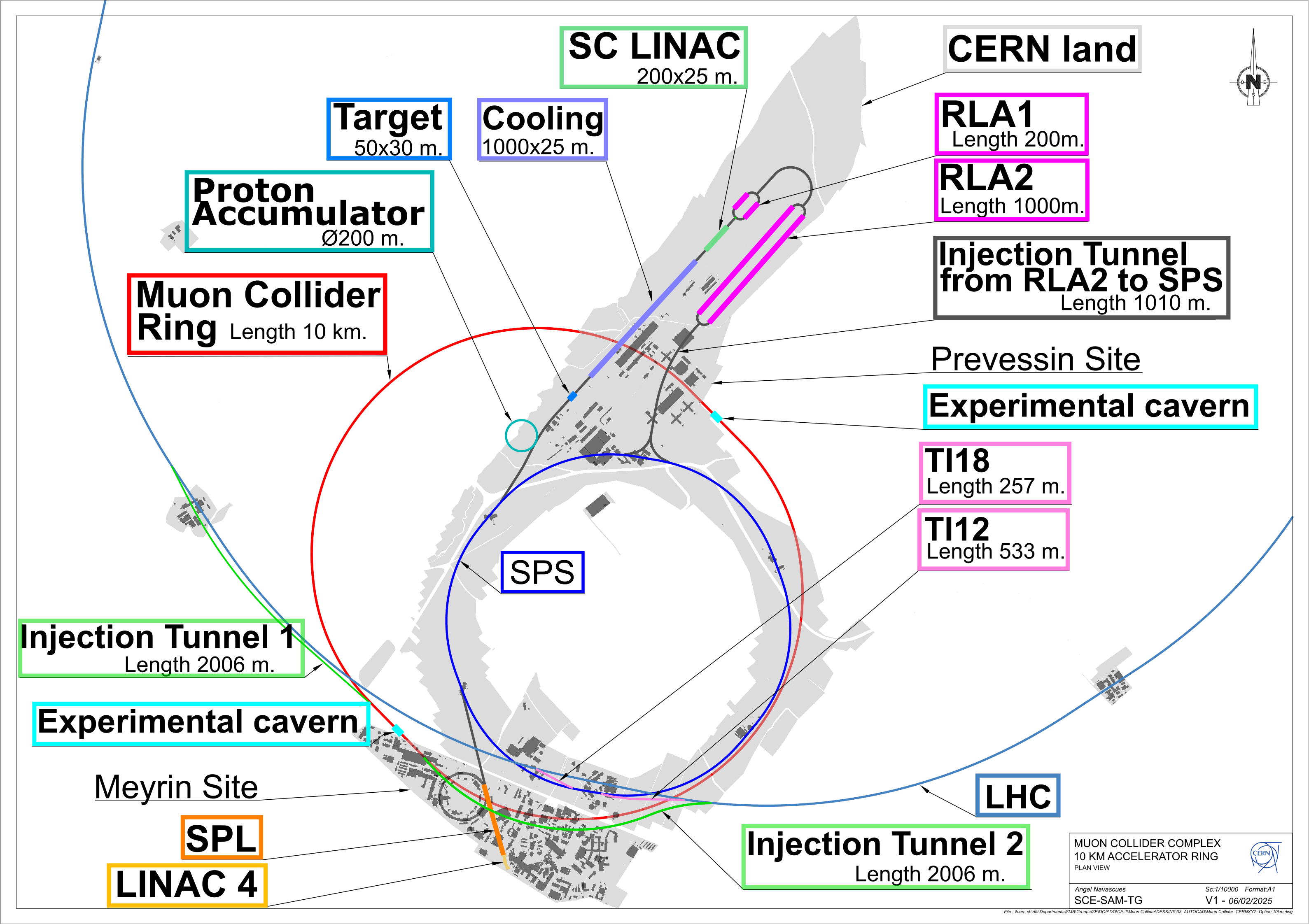}
\caption{\label{fig_Muon_Collider_Complex} Muon Collider Complex.}
\end{figure}\FloatBarrier

\printbibliography

\end{document}